\documentclass[12pt]{article}
\usepackage{amsmath}
\usepackage{enumerate}
\usepackage{graphicx}
\begin{document}
\def\be{\begin{equation}}
\def\bea{\begin{eqnarray}}
\def\ee{\end{equation}}
\def\eea{\end{eqnarray}}
\def\d{\partial}
\def\vep{\varepsilon}
\def\eps{\epsilon}
\def\la{\lambda}
\def\b{\bigskip}
\def\nn{\nonumber \\}
\def\p{\partial}
\def\t{\tilde}
\def\k{\kappa}
\def\h{{1\over 2}}
\def\be{\begin{equation}}
\def\bea{\begin{eqnarray}}
\def\ee{\end{equation}}
\def\eea{\end{eqnarray}}
\def\b{\bigskip}
\def\u{\uparrow}
\def\AdSS{AdS$_2\times$S$^2$}
\newcommand{\comment}[2]{#2}

\makeatletter
\def\blfootnote{\xdef\@thefnmark{}\@footnotetext}  
\makeatother

\begin{center}
{\LARGE Bubbling geometries for \AdSS}
\\
\vspace{18mm}
{\bf   Oleg Lunin}
\vspace{14mm}

Department of Physics,\\ University at Albany (SUNY),\\ Albany, NY 12222, USA\\ 

\vskip 10 mm

\blfootnote{email: olunin@albany.edu}

\end{center}

We construct BPS geometries describing normalizable excitations of AdS$_2\times$S$^2$. All regular  horizon--free solutions are parameterized by two harmonic functions in ${\bf R}^3$ with sources along closed curves. This local structure is reminiscent of the ``bubbling solutions'' for the other AdS$_p\times$S$^q$ cases, however, due to peculiar asymptotic properties of AdS$_2$, one copy of ${\bf R}^3$ does not cover the entire space, and we discuss the procedure for analytic continuation, which leads to a nontrivial topological structure of the new geometries. We also study supersymmetric brane probes on the new geometries, which represent the AdS$_2\times$S$^2$ counterparts of the giant gravitons.

\newpage

\tableofcontents

\section{Introduction}
\renewcommand{\theequation}{1.\arabic{equation}}
\setcounter{equation}{0}

\AdSS\ has a peculiar status in string theory: while being especially interesting as a near horizon limit of the extremal black holes in four dimensions, this space still evades a holographic description available for its counterparts in higher dimensions \cite{Mald,WittGKP,MAGOO}. Although a significant progress towards formulation of the AdS$_2$/CFT$_1$ correspondence has been made over the last two decades \cite{AdS2CFT}, the complete understanding of the `field theory side' is still missing. Nevertheless, one can apply the methods used to study the bulk side of the AdS/CFT correspondence to \AdSS, and remarkably this space shares some of the nice analytic properties with its higher dimensional counterparts. In particular, integrability of strings, which was discovered for AdS$_5\times$S$^5$ and AdS$_3\times$S$^3$ \cite{Integr5,Integr3}, persists for \AdSS\ as well \cite{STWZ,Integr2}\footnote{In fact, intergrability persists even for the eta--deformation of \AdSS\ \cite{EtaAdS2}, which is analogous to a similar modification of the higher dimensional spaces \cite{EtaGen}.}.  

In this article we will study supersymmetric excitations of \AdSS, which come in several varieties. Very light perturbations correspond to the gravity multiplet, which is included in the analysis of \cite{STWZ}. As the energy of a supersymmetric excitation increases, a better semiclassical description is given in terms of probe branes on \AdSS, and the counterparts of such branes in higher dimensions are known as giant gravitons \cite{SusskGiant,MyersGiant}. When the energy increases even further\footnote{The transition between graviton, brane, and geometry regimes is governed by the scaling of energy with the string coupling constant. This scaling can also be rewritten in terms of number of branes which created the AdS geometry: gravitons have finite energy, branes have energy that scales as $N$, and energy of classical geometries scales as $N^2$.}, the gravitational backreaction of branes becomes important, and one needs to find deviations from the AdS$_p\times$S$^q$ geometry. In the past this problem has been analyzed for $p=3,4,5,7$, where all supersymmetric branes have been classified \cite{Mikhailov,SkenTaylAdS}, all regular geometries preserving half of the supersymmetries have been constructed \cite{Parad,lmm,LLM}, and a progress towards finding 1/4--BPS geometries has been made \cite{3charge,quarter,BenaWarn}. The goal of this article is to extend these successes to supersymmetric excitations of \AdSS. 

In string theory Anti--de-Sitter spaces naturally appear as near--horizon limits of D branes, but in this context one recovers only a part of the AdS geometry known as a Poincare patch. Since this patch is geodesically incomplete, and any point of this geometry is separated from the center of AdS by an infinite distance, such near--horizon limits look somewhat singular. On the other hand, a completion of this space gives the global AdS space, which is smooth everywhere. While the geodesic completion of the bosonic sector is straightforward, this procedure modifies the boundary conditions for fermions, so the relation between string theories on the Poincare patch and on the global AdS is not trivial. Large classes of BPS excitations of the Poincare patch can be constructed by considering several parallel stacks of D branes, and such configurations correspond to the Coulomb branch of the dual field theory \cite{Mald,Coulomb}. 
Not surprisingly, all such solutions have singularities at the locations of the branes, where explicit sources are introduced, but these singularities are infinitely far from any point on the Poincare patch. Such singularity seems unavoidable if one starts from the probe branes and surrounds them by a large sphere: since such sphere carries a quantized flux of an appropriate field strength\footnote{For example, D3 branes in ten non--compact dimensions can be surrounded by $S^5$ which carries a flux of $F_5$, D1--D5 bound states in six non--compact dimensions can be surrounded by $S^3$ carrying a flux of $F_3$, and so on.}, making the sphere smaller and smaller, one arrives at a singular point. In contrast to the Poincare patch, the geometry of the global AdS is completely smooth, and we will now review the mechanism of such regularization in various dimensions.

We begin with the global AdS$_5\times$S$^5$ and its embedding into the `bubbling ansatz' \cite{LLM}. The metric contains two explicit $S^3$ factors, the time, and a three--dimensional base, then the quantized flux is carried by $S^5$, which is constricted by fibering one of the $S^3$ over a two--dimensional surface on the base. Naively, the arguments presented in the last paragraph suggest that a singularity becomes unavoidable since $S^5$ can contract to zero size, but in the bubbling geometries such contraction is avoided since the second sphere ${\tilde S}^3$ collapses causing the space to end while the size of $S^5$ is still finite. 
By taking a singular limit of the bubbling geometries, one can make the regions where ${\tilde S}^3$ collapses arbitrarily small, this leads to recovery of the Poincare patch and to singular geometries produced by multiple stacks \cite{Coulomb}. To summarize, the ten--dimensional bubbling geometries are regularized by a topological mechanism based on termination of space by collapsing one of the spheres.

While a similar mechanism can be applied to some D1--D5 geometries \cite{Vaman}, in general regularization of the D1--D5 system happens for a different reason. At infinity there is a clear separation between the three dimensional sphere $S^3$, which supports the flux, and one of the compact directions. If such separation persisted everywhere, then a singularity would be unavoidable, but, as demonstrated  in \cite{MMaoz}, it is possible to mix the sphere and a compact direction so that the sphere which collapses at the `location of the branes' is not the same $S^3$ that carries the flux. As the result of this construction, the space can end in a smooth fashion without violating the conservation of flux. The `location of the branes' has a smooth geometry of the KK--monopole \cite{MMaoz}, so the sources are completely dissolved in geometry. In \cite{lmm} this construction had been extended to all 1/2--BPS geometries with AdS$_3\times$S$^3$ asymptotics. Interestingly, in the AdS$_3\times$S$^3$ case, one can continuously connect the flat space with global AdS coordinates \cite{MMaoz,Parad}, and this fact has inspired a very promising approach to the resolution of the black hole information paradox known as the fuzzball proposal \cite{Parad,Fuzz}. Regularization of 1/4--BPS geometries is slightly different, and it comes from simultaneous change in signature of the base space and the harmonic functions. The complete picture that works for all such geometries is still missing, and we refer to \cite{3charge,BenaWarn} for further discussion.

So far we have encountered two mechanisms for resolving singularities and making the brane sources geometric: the first one relies on termination of space via the `bubbling mechanism', and the second one is based on dissolving the brane charges in the geometry of the KK-monopole. It turns out that \AdSS\ geometries are regularized in yet another way, which is based on a peculiar property of the global AdS$_2$: in contrast to its higher--dimensional counterparts, this space has a disconnected boundary. In this article we will demonstrate that a bubbling picture with flat base, which worked for AdS$_5\times$S$^5$ and AdS$_3\times$S$^3$, can be extended to the  \AdSS, but to cover this space completely, one needs two copies of the base ${\bf R}^3$, and these copies are connected through a branch cut\footnote{Interestingly, similar branch cuts had been introduced on the {\it boundary side} of the AdS$_2$/CFT$_1$ correspondence to account for the black hole entropy 
\cite{SenCut}. It would be nice to determine whether there is a relation between the cuts of \cite{SenCut} and the ones discussed in this paper.}. This feature is not very surprising since ${\bf R}^3$ has a connected boundary, while the boundary of \AdSS\ contains two disconnected pieces. Notice that by taking a near horizon limit of a brane configuration, one can obtain only a half of the space, so the Poincare patch is singular. As we will demonstrate in section \ref{SecRegul}, the branch cut introduces a new mechanism which makes all bubbling geometries with \AdSS\ asymptotics regular without compromising the conservation of flux. The branch cuts also lead to very interesting topological structures, which are discussed in section \ref{SecTopology}. 

This paper has the following organization. In section \ref{SecProbesAdS} we analyze the dynamics supersymmetric branes on \AdSS, which can be viewed as lower--dimensional counterparts of the giant gravitons. In particular, we will find that, unlike giant gravitons in AdS$_5\times$S$^5$, which can expand only on AdS or on a sphere, the branes on \AdSS\ can be placed at any point on a three dimensional space while stretching in the time direction. A similar feature is exhibited by the giant gravitons in  AdS$_3\times$S$^3$ and in all 1/2--BPS fuzzball geometries, and such branes are discussed in Appendix \ref{AppAdS3}. In section \ref{SecMain} we construct all BPS geometries with \AdSS$\times$T$^6$ asymptotics which can be viewed as backreactions of the configurations discussed in section \ref{SecProbesAdS}. We will demonstrate that regular BPS geometries (\ref{LocalSoln})--(\ref{HarmMain}) are parameterized by one complex harmonic function, which has a form (\ref{MainHarmInt}), and to recover a geodesically--complete space, one must connect at least two copies of the three--dimensional base through two--dimensional branch cuts. The details of such analytic continuation and the interesting topological structures originating from it are discussed in sections \ref{SecPlane} and \ref{SecTopology}. While we demonstrate that the task of finding the BPS geometries reduces to solving a Laplace equation with some mixed Dirichlet/Neumann boundary conditions, this linear problem is still rather nontrivial, and in section \ref{SecExamples} we present several explicit examples of regular solutions. Finally, in section \ref{SecBrane} we add probe branes to the new geometries, solve their equations of motion, and prove supersymmetry by analyzing the kappa--projection. The resulting picture is very reminiscent of the one found for branes on AdS$_3\times$S$^3$. Some technical calculations and supplementary material are presented in the appendices.

\section{Brane probes on \AdSS}
\renewcommand{\theequation}{2.\arabic{equation}}
\setcounter{equation}{0}
\label{SecProbesAdS}

According to the standard AdS/CFT dictionary, \AdSS\ must correspond to a vacuum of some quantum mechanics living on the boundary of the AdS space, and excitations of this quantum mechanics are mapped into normalizable modes on the bulk side. The light excitations are mapped into strings moving on \AdSS, and as demonstrated in \cite{STWZ}, such strings are integrable. Heavier excitations correspond to probe D branes, which will be discussed in this section, and when many such branes are put together, they produce normalizable deviations from the \AdSS\ geometry, and the resulting solutions of supergravity will be constructed in section \ref{SecMain}.

We begin with reviewing some  known properties of \AdSS. There are several ways of embedding this space in string theory, and we will mostly focus on the implementation based on D3 brane sources \cite{KlebTseytl}:
\bea\label{BraneScan}
\begin{array}{c|cccccccccc|}
&t&x_1&x_2&x_3&X_1&X_2&X_3&Y_1&Y_2&Y_3\\
\hline
D3_1&\bullet&&&&\bullet&\bullet&\bullet&\sim&\sim&\sim\\
D3_2&\bullet&&&&\bullet&\sim&\sim&\sim&\bullet&\bullet\\
D3_3&\bullet&&&&\sim&\bullet&\sim&\bullet&\sim&\bullet\\
D3_4&\bullet&&&&\sim&\sim&\bullet&\bullet&\bullet&\sim\\
\hline
\end{array}
\eea
Here bullets denote the directions wrapped by the branes and tildes denote the coordinates in which branes are smeared. We also assume that directions $(X_1,X_2,X_3,Y_1,Y_2,Y_3)$ are compactified on a torus with finite volume $V$.

Four stacks of D3 branes (\ref{BraneScan}) produce an asymptotically flat geometry constructed in \cite{KlebTseytl}\footnote{Since the main goal of this article is construction of supersymmetric geometries, we use supergravity normalization of fluxes \cite{SchwWest} throughout the paper and set  $\kappa=1$. To write the action for the probe branes, such as (\ref{ProbeAct}), one should recall that fluxes in string theory have different normalization, in particular, $F_5^{(string)}=\frac{4\kappa}{g_s}F_5$. This is the origin of the additional factor of $4$ in (\ref{ProbeAct}) and in other actions for the brane probes. See \cite{GranPolch} for the detailed discussion of the map between string and supergravity normalizations for various fluxes.\label{ftnOne}
}
\bea\label{D3stackGeom}
ds^2&=&-\frac{dt^2}{f}+f\left[dr^2+r^2 d\Omega_2^2+\frac{dX_1^2}{h_1h_2}+\frac{dX_2^2}{h_1h_3}+
\frac{dX_3^2}{h_1h_4}+\frac{dY_1^2}{h_3h_4}+\frac{dY_2^2}{h_2h_4}+\frac{dY_3^2}{h_2h_3}\right]\nn
F_5&=&\frac{1}{4}dt\wedge d\left[-\frac{1}{h_1} dX_{123}+\frac{1}{h_2}dX_1dY_{23}-\frac{1}{h_3}dX_2dY_{13}+\frac{1}{h_4}dX_3 dY_{12}\right]+dual,\\
h_i&=&1+\frac{Q_i}{r},\quad f=\sqrt{h_1h_2h_3h_4},\nonumber
\eea
which describes a BPS black hole with an area
\bea
A_8=4\pi V\sqrt{Q_1Q_2Q_3Q_4}. 
\eea
The near horizon limit of (\ref{D3stackGeom}) is the \AdSS$\times$T$^6$, where the AdS and the sphere have the same radius 
\bea
L=\left[Q_1Q_2Q_3Q_4\right]^{1/4}.
\eea
In this article we will focus on configurations with $Q_1=Q_2=Q_3=Q_4$, then the near--horizon geometry can be written in global coordinates as
\bea\label{GlobAdS1}
ds^2&=&L^2\left[-(\rho^2+1)d{t}^2+\frac{d\rho^2}{\rho^2+1}+d\theta^2+
\sin^2\theta d{\phi}^2\right]+dz^ad{\bar z}_a\\
F_5&=&\frac{L}{4}d\rho dt\wedge\mbox{Re}(dz_{123})+dual,\nonumber
\eea
where
\bea\label{XYdefZ}
z_a=X_a+i Y_a.
\eea

Supersymmetric excitations of the metric (\ref{GlobAdS1}) fall into several categories:
perturbative gravitons, D-branes, and topologically nontrivial deformation of the metric, and one moves between these three cases as a mass of the excitation grows. The excitations whose energy scales as $N^0$ appear as perturbative gravitons, and they were studied  in \cite{gravSpectrAdS2} using  the standard analysis of spectrum which had previously been applied to other AdS$\times$S spaces \cite{gravSpectr}. Semiclassical excitations with energies of order $N^1$ behave as supersymmetric D branes, which are studied in this section. Once the energy reaches $N^2$, gravitational backreaction of branes becomes important, and the resulting solutions of supergravity are constructed in the next section. 

Supersymmetric branes on AdS$_p\times$S$^p$ are known as `giant gravitons' \cite{SusskGiant,MyersGiant}: they expand on contractible cycles on AdS or the sphere, and they are prevented from collapsing by angular momentum. If $p>3$, then the size of the giant graviton is fixed by the angular momentum, while for $p=3$ the giant gravitons may expand to an arbitrary size. Moreover, on AdS$_3\times$S$^3$ one can generalize giant gravitons to branes wrapping cycles on both AdS and the sphere, and this construction is presented in Appendix \ref{AppAdS3}. The \AdSS\ case is very similar with a small caveat: the `giant gravitons' are pointlike. As we will see in the next section, this similarity between $p=3$ and $p=2$ gives rise to a similarity in classifying gravity solutions for AdS$_3\times$S$^3$ and \AdSS.

To study supersymmetric D3 branes on (\ref{GlobAdS1}), we impose the static gauge for the worldvolume:
\bea\label{StaticGauge}
t=\tau,\quad X_1=\xi_1\cos\beta,\quad Y_1=\xi_1\sin\beta,\quad X_2=\xi_2,\quad X_3=\xi_3,\quad Y_2=Y_3=0,
\eea
and assume that $(\rho,\theta,\phi)$ are functions of $\tau$. Then the action for the D brane\footnote{The origin of the factor of four in the Chern--Simons term is explained in the footnote \ref{ftnOne}.},
\bea\label{ProbeAct}
S=-T\int d^4\xi\sqrt{-\mbox{det}\left[g_{mn}\frac{\d x^m}{\d\xi^a}\frac{\d x^n}{\d\xi^b}\right]}+
4T\int P[C_4],
\eea
becomes\footnote{To shorten numerous formulas appearing throughout this article, we introduce a convenient shorthand notation for trigonometric functions: $s_\alpha\equiv \sin\alpha$, $c_\alpha\equiv\cos\alpha$.}
\bea\label{AdS2GiantAct}
S=-TL\int d^4\xi\sqrt{(\rho^2+1)-\frac{{\dot\rho}^2}{\rho^2+1}-{\dot\theta}^2-s_\theta^2 {\dot\phi}^2}+TL\int d^4\xi [c_\beta\rho+s_\beta c_\theta{\dot\phi}]
\eea
Equations of motion are solved by constant $(\rho,\theta,{\dot\phi})$ as long as two relations are satisfied:
\bea
\frac{\rho}{\sqrt{\rho^2+1-s_\theta^2{\dot\phi}^2}}=c_\beta,\quad
\frac{c_\theta{\dot\phi}}{\sqrt{\rho^2+1-s_\theta^2{\dot\phi}^2}}=s_\beta.\nonumber
\eea
Combining these relations, we conclude that  
\bea\label{RotBrnSoln}
\dot\phi=1,\qquad c_\theta=\rho\tan\beta.
\eea
In particular, $\beta=\frac{\pi}{2}$ corresponds to $\rho=0$ and an arbitrary $\theta$, giving the \AdSS\ counterpart of the giant graviton present in the higher dimensions \cite{SusskGiant}. Another special case, $\beta=0$, gives $\theta=\frac{\pi}{2}$ and an arbitrary $\rho$, which corresponds to the dual giant \cite{MyersGiant}. Interpolating values of $\beta$ have counterparts only in the AdS$_3\times$S$^3$ case, which is discussed in the Appendix \ref{AppAdS3}, and they correspond to arbitrary points in the $(\rho,\theta)$ space. 
For completeness we also give the expressions for the angular momentum and the energy densities of the branes:
\bea
{\cal J}&=&\frac{\d{\cal L}}{\d{\dot\phi}}=\frac{TLs_\theta^2}{\sqrt{\rho^2+c_\theta^2}}+
TLs_\beta c_\theta=\frac{TL}{\sqrt{\rho^2+c_\theta^2}}\,,\\
\label{EnergGiant}
{\cal E}&=&\left[\frac{TLs_\theta^2}{\sqrt{\rho^2+c_\theta^2}}+
TLs_\beta c_\theta\right]{\dot\phi}+TL\left[\sqrt{\rho^2+c_\theta^2}-c_\beta\rho-s_\beta c_\theta\right]
={\cal J}.
\eea
Notice that fixing ${\cal J}$ and ${\cal E}$ still leaves some freedom in the location of the brane, and this feature distinguishes \AdSS\ and AdS$_3\times$S$^3$  from the higher--dimensional cases \cite{MyersGiant}. 
Relation ${\cal E}={\cal J}$ implies that the branes saturate the BPS bound, so they are supersymmetric. 

To identify the supersymmetries preserved by the rotating branes, we recall the kappa--symmetry projection associated with a D3 brane \cite{KappaProj}:
\bea
\Gamma\eps=\eps,\quad \Gamma=i\sigma_2\otimes\left[{\cal L}^{-1}\left(\prod_{a=0}^3\frac{\d x^{m_a}}{\d\xi^a}\right)\gamma_{m_0\dots m_3}\right],\quad
{\cal L}=\sqrt{-\mbox{det}\left[g_{mn}\frac{\d x^m}{\d\xi^a}\frac{\d x^n}{\d\xi^b}\right]}
\eea
and apply it to the rotating branes (\ref{StaticGauge}), (\ref{RotBrnSoln}):
\bea\label{GammaProjInterm}
\Gamma&=&i\sigma_2\otimes\frac{1}{\sqrt{\rho^2+c_\theta^2}}\left[\sqrt{\rho^2+1}\Gamma_{\hat t}+s_\theta \Gamma_{\hat\phi}\right]
(c_\beta \Gamma_{\hat X_1}+s_\beta\Gamma_{\hat Y_1})\Gamma_{\hat X_2}\Gamma_{\hat X_3}\nn
&=&i\sigma_2\otimes\left[\cosh_\sigma\Gamma_{\hat t}+\sinh_\sigma \Gamma_{\hat\phi}\right]
(c_\beta \Gamma_{\hat X_1}+s_\beta\Gamma_{\hat Y_1})\Gamma_{\hat X_2}\Gamma_{\hat X_3}\\
&=&e^{\h\sigma\Gamma_{\hat t\hat \phi}}e^{-\h\beta\Gamma_{\hat X_1\hat Y_1}} 
\left[i\sigma_2\otimes  \Gamma_{\hat t} 
\Gamma_{\hat X_1}\Gamma_{\hat X_2}\Gamma_{\hat X_3}
\right]e^{-\h\sigma\Gamma_{\hat t\hat \phi}}e^{\h\beta\Gamma_{\hat X_1\hat Y_1}}\nonumber
\eea
Here $\Gamma_{\hat m}$ denote the gamma matrices with flat indices,
\bea\label{FlatDirac}
\Gamma_{\hat m}=e_{\hat m}^n\gamma_n,\quad 
\{\Gamma_{\hat m},\Gamma_{\hat n}\}=2\eta_{\hat m\hat n},\quad
\{\gamma_m,\gamma_n\}=2g_{mn},
\eea
and parameter $\sigma$ is defined by
\bea
\sinh\sigma\equiv \frac{s_\theta}{\sqrt{\rho^2+c_\theta^2}},\quad
\cosh\sigma\equiv \frac{\sqrt{\rho^2+1}}{\sqrt{\rho^2+c_\theta^2}}
\eea
Relation (\ref{GammaProjInterm}) implies that the brane preserves supersymmetry satisfying a projection that depends on $\beta$:
\bea\label{GammaProjProbeA}
\Gamma\eps=\eps:\qquad 
\eps=e^{\h\sigma\Gamma_{\hat t\hat \phi}}e^{-\h\beta\Gamma_{\hat X_1\hat Y_1}}\eps_0,\quad
i\sigma_2\otimes  \Gamma_{\hat t} 
\Gamma_{\hat X_1}\Gamma_{\hat X_2}\Gamma_{\hat X_3}\eps_0=\eps_0.
\eea
As we will see in the next section, supersymmetric geometries preserve Killing spinors which are proportional to $\eps_0$ as well, and the prefactor varies in space. Moreover, in section \ref{SecBrane} we will demonstrate that Killing spinors for the solutions corresponding to rotating branes reduce to (\ref{GammaProjProbeA}) in the vicinity of a probe brane. 

We conclude this section by observing that the ansatz (\ref{StaticGauge}) describing supersymmetric branes can be generalized by applying an $SO(3)$ rotation on two vectors $(X_1,X_2,X_3)$ and $(Y_1,Y_2,Y_3)$. The reduced action (\ref{AdS2GiantAct}) remains unchanged, and the supersymmetry analysis is modified by the appropriate rotation matrices.

\section{Supergravity solutions}
\label{SecMain}
\renewcommand{\theequation}{3.\arabic{equation}}
\setcounter{equation}{0}
\subsection{Local structure}
\label{SecLocal}

Our goal is to construct a family of supersymmetric solutions of ten--dimensional supergravity which approach
\bea\label{VacAdS}
\mbox{AdS}_2\times\mbox{S}^2\times\mbox{T}^6
\eea
at infinity.  The `vacuum' (\ref{VacAdS}) can be lifted to a supersymmetric ten--dimensional geometry in several ways. We will mostly focus on the embedding (\ref{GlobAdS1}) into type IIB SUGRA, and some alternative options will be discussed in subsection \ref{SecEmbed}.  

We are looking for supersymmetric excitations of (\ref{GlobAdS1}) which preserve the torus and the structure of $F_5$:
\bea\label{Ansatz}
ds^2&=&g_{mn}dx^mdx^n+dz^ad{\bar z}_a\nonumber\\
F_5&=&\frac{1}{2}F_{mn}dx^{mn}\wedge\mbox{Re}\,\Omega_3-
\frac{1}{2}{\tilde F}_{mn}dx^{mn}\wedge\mbox{Im}\,\Omega_3\\
\Omega_3&=&dz_{123},\qquad \star_6\Omega_3=i\Omega_3,\quad \star_4 F={\tilde F}.\nonumber
\eea
Here $g_{mn}$, $F$, and ${\tilde F}$ are undetermined ingredients, which can be found by requiring the state (\ref{Ansatz}) to be supersymmetric. This implies that the gravitino equation,
\bea\label{temp5}
\nabla_M\eta+\frac{i}{480}{\not F}_5\Gamma_M\eta=0,
\eea
must have a nontrivial solution, and combining (\ref{temp5}) with the equation of motion for $F_5$, one can determine all functions appearing in (\ref{Ansatz}). The details of this analysis are presented in the Appendix \ref{AppMain}, and here we only mention one interesting feature: the ten--dimensional equation (\ref{temp5}) reduces to equation (\ref{4DspinorTil}) for an effective four--component spinor ${\tilde\eta}$ in four dimensions spanned by $x^m$:
\bea
\nabla_m{\tilde\eta}+{i}\not F
{\gamma}_m{\tilde\eta}=0.
\eea

The complete solution of the gravitino equation and equations of motion solution is derived in Appendix \ref{AppMain}, and it reads (see (\ref{A3metric}), (\ref{Gam5Proj}),
(\ref{Vdual}),  (\ref{Strenght}))
\bea\label{LocalSoln}
ds^2&=&-h^{-2}(dt+V)^2+h^2 dx_adx_a+dz^{\dot a}d{\bar z}_{\dot a}\nonumber\\
F_5&=&F\wedge\mbox{Re}\,\Omega_3-
{\tilde F}\wedge\mbox{Im}\,\Omega_3\nonumber\\
F&=&-\d_a A_t (dt+V)\wedge dx^a+h^2\star_3 d{\tilde A}_t\,,\quad {\tilde A}_t+iA_t=-\frac{1}{4h}e^{i\alpha}\\
{\tilde F}&=&-\d_a {\tilde A}_t (dt+V)\wedge dx^a-h^2\star_3 d{A}_t\,,\quad 
dV=-2h^2 \star_3 d\alpha\,,\nonumber\\
{\tilde\eta}&=&h^{-1/2}e^{i\alpha\Gamma_5/2}\eps,\qquad \Gamma^{\bf t}\Gamma_5\eps=\eps.
\nonumber
\eea
This geometry is parameterized by two functions $h$ and $\alpha$, which in turn can be extracted from two harmonic functions $H_1$ and $H_2$:
\bea\label{HarmMain}
&&H_1=h\sin\alpha,\quad H_2=h\cos\alpha,\quad d\star_3d H_a=0,\\
&&dV=-2\star_3[H_2 dH_1-H_1 dH_2],\quad {\tilde A}_t+iA_t=-\frac{1}{4(H_2-iH_1)}\,.\nonumber
\eea
Notice that solution (\ref{D3stackGeom}) with $h_1=h_2=h_3=h_4$ is recovered by setting $H_1=h_1$, $H_2=0$.

Equations (\ref{LocalSoln})--(\ref{HarmMain}) completely specify the local structure of the solution at regular points of $(H_1,H_2)$, but the harmonic functions must have sources, where solution may become singular. One encounters a similar problem in the AdS$_3\times$S$^3$ case, where it was shown that the singularity is absent for the sources allowed in string theory \cite{Parad,lmm}. Unfortunately in the present case the condition on the sources is less intuitive, so we begin with studying it for the vacuum (\ref{VacAdS}) before extending it to a general state in subsection \ref{SecRegul}. 

\subsection{An example of a regular solution: AdS$_2\times$S$^2$}
\label{SecRegulAdS}

The easiest way to recover \AdSS\ from (\ref{LocalSoln})--(\ref{HarmMain}) is to set 
\bea\label{PoincAdS}
H_1=\frac{Q}{\sqrt{x_ax_a}}, \quad H_2=0,
\eea 
but the resulting geometry covers only the Poincare patch, so it is not geodesically complete. Since relations  (\ref{LocalSoln})--(\ref{HarmMain})  describe all supersymmetric solutions which have the form (\ref{Ansatz}), they must include the global \AdSS\ as well, and in this subsection we will discuss such embedding and analyze the mechanism that makes \AdSS\ regular in spite of singularities in $(H_1,H_2)$. In the next subsection we will use the intuition acquired from the \AdSS\ solution to classify all regular geometries covered by (\ref{LocalSoln})--(\ref{HarmMain}).

The geometry of the global \AdSS\ is given by
\bea\label{GlobAdS}
ds^2&=&L^2\left[-(\rho^2+1)d{\tilde t}^2+\frac{d\rho^2}{\rho^2+1}+d\theta^2+
\sin^2\theta d{\tilde\phi}^2\right]+dz^ad{\bar z}_a\\
F_5&=&\frac{L}{4}d\rho d{\tilde t}\wedge\mbox{Re}(dz_{123})+dual,\nonumber
\eea
and to put it in the form (\ref{LocalSoln}) we shift the angular coordinate as ${\tilde\phi}=\phi+{\tilde t}$. For convenience we also rescale time, ${\tilde t}=t/L$, to make the harmonic functions dimensionless. This gives the metric
\bea\label{AdSemb1}
ds^2&=&-h^{-2}\left(dt-
\frac{L\sin^2\theta d\phi}{\rho^2+\cos^2\theta}\right)^2+L^2\left[\frac{d\rho^2}{\rho^2+1}+d\theta^2+
\frac{\sin^2\theta (\rho^2+1)d\phi^2}{\rho^2+\cos^2\theta}\right]+dz^ad{\bar z}_a\nonumber\\
h^{-2}&\equiv&\rho^2+\cos^2\theta.
\eea
Geometry (\ref{AdSemb1}) has the form (\ref{LocalSoln}), in particular, the metric on the base, 
\bea\label{AdSbase}
ds_{base}^2&=&L^2\left[\frac{(\rho^2+\cos^2\theta)d\rho^2}{\rho^2+1}+(\rho^2+\cos^2\theta)d\theta^2+
{\sin^2\theta (\rho^2+1)d\phi^2}\right],
\eea
is flat, as can be seen by going to cylindrical coordinates $(r,\phi,y)$:
\bea\label{AdSRy}
r=L\sqrt{\rho^2+1}\sin\theta,\quad y=L\rho \cos\theta,\ \Rightarrow\
ds_{base}^2=dy^2+dr^2+r^2d\phi^2.
\eea
Notice that  transformation 
\bea\label{DoubleMapAdS}
\rho\rightarrow-\rho,\quad \theta\rightarrow \pi-\theta
\eea
does not affect the new coordinates, so every point in $(r,y)$ half--plane, with an exception of $(r,y)=(L,0)$, corresponds to two points in the $(\rho,\theta)$ space. As we will see, this double cover plays an important role in ensuring regularity of the solution. 

The harmonic functions and the vector field $V$ corresponding to the solution (\ref{AdSemb1}) can be extracted by a direct comparison with (\ref{LocalSoln}):
\bea\label{H12AdS}
&&H_1=\frac{L\sqrt{r^2 + y^2 -L^2 + f}}{\sqrt{2}f},
\quad
H_2=\frac{L^2\sqrt{2}y}{f \sqrt{r^2 + y^2 -L^2 + f}},
\quad h^2=\frac{L^2}{f},\\
&&V=-\frac{L}{2}\left[\frac{r^2+y^2+L^2}{f}-1\right]d\phi,\quad
f=\sqrt{4L^2y^2 + (r^2 + y^2-L^2)^2}.\nonumber
\eea
Expressions for $H_1$ and $H_2$ can be encoded in terms of a complex valued harmonic function
\bea\label{ComplHarm}
H\equiv H_1+iH_2, 
\eea
which for (\ref{H12AdS}) has a very simple form:
\bea\label{ComplAdS}
H=\frac{L}{\sqrt{r^2+(y-iL)^2}}\,.
\eea
We will use the complex function (\ref{ComplHarm}) to parameterize the general solution (\ref{HarmMain}) as well.

\begin{figure}[t]
\centering
\includegraphics[width=0.8\textwidth]{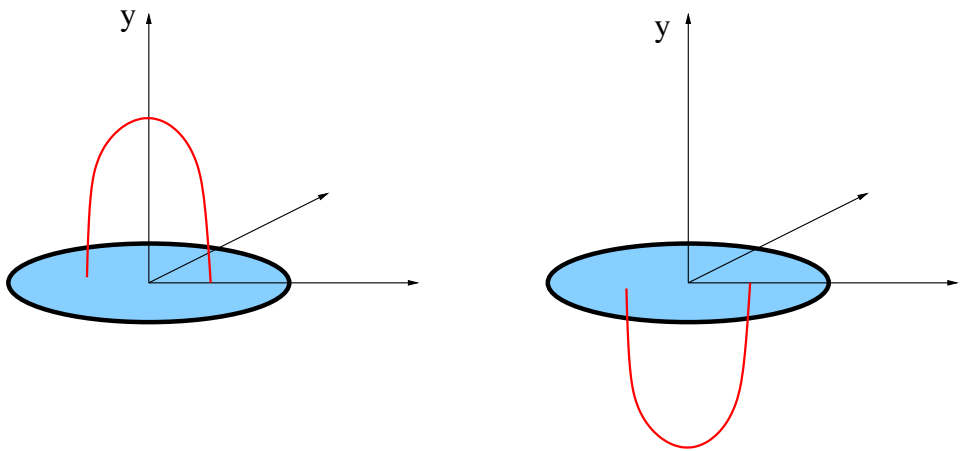} 
\caption{Sources corresponding to the \AdSS: two copies of ${\bf R}^3$ are connected through a branch cut. The open lines appearing on two copies are joint into one closed loop on the \AdSS.}
\label{FigAdSCopies}
\end{figure}

Expressions (\ref{H12AdS}) become singular on the $r=L$ circle in the $y=0$ plane, and this curve coincides with the set of fixed points of the transformation (\ref{DoubleMapAdS}). To analyze the behavior of $(H_1,H_2)$ in the vicinity of the circle, we introduce polar coordinates $(R,\zeta)$ in the plane orthogonal to the singular curve:
\bea\label{RyTh}
y=R\sin\zeta,\qquad r-L=R\cos\zeta,
\eea
The leading order near $R=0$ gives the harmonic functions
\bea\label{NearSingAdSc}
f\simeq 2LR,~
H_1\simeq\frac{L\cos\frac{\zeta}{2}}{\sqrt{2LR}},~
H_2\simeq\frac{L\sin\frac{\zeta}{2}}{\sqrt{2LR}},\quad
V\simeq -\frac{L^2}{2R}d\phi,~
-\frac{(V_\phi)^2}{h^2}+h^2r^2\simeq L^2
\eea
and the metric
\bea\label{apprAdS}
ds^2&\simeq&-\frac{2R}{L}(dt-\frac{L^2}{2R}d\phi)^2+\frac{L}{2R}[dR^2+R^2 d\zeta^2+(L+R\cos\zeta)^2d\phi^2]
\nonumber\\
&\simeq&2Ldtd\phi+L^2 d\phi^2+
2L[(d\sqrt{R})^2+\frac{R}{4} d\zeta^2]\,.
\eea
Naively expressions (\ref{RyTh}) suggest that $\zeta\in [0,2\pi)$, then the metric (\ref{apprAdS}) appears to have a conical singularity at $R=0$. However, we recall that a point in the $(r,y)$ plane corresponds to two points in the global AdS (see (\ref{AdSRy})), and this double cover breaks down precisely at $R=0$. For small values $R$ we find approximate expressions
\bea
\rho\simeq\sqrt{\frac{2R}{L}}\cos\frac{\zeta}{2},\quad 
\theta\simeq\frac{\pi}{2}-\sqrt{\frac{2R}{L}}\sin\frac{\zeta}{2},
\eea
so to cover the full vicinity of $\rho=0$, coordinate $\zeta$ must vary between zero and $4\pi$:
\bea\label{ZetaRange}
R>0,\quad 0\le \zeta< 4\pi.
\eea
This range ensures regularity of (\ref{apprAdS}) at $R=0$ and 
provides a double cover of the $(r,y)$ plane (\ref{RyTh}). As $\zeta$ changes from $0$ to $2\pi$ a point goes from one copy $(r,y)$ to another, so a branch cut must be introduced on the way. As in the case of multivalued functions in a complex plane, the location of such branch cut is ambiguous, but it has to be a surface bounded by the singular curve\footnote{Recall that for analytic functions the branch cut must begin and end on branching points. For example, the branch cut for the function $f(z)=\sqrt{z^2-1}$ must go from $z=-1$ to $z=1$, but the path is ambiguous. In a small vicinity of $z=1$ there are always two points with the same value of $f$, and $f=0$ is the only exception. A similar situation is encountered in the neighborhood of the singular curve.}. An example of such branch cut and images of one closed loop on two copies of ${\bf R}^3$ are depicted in figure \ref{FigAdSCopies}. 

We conclude this section by summarizing the mechanism of regularization for \AdSS:
\begin{itemize}
\item{Geometry (\ref{LocalSoln})--(\ref{HarmMain}) is regular in the Cartesian coordinates $x_a$ everywhere away from the singular points of $H$. To cover the entire \AdSS\ we need two copies of the flat base, as depicted in figure \ref{FigAdSCopies}.}
\item{On the curve where $H$ has sources, the geometry remains regular, but the Cartesian coordinates break down, and they should be replaced by $(R,\zeta)$ defined by (\ref{RyTh}). The range (\ref{ZetaRange}) covers the vicinity of the singular curve on both copies, but the curve itself ($R=0$) is covered only once, so two copies are glued along this curve.}
\item{By going around the singular curve, a point moves from one copy to another, so a ``branch cut surface" must be introduced. As in the case of Riemann surfaces, the precise location of this branch cut is ambiguous, as long as it is bounded by the singular curve.}
\end{itemize}
In the next subsection we will use the insights from the \AdSS\ geometry to formulate the regularity conditions for an arbitrary closed curve.

\subsection{Regularity conditions}
\label{SecRegul}

Solutions presented in section \ref{SecLocal} provide a local description of supersymmetric geometries, but a generic harmonic function $H$ gives rise to a singular metric (\ref{LocalSoln}). The simplest example of such singular solution is the Poincare patch of the AdS space, which corresponds to harmonic functions (\ref{PoincAdS}). Introducing several point--like sources for $H_1$ while keeping $H_2=0$, one would describe a geometry produced by several stacks of D3 branes, which corresponds to the AdS$_2$ counterpart of the Coulomb branch discussed in \cite{Coulomb}. In this paper we are interested in regular solutions, which are analogous to the bubbling geometries of \cite{LLM}, and as we will show in this subsection, the requirement of regularity imposes severe constraints on the allowed sources of $H_1$ and $H_2$. We will demonstrate that once such constraints are satisfied, the geometries  
(\ref{LocalSoln})--(\ref{HarmMain}) are guaranteed to be regular, as long as the appropriate analytic continuation is performed.

To preserve the \AdSS\ asymptotics, harmonic functions $H_1$ and $H_2$ must vanish at infinity, this implies that they must have sources at finite points in ${\bf R}^3$, rendering the coordinate system (\ref{LocalSoln}) singular. Generic sources lead to curvature singularities in (\ref{LocalSoln}), but for some special configurations geometry may remain regular, as we saw in the last subsection. A similar situation has been encountered in the AdS$_3\times$S$^3$ case, where sources parameterized by string profiles on the base space led to regular solutions \cite{Parad,lmm}, but in the present case there is an important caveat: while the geometry may remain regular, the patch covered by the coordinate system (\ref{LocalSoln}) cannot be geodesically complete. 
We have already encountered this phenomenon in the last subsection, where coordinates (\ref{LocalSoln}) covered only a half of the \AdSS\ parameterized by $(\rho,\theta)$, and a second copy of ${\bf R}^3$ had to be attached to 
describe the full geometry. This is not very surprising since the global AdS$_2$ space is known to have two boundaries, and the asymptotic region of ${\bf R}^3$ described by large $(x_nx_n)$ can only describe a vicinity of one boundary. In this subsection we will identify the sources of $H_1$ and $H_2$ that lead to regular geometries and describe the procedure for extending a patch (\ref{LocalSoln}) to a geodesically complete space.

First we recall that in the  AdS$_3\times$S$^3$ case all 1/2--BPS solutions are parameterized by several harmonic functions defined on a flat four--dimensional base \cite{Parad}, and all regular geometries share the same mechanism for resolving singularities at the location of the sources \cite{lmm}. Using that case as a guide, we expect the mechanism of regularization described in the last subsection to be generic for all metrics (\ref{LocalSoln}). Specifically, we focus on complex harmonic functions $H$ (\ref{ComplHarm}) which satisfy four conditions:
\begin{enumerate}[(a)]
\item $H$ can have sources only on closed curves\footnote{It is also possible to have patches with sources at isolated points, such as the Poincare patch of \AdSS, but such sources can be viewed as singular limits of curves.}, and the space (\ref{LocalSoln}) develops a conical defect $\pi$ in the vicinity of every point on the curve, so branch cuts have to be introduced.
\item Once an analytic continuation to the second sheet is performed, the metric remains regular in a vicinity of the curve.
\item To ensure that the metric is regular away from the curve, the harmonic function $H$ cannot vanish at finite points in ${\bf R}^3$.
\item For asymptotically \AdSS\ geometries the harmonic function approaches 
\bea\label{AsympH0}
H_0=\frac{L}{\sqrt{x_1^2+x_2^2+y^2}}
\eea
at infinity. 
\end{enumerate}
We will now present a procedure for constructing the functions satisfying conditions (a)--(d) and demonstrate that they lead to regular solutions after an appropriate analytic continuation. The regular geometry will be parameterized by a closed contour, by a charge density, and by an additional vector field. 

\bigskip

\begin{figure}[t]
\centering
\includegraphics[width=0.3\textwidth]{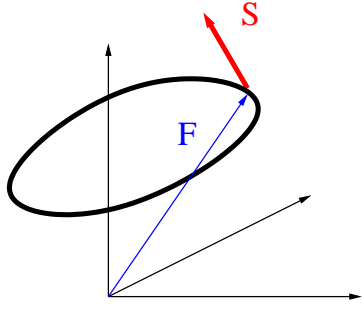} 
\caption{An example of a singular curve and the corresponding vectors ${\bf F}$ and ${\bf S}$, which lead to a regular geometry.}
\label{FigCurveVec}
\end{figure}

As demonstrated in Appendix \ref{AppRegul}, requirement (a) determines the leading contribution to the metric in a vicinity of a curve. Selecting an arbitrary point on a curve and introducing cylindrical coordinates $(R,\zeta,x_3)$ with an origin at that point and with $x_3$ axis pointing along the curve, we find
\bea\label{RequireH}
ds^2&=&-h^{-2}(dt+V)^2+h^2 \left[dR^2+R^2 d\zeta^2+dx_3^2\right],\quad
h^2=H{\bar H},\nonumber\\
H&\simeq&\frac{ae^{i\zeta}}{\sqrt{R}}
\eea
with a complex parameter $a$ which can vary along the curve. Clearly, the space (\ref{NearSingAdSc}), (\ref{apprAdS}) has this form. 
Additional analysis presented in Appendix \ref{AppRegul} demonstrates that the most general harmonic function  with properties (\ref{RequireH}) in the vicinity of the sources has the form 
\bea\label{MainHarmInt}
H=H_1+iH_2=\frac{1}{2\pi}\int 
\frac{\sigma\sqrt{({\bf r}-{\bf F})\cdot ({\bf r}-{\bf F}+{\bf A})}}{
({\bf r}-{\bf F})^2}dv+H_{reg},
\eea
Here  ${\bf F}(v)$ is the location of the profile, $\sigma(v)$ is the `charge density', $H_{reg}$ is a harmonic function that remains regular everywhere, and  ${\bf A}(v)$ is a complex vector field subject to two constraints:
\bea
{\bf A}\cdot{\dot{\bf F}}=0,\qquad {\bf A}\cdot{{\bf A}}=0.
\eea
Such field can be expressed in terms of one {\it real} vector ${\bf S}$, and in the natural 
parameterization of the curve, where
\bea\label{Sconstr1}
(\dot{\bf F})^2=1,
\eea
the answer becomes especially simple:
\bea\label{Sconstr2}
{\bf A}\equiv{\bf S}+i{\dot{\bf F}}\times {\bf S},\quad ({\bf S {\dot F}})=0.
\eea
A pictorial representation of vectors ${\bf F}$ and ${\bf S}$ in shown in figure \ref{FigCurveVec}.
Expression (\ref{MainHarmInt}) can be used for several closed curves as well, but the integral should be understood as integration over every connected piece and summation over such pieces. 

Harmonic function (\ref{MainHarmInt}) satisfies the condition (a), but to ensure the regularity condition (b) one needs subleading contributions to (\ref{RequireH}). Moreover, one has to impose the requirement  (c) since vanishing of $H$ at any finite point leads to singularities in the metric. Conditions (b) and (c) are enforced by a specific choice of the branch cuts and the regular function $H_{reg}$, which we will now describe.

\begin{figure}[t]
\centering
\includegraphics[width=0.9\textwidth]{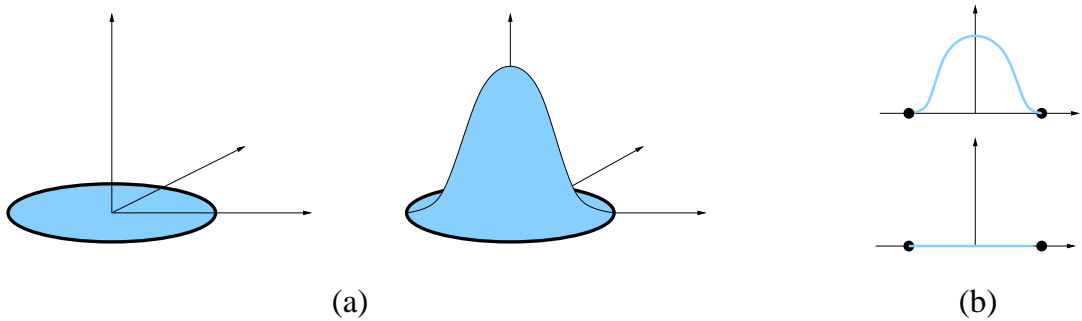} 
\caption{Discussion after equation (\ref{HasympTmp}) requires the cuts to approach the singular curve from a particular direction, but this still leaves some ambiguity depicted in figure (a). This is analogous to an ambiguity for the cuts in a complex plane that remains after imposing a particular direction for the cut at the branching points, as shown in figure (b).}
\label{FigComplCont}
\end{figure}
In a vicinity of every point on a singular curve function $H$ behaves as (\ref{RequireH}):
\bea\label{HasympTmp}
H&=&\frac{ae^{i\zeta/2}}{\sqrt{R}}+O\left(1\right).
\eea
For a given complex $a$ we can choose the range $\zeta_0< \zeta<\zeta_0+2\pi$ where the real part of $H$ remains positive and introduce a branch cut at $\zeta=\zeta_0$, where the first term in (\ref{HasympTmp}) is purely imaginary. This determines the direction of the branch cut in the vicinity of every point on the singular curve, but still leaves an ambiguity in the complete location of the cut, which will not affect our discussion. One encounters an analogous ambiguity for the holomorphic function $f(z)=\sqrt{z^2-1}$ by requiring that the branch cut goes in the real direction from $z=\pm 1$ (see figure \ref{FigComplCont}). Our choice of the branch cut guarantees that the {\it real part} of the function (\ref{MainHarmInt}) remains finite on the cut, then we can determine the harmonic function $H_{reg}$ by requiring
\bea\label{HarmCut}
\mbox{Re}\, H|_{cut}=0.
\eea
We will now demonstrate that this construction leads to regular solutions satisfying conditions (a)--(d). Moreover, once functions $({\bf F},{\bf A},\sigma)$ and the location of the branch cut are chosen, function $H$ exists, and its real part is unique. The analysis contains two ingredients:
\begin{enumerate}
\item{Regularity in the vicinity of the curve}
\item{Regularity away from the curve}
\end{enumerate}
and we will now present the relevant arguments.

~\\
\noindent
{\bf 1. Regularity in the vicinity of the curve}

To prove regularity of the metric at the location of the singular curve, we should analyze the subleading contributions to $H$. Let us pick a point on a curve and introduce local Cartesian coordinates $(x,y,z)$ by 
choosing $x$ direction along ${\dot{\bf F}}$, $z$ direction along $\ddot{\bf F}$, and $y$ direction along 
$\ddot{\bf F}\times {\dot{\bf F}}$. According to (\ref{HasympTmp}), the leading contribution to the harmonic function is
\bea\label{HasympTmp1}
H&\simeq&\frac{A}{\sqrt{z+iy}},
\eea
and the next order can be written as
\bea\label{hhOne}
H=\frac{A}{\sqrt{f}},\quad f\equiv z+iy+e_1 x^2+e_2 z^2+e_3 xz+e_4y^2+iy(e_5 x+e_6 z)+\dots
\eea
In this approximation the curve is contained in the $(x,z)$ plane, and the branch cut is given by 
an open surface
\bea
y=0,\quad f\simeq z+e_1 x^2+e_2 z^2+e_3 xz<0,
\eea
In particular, equation $f=0$ describes the curve in the $y=0$ plane, so coefficients $(e_1,e_2,e_3)$ must be real. Laplace equation for function $H$ determines $(e_4,e_5, e_6)$ and leads to the final expression 
\bea\label{hhTwo}
f\simeq(z+iy)[1+e_3 x+e_2(z+i y)]+e_1[x^2+2iy(z+iy)]
\eea
The gauge field can be found by integrating the defining relation
\bea
dV=i\star_3\left[Hd{\bar H}-{\bar H}dH\right],
\eea
and in a convenient gauge $V_y=0$ the result is
\bea\label{hhThree}
V\simeq A^2\left[c_1+\frac{e_1 x^2 z}{(y^2+z^2)^{3/2}}-\frac{(1-e_3 x-e_2 z)}{\sqrt{y^2+z^2}}\right]dx+A^2\left[c_2+\frac{2 e_1 x}{\sqrt{y^2+z^2}}\right]dz
\eea
Substituting the harmonic functions (\ref{hhOne}), (\ref{hhTwo}), (\ref{hhThree}) into the metric (\ref{LocalSoln}) and removing the cross terms between $dx$ and other coordinates on the base by shifting the $z$ coordinate as
\bea
z=v-e_1 x^2-c_2 x\sqrt{v^2+y^2},
\eea
we arrive at the final expression for the metric in the vicinity of the singular curve:
\bea\label{SingMetrTmp}
ds^2=-2h^{-2}dt V+\frac{A^2}{\sqrt{v^2+y^2}}\left[dv^2+dy^2+O(R^3)\right]+2A^2c_1 dx^2+O(R^3)
\eea
In the leading order
\bea
h^2\simeq\frac{A^2}{\sqrt{v^2+y^2}},\quad V\simeq -\frac{A}{\sqrt{v^2+y^2}}dx,\quad
h^{-2}V\simeq -\frac{1}{A}dx,
\eea
and metric (\ref{SingMetrTmp}) becomes
\bea\label{SingMetrTmp1}
ds^2&=&\frac{2}{A}dt dx+\frac{A^2}{R}\left[dR^2+R^2d\zeta^2+O(R^3)\right]+2A^2c_1 dx^2+O(R^3),
\\
&=&\frac{2}{A}dt dx+4A^2\left[(d\sqrt{R})^2+(\sqrt{R})^2\left(d\frac{\zeta}{2}\right)^2+O(R^3)\right]+2A^2c_1 dx^2+O(R^3)
\nonumber
\eea
where
\bea
v+iy\equiv Re^{-i\zeta}.
\eea
Metric (\ref{SingMetrTmp1}) remains regular for arbitrary values of $(c_1,c_2)$, as long as angle $\zeta$ is identified with periodicity  $4\pi$. As in the \AdSS\ example, we observe that the branch cut and introduction of a second copy of ${\bf R}^3$ plays a crucial role in making the solution regular.

To summarize, we have demonstrated that the prescription (\ref{MainHarmInt}), (\ref{HarmCut}) ensures regularity of the solution in the vicinity of the singular curve. We will now show that the geometry (\ref{LocalSoln}) does not develop singularities elsewhere.

~\\
\noindent
{\bf 2. Regularity away from the curve}

Since the complex harmonic function $H$ remains finite and differentiable away from the singular curve, the metric (\ref{LocalSoln}) can become singular if and only if $H$ vanishes at some point. This can only happen when the real and imaginary parts of this function vanish at the same point. Since condition (\ref{HarmCut}) imposes a restriction only on the real part of the regular harmonic function $H_{reg}$ (see (\ref{MainHarmInt})), one can always shift the imaginary part of this object to ensure that $\mbox{Im}[H]$ never vanishes on the branch cut:
\bea\label{ImCutA}
\mbox{Im}\, H|_{cut}\ne 0.
\eea
This does not fix $\mbox{Im}[H_{reg}]$ completely, and in section \ref{SecPlane} we will impose additional restrictions which lead to a convenient analytic continuation. For regularity it is sufficient to require (\ref{ImCutA}) and to prove that $\mbox{Re}[H]>0$ away from the cut. This would guarantee that $|H|^2$ never vanishes. 

\begin{figure}[t]
\centering
\includegraphics[width=0.8\textwidth]{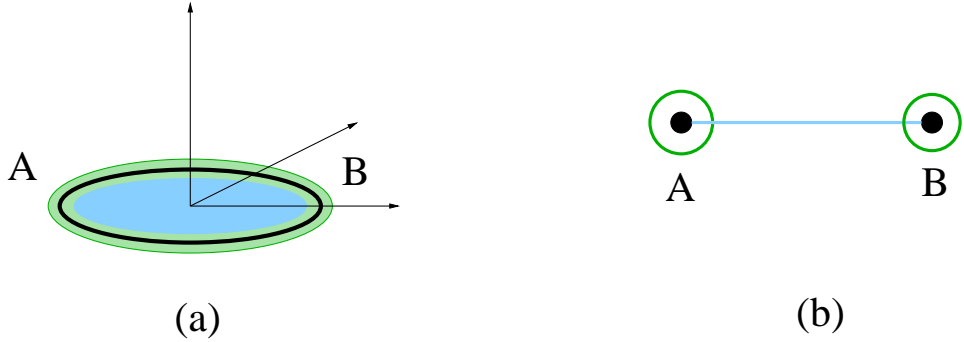} 
\caption{Combination of the original branch cut (blue) and the tubular cuts introduced around the singular curve (green). Figure (a) shows a picture on the three dimensional base, and figure (b) presents a projection on a plane going through points A and B.}
\label{FigTubeCuts}
\end{figure}

To demonstrate positivity of the real part of $H$, we introduce additional cuts shaped as thin tubes around singular curves, as depicted in figure \ref{FigTubeCuts}. These tubes begin and end on the cuts introduced earlier. We also remove the infinity by focusing on the interior of a very large sphere. The construction presented after equation (\ref{HasympTmp}) guarantees that the harmonic function 
$H_1=\mbox{Re}[H]$ satisfies several conditions:
\begin{enumerate}[(a)]
\item $H_1=0$ on the disk-shaped `standard' cuts.
\item $H_1>0$ on the tubular cuts.
\item $|H_1|<\eps$ on the large sphere, and $\eps$ can be made arbitrarily small by increasing the radius of the sphere.
\item $H_1$ is harmonic and finite in the region bounded by the cuts. 
\end{enumerate}
These conditions imply that function $H_1$ is non--negative on the boundary of a finite region surrounded by the cuts, and application of the strong maximum principle for harmonic functions leads to the conclusion that  $H_1$ must be positive away from the cuts. Hence $H$ cannot vanish anywhere, and the metric (\ref{LocalSoln}) cannot have singularities away from the curves ${\bf r}={\bf F}$. As we have already demonstrated, the solution remain regular near such curves as well. 

\bigskip

As a byproduct of the analysis presented above, we also conclude that functions $({\bf F},{\bf S},\sigma)$ and the choice of the branch cuts lead to the unique harmonic function $H_1$. Indeed, if two such functions were possible, their difference $\Delta H_1$ would remain finite on all cuts, and by making the tubes sufficiently small and the sphere sufficiently large, one can ensure that $|\Delta H_1|<\eps$ on all cuts. Then using the maximum principle, one concludes that $\Delta H_1=0$, proving uniqueness of $H_1$. Existence of $H_1$ and $H_2$ follows from the standard arguments for the Dirichlet problem for the Laplace equation. Notice that the construction presented here does not lead to a unique function $H_2$, and the freedom in selecting this function will be fixed by performing an analytic continuation and by requiring regularity on the additional sheets. This will be discussed further in section \ref{SecTopology}.

To summarize, we have demonstrated that a regular solution can be constructed by performing the following steps:
\begin{enumerate}[(1)]
\item Starting with functions $({\bf F},{\bf S},\sigma)$ parameterizing the profile, construct the harmonic function (\ref{MainHarmInt}) with undetermined $H_{reg}$.
\item Select branch cuts terminating on the singular curve and ensure that 
$\mbox{Re}[H]\ge 0$ in the vicinity of the singular curves on one of the sheets (see the discussion following equation (\ref{HasympTmp})).  
\item Determine the regular part of the harmonic function $H_{reg}$ by enforcing (\ref{HarmCut}) and  (\ref{ImCutA}), as well as the asymptotic behavior (\ref{AsympH0}). 
\end{enumerate}
This construction guaranties that the resulting solution remains regular in the vicinity of the singular curve and at all points on the selected sheet. In addition, one has to perform an analytic continuation and to enforce regularity and an appropriate asymptotic behavior on the second sheet, but conditions (\ref{HarmCut}) and  (\ref{ImCutA}) are not sufficient to guarantee uniqueness of the analytic continuation through the branch cut. Moreover, it might be possible to have more than two sheet, and such solutions would have several asymptotic \AdSS\ regions. The detailed discussion of such interesting geometries is beyond the scope of this article, and in the next subsection we will focus on describing the simplest analytic continuation for a large class of regular solutions. 

\begin{figure}[t]
\centering
\includegraphics[width=0.9\textwidth]{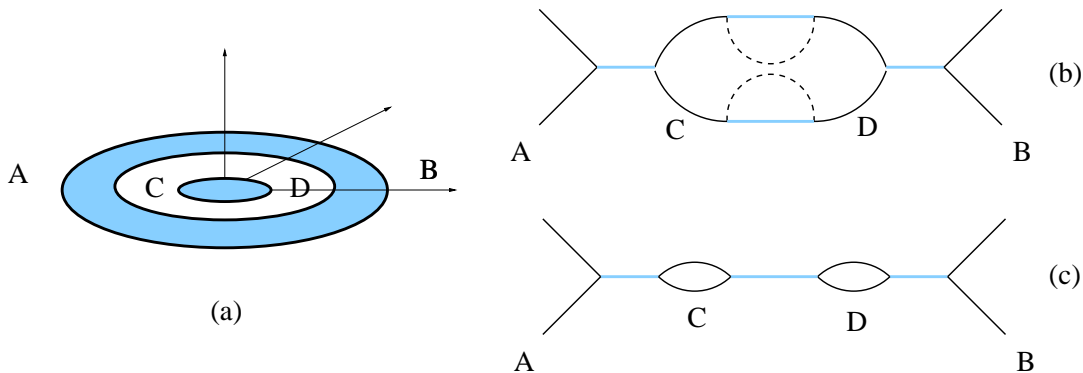} 
\caption{A solution with a sequence of branch cuts depicted in figure (a) can be continued to other sheets in several different ways.
In figures (b) and (c) we show two such continuations focusing on the images of the AB line. The geometry in figure (b) contains four sheets: two infinite ones are denoted by solid lines, and two compact 'handles' are denoted by the dashed lines. Although continuations depicted in figure (b) could lead to interesting solutions with \AdSS\ asymptotics, in this paper we focus on the simplest continuations involving only two sheets, as depicted in figure (c). 
}
\label{FigHandles}
\end{figure}

\subsection{Special case: planar curves}
\label{SecPlane}

While geometries with several \AdSS\ regions are very interesting, in this article we are focusing on solutions describing the backreaction of supersymmetric branes discussed in section \ref{SecProbesAdS}. In particular, such branes are not expected to introduce drastic changes far away from the sources, so we expect to have only two sheets with asymptotic \AdSS\ regions, as happened for the vacuum solution (\ref{GlobAdS}), (\ref{H12AdS}). In principle, this does not eliminate a possibility of having `handles'\footnote{Unlike 
${\bf R}^3$, the handle is a compact manifold, and solutions of the Laplace equation on such spaces are more complicated than (\ref{MainHarmInt}). It would be interesting to study such solutions in detail.}, such as one depicted in figure \ref{FigHandles}(b), but we will focus on the simplest case of two sheets connected through a series of branch cuts as in figure \ref{FigHandles}(c). To find the explicit expression for the full geometry, we will also require all curves to be in the $(x_1,x_2)$ plane.

\begin{figure}[t]
\centering
\includegraphics[width=0.3\textwidth]{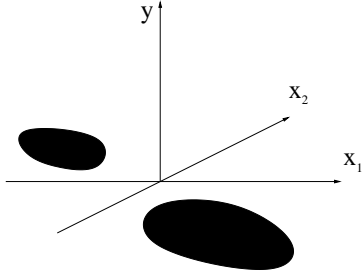} 
\caption{An example of a droplet configuration in the $(x_1,x_2)$ plane. Black and white regions correspond to boundary conditions (\ref{BubblBC}) for the complex harmonic function $H$.}
\label{FigPlaneShape}
\end{figure}

Let us introduce Cartesian coordinates $(x_1,x_2,y)$ in ${\bf R}^3$ and assume that all profiles are drawn in the $(x_1,x_2)$ plane:  ${\bf F}=(F_1,F_2,0)$. Since ${\dot{\bf F}}$ and ${\ddot{\bf F}}$ belong to the same plane, vector ${\bf S}$ parameterizing the profile through (\ref{MainHarmInt}) and (\ref{Sconstr2}) must point along $y$ direction. Then approximation (\ref{HasympTmp}) for the integral (\ref{MainHarmInt}) implies that in a small vicinity of the curve, $\mbox{Re}[H]$ can vanish only at $y=0$, so one can choose the branch cuts to be in the  $(x_1,x_2,y)$ plane. To perform an analytic continuation, we remove the space with $y<0$ and introduce a boundary at $y=0$. Part of this boundary (black regions in figure \ref{FigPlaneShape}) is formed by the branch cuts where 
\bea\label{PartBC1}
\mbox{Re}\, H|_{black}=0,
\eea
and another part (white regions) extends to infinity. According to (\ref{HasympTmp}), $\mbox{Im}\, H$ remains finite in the white region, so one can always choose function $H_{reg}$ by requiring
\bea\label{PartBC2}
\mbox{Im}\, H|_{white}=0.
\eea
Since now we have a space with a boundary at $y=0$, conditions (\ref{PartBC1}), (\ref{PartBC2}) do not determine $H$ completely. The leading contribution (\ref{HasympTmp}) ensures that $\mbox{Re}\, [\d_yH]$ remains finite in the white region, and $\mbox{Im}\, [\d_yH]$ remains finite in the black regions, so the regular part of the harmonic function $H_{reg}$ can be chosen to enforce the boundary conditions:
\bea\label{BubblBC}
\mbox{black}:&&\quad \mbox{Re}\, H=0,\quad \mbox{Im}\, \d_yH=0,\nn
\mbox{white}:&&\quad \mbox{Im}\, H=0,\quad \mbox{Re}\, \d_yH=0
\eea
Decomposition of the plane into black and white regions and the boundary conditions (\ref{BubblBC}) are reminiscent of the construction of the 1/2--BPS bubbling solutions \cite{LLM}, however, there are two important caveats. First, the $y>0$ region of the 1/2--BPS bubbling solutions covered the full space, while now this region is not geodesically complete and an analytic continuation is required. The second difference is technical: while the harmonic function describing the 1/2--BPS bubbling solutions in type IIB supergravity had the Dirichlet boundary conditions in the plane, the conditions (\ref{BubblBC}) are mixed, so finding explicit solutions becomes more difficult. Nevertheless, conditions (\ref{BubblBC}) describe a standard electrostatic problem, so the solution for $H_{reg}$ exists, and it is unique\footnote{Of course, one also has to impose the \AdSS\ asymptotics (\ref{AsympH0}). Introducing tubular cuts and using the maximum principle, as in subsection \ref{SecRegul}, one can demonstrate that boundary conditions (\ref{BubblBC}) lead to the unique solution, and existence follows from the standard theory of harmonic functions.}. Moreover, the analysis presented in the last subsection guarantees that the harmonic function (\ref{MainHarmInt}) with boundary conditions (\ref{BubblBC}) is regular at $y\ge 0$, and that the resulting geometry 
(\ref{LocalSoln}) has a conical singularity with a deficit angle $\frac{3\pi}{2}$ at the location of the singular curve. Thus gluing three more sheets along such curves would produce regular geometries. 

\begin{figure}[t]
\centering
\includegraphics[width=0.8\textwidth]{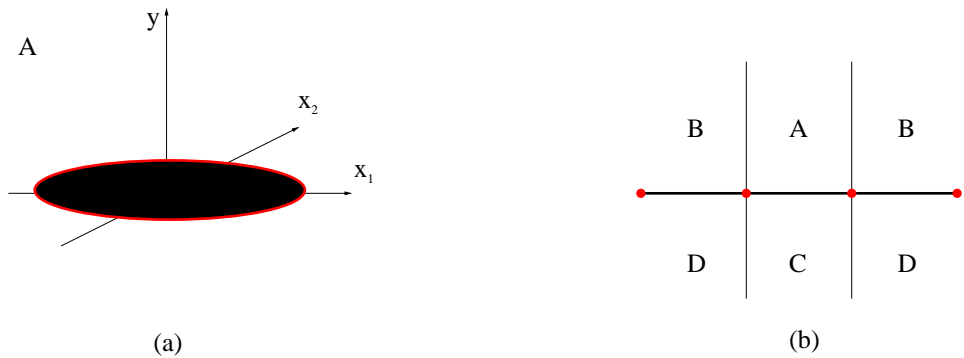} 
\caption{Analytic continuation (\ref{HonSheets})--(\ref{gluing}) for planar curves: starting with a complex solution of the 
Laplace equation on sheet A (figure (a)), and performing the analytic continuation (\ref{HonSheets}), one should glue the other three sheets as shown in figure (b). Each sheet appears only once, so the left and the right sides of figure (b) should be identified. 
}
\label{FigContin}
\end{figure}

Conditions (\ref{BubblBC}) make the analytic continuation rather simple. Starting with a harmonic function $H$ defined at $y\ge 0$, we introduce four sheets:
\bea\label{HonSheets}
H_A(x_1,x_2,y)&=&H(x_1,x_2,y)\nn
H_B(x_1,x_2,y)&=&{\overline{H(x_1,x_2,-y)}}\\
H_C(x_1,x_2,y)&=&-{\overline{H(x_1,x_2,-y)}}\nn
H_D(x_1,x_2,y)&=&-{{H(x_1,x_2,y)}}\nonumber
\eea
Then the gluing across the cuts is performed using the following rules:
\bea\label{gluing}
\mbox{white}:\begin{array}{c}
H_A\leftrightarrow H_B\\
H_C\leftrightarrow H_D
\end{array}\, ,
\qquad\qquad
\mbox{black}:\begin{array}{c}
H_A\leftrightarrow H_C\\
H_B\leftrightarrow H_D
\end{array}\, .
\eea
A pictorial representation of the continuation (\ref{gluing}) is shown in figure \ref{FigContin}. 

All four sheets converge at the location of the profile ${\bf F}$, and since each sheet describes a wedge of a flat space with an opening angle $\frac{\pi}{2}$, the total angle around the curve adds to $2\pi$, so the arguments presented in section \ref{SecRegul} guarantee regularity in the vicinity of the curve. Conditions (\ref{gluing}) ensure that function $H$ and its derivatives are continuous across all branch cuts, so the geometry remains regular on the cuts as well. Finally, to demonstrate regularity at a generic point we have to show that $|H|$ never vanishes. To do so, we combine sheets $A$ and $B$  to produce an ${\bf R}^3$ with branch cuts along the black disks. Then $H$ approaches (\ref{AsympH0}) at infinity, and arguments presented in the last subsection prove that $|H|$ does not vanish on $A$ or $B$ sheets. Then the explicit analytic continuation (\ref{gluing}) ensures that $|H|$ never vanishes, and the geometry is regular everywhere. 

To analyze the asymptotic behavior of the geometry, it is convenient to construct two copies of ${\bf R}^3$ by combining $(A,B)$ and $(C,D)$ sheets. These two copies are glued through the black region, as expected from the general analysis presented in section \ref{SecRegul}. At infinity functions $(H_A,H_B)$ approach $H_0$ given by (\ref{AsympH0}), while functions $(H_C,H_D)$ approach $(-H_0)$. In both cases the geometry approaches \AdSS, but the two asymptotic regions are disconnected. We have already encountered this situation in section \ref{SecRegulAdS}, and now we see that backreaction of the branes modified the structure of the black regions, but it preserves the asymptotic behavior, as expected for the normalizable excitations. 

To summarize, in this subsection we have focused on planar curves and we found an explicit construction for the global geometry which preserves the asymptotic structure of \AdSS. Starting from the general solution (\ref{MainHarmInt}), (\ref{Sconstr2}) with planar curves, one should divide the $y=0$ plane into black and white regions and impose the `bubbling boundary conditions' (\ref{BubblBC}) along with  asymptotic behavior (\ref{AsympH0}) to determine the unique harmonic function $H_{reg}$. Then analytic continuation (\ref{HonSheets}), (\ref{gluing}) leads to the harmonic function which describes the global geometry, and the resulting metric is regular. In the next subsection we will discuss the topological structure of the new solutions. 

\subsection{Topology and fluxes}
\label{SecTopology}

The branch cuts and analytic continuations, which make solutions constructed in section \ref{SecLocal} regular, also introduce some interesting topological structures. In particular, the four--dimensional part of the geometry (\ref{LocalSoln}) acquires some non--contractible two--cycles, which can support quantized fluxes of $F_2$. Upon lifting to ten dimensions these fluxes can be interpreted as dissolved D3 branes\footnote{Notice that since the geometry remains regular everywhere, it does not have brane sources. The encoding of dissolved D3 branes in regular geometries has been already encountered in \cite{LLM,quarter}.}. In this subsection we will analyze the topological structure of (\ref{LocalSoln}) and the associated fluxes. We will first focus on  planar curves, for which the explicit analytic continuation is known, and then extend the discussion to more general solutions.

\begin{figure}[t]
\centering
\includegraphics[width=0.9\textwidth]{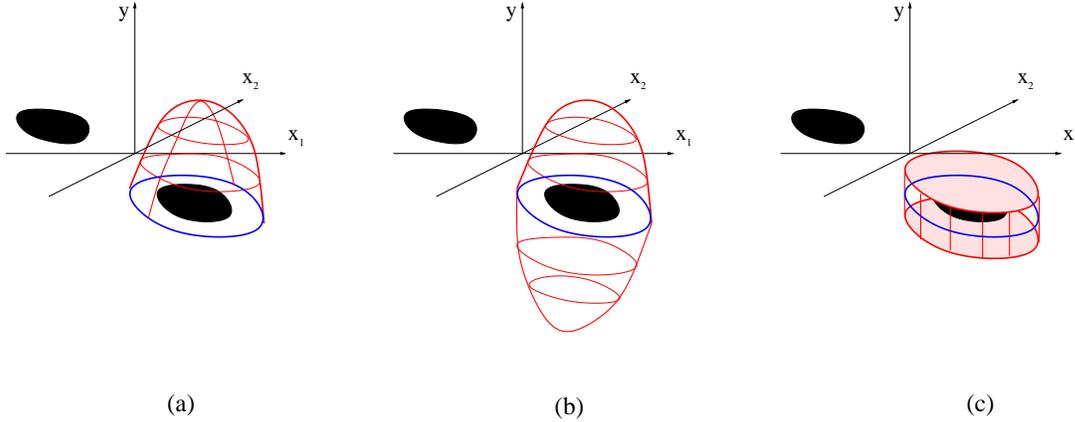} 
\caption{Topology and fluxes. Starting from an arbitrary curve ${\cal C}$ in the white region, one can construct a closed surface 
${\cal D}$ by combining an open cap in region A (figure (a)) with its mirror image in region B (figure (b)). The resulting surface can be deformed into two disks above and below the $y=0$ plane (figure (c)) to make evaluation of (\ref{FluxOint}) easier.}
\label{FigCap}
\end{figure}

Let us consider a plane divided into black and white regions and draw a curve ${\cal C}$ (not to be confused with a `singular curve' separating the regions) that lies entirely in the white area. Then we attach a cap ending on this curve and approaching the curve vertically, as shown in figure \ref{FigCap}(a). Next we construct a smooth closed surface ${\cal D}$ by combining the cap on the sheet $A$ and its image under (\ref{HonSheets}) on the sheet $B$ (see figure \ref{FigCap}(b)). If the curve ${\cal C}$  can be contracted without leaving the white region, then ${\cal D}$ is contractible. On the other hand, if one tries to contract the curve ${\cal C}$ by moving it through a black region, then ${\cal D}$ would develop a cusp when ${\cal C}$ approaches a singular curve\footnote{Recall that in the ${\bf R}^3$ spanning regions $A$ and $B$ there is branch cut along the black region and a conical singularity on its boundary.}, and smooth continuation beyond this point would not be possible. This implies that curve ${\cal C}$ circling a black region gives rise to a non--contractible surface ${\cal D}$, which is topologically equivalent to $S^2$, and this surface lies entirely on sheets $A$ and $B$. There is a `mirror image' of this sphere on sheets $C$ and $D$, and the two surfaces go into each other by passing through the the cut, but they never collapse. We have already encountered this phenomenon for the \AdSS\ example in section \ref{SecRegulAdS}, where the surface can be taken to be 
\bea
\rho=\mbox{const},\quad 0\le \theta<\pi,\quad 0\le \phi<2\pi 
\eea
in parameterization (\ref{GlobAdS}). For positive values of $\rho$ this sphere remains on $A$ and $B$ sheets, for negative values of $\rho$ it belongs to $C$ and $D$ sheets, and at $\rho=0$ the surface goes through the branch cut without collapsing. Similarly, a curve in black region that circles around a white droplet gives rise to a non--contractible surface, which lies either on $A$ and $C$ sheets or on $B$ and $D$ sheets. 

Every non-contractible surface ${\cal D}$ based on a contour ${\cal C}$ in a white region carries a flux 
of the field 
${\tilde F}$ from (\ref{LocalSoln}). To evaluate $\oint_{\cal D} {\tilde F}$, we deform the surface into two disks on sheets $A$ and $B$ located very close to the $y=0$ plane (see figure \ref{FigCap}(c)):
\bea\label{FluxOint}
\oint_{\cal D} {\tilde F}=\int_{y=\eps} {\tilde F}-\int_{y=-\eps} {\tilde F}
\eea
We will now demonstrate that the integrals in the right hand side receive contribution only from the parts of the disk immediately above or below the black droplet, so the left hand side does not change is one varies the contour ${\cal C}$  within the white region. 

To treat the while and black regions symmetrically, we define a complex two--form
\bea
{\cal F}\equiv F-i{\tilde F}=i(dt+V)\wedge d[{\tilde A}_t+iA_t]+h^2\star_3 d[{\tilde A}_t+iA_t]
\eea
Using relations
\bea
h^2\star_3 d[{\tilde A}_t+iA_t]&=&-(H{\bar H})\star_3 d\left[\frac{i}{4H}\right]=\frac{i{\bar H}}{4H}\star_3 dH,\\
iV\wedge d[{\tilde A}_t+iA_t]&=&-\frac{1}{4H^2}V\wedge dH,
\nonumber
\eea
which follow from equations (\ref{LocalSoln})--(\ref{HarmMain}), and boundary conditions (\ref{BubblBC}), we conclude that the integral
\bea
\int_{y=\eps} {\cal F}=\int_{y=\eps} \left[-\frac{1}{4H^2}V\wedge dH+\frac{i{\bar H}}{4H}\star_3 dH
\right]
\eea
is real in the white region and pure imaginary in the black region, so expression (\ref{FluxOint}) receives contributions only from integration over the black droplets. The right hand side of (\ref{FluxOint}) can be viewed as a jump of a relevant function across the branch cuts going through the black droplets, and this interpretation leads to the final expression
\bea\label{FluxBlack}
&&\oint_{\cal D} {\tilde F}=\frac{i}{4}\int_{cut} \left[V\wedge d(\Delta H^{-1})-i\star_3 d(\Delta H)
\right].
\eea
Notice that the second term in (\ref{FluxBlack}) picks up only $\Delta(\d_y H)$, so the boundary conditions (\ref{BubblBC}) guarantee reality of the last equation. 
Similarly, starting with contour ${\cal C}$ in a black region and attaching caps to it, one finds a manifold the has a topology of a two--sphere, which is spanned by the flux
\bea\label{FluxWhite}
&&\oint_{\cal D} {F}=\frac{1}{4}\int_{cut'} \left[V\wedge d(\Delta H^{-1})+i\star_3 d(\Delta H)
\right],
\eea
where $cut'$ denotes a cut along a white region encompassed by ${\cal D}$. Notice that integral in  (\ref{FluxBlack}) involves sheets $A$ and $B$, while integral in (\ref{FluxWhite}) involves sheets $A$ and $C$. For the symmetric analytic continuation (\ref{HonSheets}), all integrals can be expressed in terms of the sheet $A$:
\bea\label{FluxAsheet}
\oint_{\cal D} {\tilde F}&=&-\frac{1}{2}\int_{y=0} \mbox{Im}\left[V\wedge d(H^{-1})-i\star_3 dH
\right],\nn
\oint_{\cal D} {F}&=&\frac{1}{2}\int_{y=0} \mbox{Re}\left[V\wedge d(H^{-1})+i\star_3 dH
\right],
\eea
and boundary conditions  (\ref{BubblBC}) ensure that the first integral receives contribution only from the black regions, while the second integral is supported only by the white ones. The integrations are performed only over the interior of a defining curve ${\cal C}$. 

\begin{figure}[t]
\centering
\includegraphics[width=0.5\textwidth]{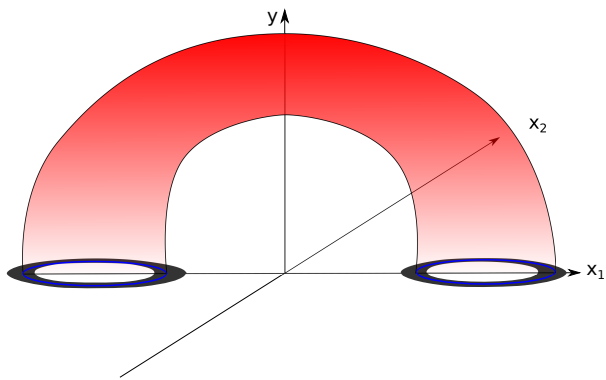}
\caption{An example of non--contractible torus on a bubbling geometry.}
\label{FigTorus}
\end{figure}

Nontrivial integrals (\ref{FluxAsheet}) of $F$ and ${\tilde F}$ give rise to fluxes of the five--form $F_5$ over the relevant five--cycles. For instance, starting with a surface ${\cal D}$ with a non--vanishing integral of ${F}$ and combining it with various circles on the torus, one can construct several closed five--cycles 
${\cal D}_5$ with 
\bea
\oint_{{\cal D}_5} F_5=l_T^3 \oint_{\cal D} {F},
\eea
where $l_T$ is a linear size of the torus $T^6$. Since the last integral must be quantized in the units of $2\pi^2 l_p^4$, the natural unit for fluxes (\ref{FluxAsheet}) is $(2\pi^2 l_p^4)/l_T^3$. Some examples
of ${\cal D}_5$ are given by
\bea
{\cal D}_5:\quad 
\begin{array}{l}
{\cal D}\times S^1_{X_1}\times S^1_{X_2}\times S^1_{X_3}\\
{\cal D}\times S^1_{X_1}\times S^1_{Y_2}\times S^1_{Y_3}\\
{\cal D}\times S^1_{Y_1}\times S^1_{X_2}\times S^1_{Y_3}\\
{\cal D}\times S^1_{Y_1}\times S^1_{Y_2}\times S^1_{X_3}\\
\end{array}\,,
\eea
where $X_a$ and $Y_a$ are defined in (\ref{XYdefZ}). Nontrivial integrals of ${\tilde F}$ give rise to similar fluxes of $F_5$. 

To summarize, we have demonstrated that planar curves give rise to a rich topological structure of bubbling geometries (\ref{LocalSoln})--(\ref{HarmMain}) through connections between different branches. Any non--contractible curve ${\cal C}$ in a white or a black region gives rise to a nontrivial $S^2$, which is supported by fluxes (\ref{FluxAsheet}). It is also possible to construct surfaces with more interesting topology (for example, figure \ref{FigTorus} depicts a non-contractible torus), but the fluxes are always given by (\ref{FluxAsheet}).

\begin{figure}[t]
\centering
\includegraphics[width=0.8\textwidth]{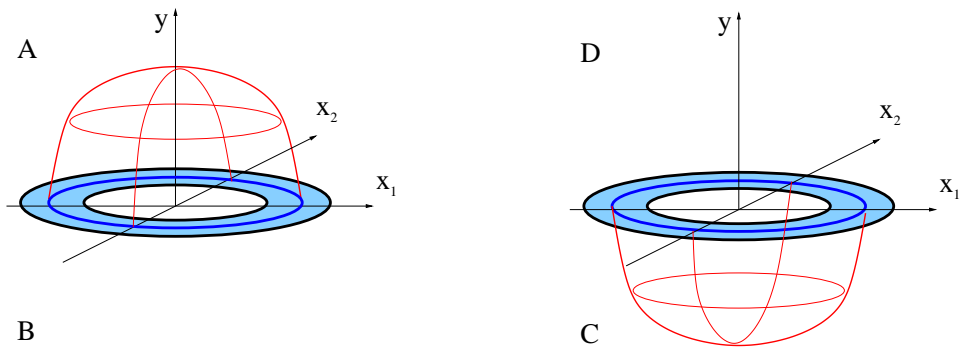} 
\caption{An example of a closed surface for a genetic singular curve constructed from a contour ${\cal C}$ on the branch cut.}
\label{FigCapsNonPlan}
\end{figure}

This construction can be extended to non--planar curved discussed in section \ref{SecRegul}, although in this case the situation is slightly less symmetric due to the absence of white regions and a lack of explicit formulas for the analytic continuation. Let us consider a collection of branch cuts in ${\bf R}^3$ associated with some number of singular curves and draw a surface ${\cal D}$ that does not touch the cuts. Two such surfaces are homotopic if they can be transformed into each other without crossing the cuts. On the other hand, a surface that cannot be collapsed to a point without crossing a cut has a nontrivial topology, and it is supported by the flux (\ref{FluxBlack}). Notice that the integrals in (\ref{FluxBlack}) involve only one copy of ${\bf R}^3$ (previously they were written in terms of sheets $A$ and $B$ which form this copy), so the details of the analytic continuation are not important. The second type of surfaces is constructed by choosing contours ${\cal C}$ in the branch cuts and attaching two caps to them. One of this caps extends to ${\bf R}^3$, and the other cap goes to the second branch, as shown in figure \ref{FigCapsNonPlan}, so the details of the analytic continuation are important for constructing such surfaces. If one focuses only on the first copy of 
${\bf R}^3$, as we did in section \ref{SecRegul}, then the non--contractible surfaces of the the second type look open 
(see figure \ref{FigCapsNonPlan}). Such surfaces are supported by the flux of ${\tilde F}$. A better understanding of the analytic continuation for arbitrary branch cuts would shed more light on structure of such non--contractible surfaces and fluxes supported by them. It would also make the treatment of $F$ and ${\tilde F}$ more symmetric, as in the case of the planar droplets. 

We conclude this subsection with a brief comment concerning angular momentum of the bubbling solutions\footnote{I am grateful to Ashoke Sen for the suggestion to add this discussion.}. Although geometries (\ref{LocalSoln}) are not static, they do not give rise to a nontrivial ADM angular momentum if one insists on preserving the \AdSS\ asymptotics. To see this, we recall that the ADM charges on AdS$_p$ contain a multiplicative factor $(p-2)$ (as discussed, for example, in \cite{ChargedAdS}), so such charges always vanish for the AdS$_2$. This absence of angular momentum plays a very important role in the counting of states for the four--dimensional black holes \cite{SenCount}. Vanishing of the angular momentum is consistent with   the statement that geometries (\ref{LocalSoln}) describe the backreaction of the giant gravitons discussed in section \ref{SecProbesAdS}: as in the AdS$_3\times$S$^3$ case, the angular momentum of such objects comes from the flat connection associated with the spectral flow operation \cite{MMaoz,MathurFlat} rather than with ADM construction. In particular, writing \AdSS\ in coordinates  (\ref{LocalSoln}), one finds that the probe branes have vanishing angular momentum\footnote{See section \ref{SecBrane} and Appendix \ref{AppAdS3} for the detailed discussion of this issue.}, and so do the geometries (\ref{LocalSoln})  produced by them. 

\subsection{Embeddings into type IIA supergravity}
\label{SecEmbed}

In this article we are focusing on \AdSS\ solutions in type IIB supergravity, but solutions (\ref{LocalSoln})--(\ref{HarmMain}) can also be lifted to type IIA SUGRA and to M theory. In this subsection we will briefly discuss such embeddings following the duality chains described in \cite{Tseytl96,KlebTseytl}. 

We begin with defining real coordinates $(X_a,Y_a)$ on the torus,
\bea
z_a=X_a+iY_a,
\eea
and rewriting the field strength appearing in (\ref{LocalSoln}) in terms of them:
\bea\label{F5flux}
F_5&=&F\wedge\left[dX_{123}-dX_1dY_{23}-dX_3 dY_{12}+dX_2dY_{13}\right]\nn
&&-{\tilde F}\wedge\left[-dY_{123}+dY_1dX_{23}+dY_3 dX_{12}-dY_2dX_{13}\right]
\eea
The system (\ref{LocalSoln})--(\ref{HarmMain}) can be mapped to type IIA theory in several ways, and we will focus on three of them:
\begin{enumerate}
\item T dualities along $(Y_1,Y_2,X_2)$ directions lead to a D4--D4--D2--D2 system, which lifts to M theory as 
an M5--M5--M2--M2 configuration.
\item T dualities along $(Y_1,Y_2,Y_3)$ directions lead to a D6--D2--D2--D2 system, which lifts to M theory as a set of three orthogonal stacks of M2 branes on a background of a KK--monopole.
\item T dualities along $(Y_1,Y_2,X_3)$ directions lead to a D4--D4--D4--D0 system, which lifts to M theory as 
three stacks of M5 branes on a plane wave background.
\end{enumerate}
None of the T dualities affect the nontrivial part of the metric (\ref{LocalSoln}). Let us briefly discuss all three options.

\bigskip
\noindent
{\bf 1. D4--D4--D2--D2 intersection.}

T dualities along $(Y_1,Y_2,X_2)$ directions transform the flux (\ref{F5flux}) into\footnote{An extra factor of two comes from combining the T duality rules with supergravity normalization of $F_5$. See \cite{GranPolch} for the detailed discussion of the relation between normalization of fluxes in string theory and in SUGRA.}  
\bea
{\cal F}&=&2F\wedge\left[-dX_{13}dY_{12}-dX_{12}dY_{31}+dX_{32}+dY_{23}\right]\nn
&&-2{\tilde F}\wedge\left[dY_{3}dX_2-dY_2dX_{3}+dY_{312} dX_{1}+dY_1dX_{132}\right]
\eea
This is a mixture of $4$-- and $6$--forms, and electromagnetic duality leads to the final solution in terms of $F_4$ only:
\bea\label{LocalSolnIIA1}
ds^2&=&-h^{-2}(dt+V)^2+h^2 dx_adx_a+dX_mdX_m+dY_mdY_n\\
F_4&=&4{\tilde F}[dX_2dY_3-dX_3 dY_2]+4F[dY_2dY_3-dX_2 dX_3]
\nonumber
\eea
All ingredients of (\ref{LocalSolnIIA1}) can be written in terms of one complex harmonic function $H$: 
\bea\label{LocalSolnIIAfields}
&&F-i{\tilde F}=i(dt+V)\wedge ({\tilde A}_t+i{A}_t)+h^2\star_3 d({\tilde A}_t+i{A}_t),\\
&&dV=-2\star_3\mbox{Im}[H d{\bar H}],\quad 
{\tilde A}_t+iA_t=-\frac{i}{4H},\quad h^2=H{\bar H}.\nonumber
\eea
Geometry (\ref{LocalSolnIIA1}) corresponds to a brane configuration, which can be obtained from (\ref{BraneScan}) by application of the T dualities: 
\bea\label{BraneScanIIA1}
\begin{array}{c|cccccccccc|}
&t&x_1&x_2&x_3&X_1&X_2&X_3&Y_1&Y_2&Y_3\\
\hline
D4_1&\bullet&&&&\bullet&\sim&\bullet&\bullet&\bullet&\sim\\
D4_2&\bullet&&&&\bullet&\bullet&\sim&\bullet&\sim&\bullet\\
D2_3&\bullet&&&&\sim&\sim&\sim&\sim&\bullet&\bullet\\
D2_4&\bullet&&&&\sim&\bullet&\bullet&\sim&\sim&\sim\\
\hline
\end{array}
\eea
This picture in terms of branes becomes useful only if one focuses on a real harmonic function $H$: in this case the geometry does have sources. For a complex harmonic function satisfying regularity conditions, the metric is source--free, so the representation (\ref{BraneScanIIA1}) is rather schematic. 

Configuration (\ref{LocalSolnIIA1}) trivially lifts to eleven dimensions by adding one more flat direction to the metric and identifying $F_4$ with a four--form in M theory. The special case of  (\ref{LocalSolnIIA1}) with real harmonic function $H$ (which corresponds to a singular near horizon limit of four stacks (\ref{BraneScanIIA1})) was discussed in \cite{Tseytl96,KlebTseytl}.

\bigskip
\noindent
{\bf 2. D6--D2--D2--D2 intersection}

Next we apply T dualities along $(Y_1,Y_2,Y_3)$ to (\ref{F5flux}), this leads to a solution of type IIA supergravity with fluxes
\bea
{\cal F}&=&2F\wedge\left[dX_{123}dY_{123}+dX_1dY_{1}+dX_3 dY_{3}+dX_2dY_{2}\right]\nn
&&-2{\tilde F}\wedge\left[1+dY_{23}dX_{23}+dY_{12} dX_{12}+dY_{13}dX_{13}\right]
\eea
Application of the electromagnetic duality to this mixture of $2,4,6,8$--forms leads to the final solution in terms of $F_2$ and $F_4$ only:
\bea\label{LocalSolnIIA2}
ds^2&=&-h^{-2}(dt+V)^2+h^2 dx_adx_a+dz^{\dot a}d{\bar z}_{\dot a}\\
F_2&=&-4{\tilde F},\quad F_4=2iF\wedge dz^{\dot a}\wedge d{\bar z}_{\dot a}
\nonumber
\eea
Once again, all ingredients can be expressed in terms of the harmonic function $H$ using (\ref{LocalSolnIIAfields}). 
The brane picture corresponding to (\ref{LocalSolnIIA2}) is a T--dual version of (\ref{BraneScan}): 
\bea\label{BraneScanIIA2}
\begin{array}{c|cccccccccc|}
&t&x_1&x_2&x_3&X_1&X_2&X_3&Y_1&Y_2&Y_3\\
\hline
D6_1&\bullet&&&&\bullet&\bullet&\bullet&\bullet&\bullet&\bullet\\
D2_2&\bullet&&&&\bullet&\sim&\sim&\bullet&\sim&\sim\\
D2_3&\bullet&&&&\sim&\bullet&\sim&\sim&\bullet&\sim\\
D2_4&\bullet&&&&\sim&\sim&\bullet&\sim&\sim&\bullet\\
\hline
\end{array}
\eea

Although the geometry (\ref{LocalSolnIIA2}) can be lifted to M theory using the standard embedding
\bea
ds^2&=&-h^{-2}(dt+V)^2+h^2 dx_adx_a+dz^{\dot a}d{\bar z}_{\dot a}+(dy+C)^2,\\
F_4&=&2iF\wedge dz^{\dot a}\wedge d{\bar z}_{\dot a}\wedge dy,\nonumber
\eea
to do this explicitly, one needs to determine the one--form $C$ by solving the defining equation
\bea
dC=-4{\tilde F}. 
\eea
Unfortunately we were not able to find a nice expression for $C$ for the general solution (\ref{LocalSolnIIA2}).
In a special (albeit singular) case ${\mbox{Re}[H]}=0$, we find
\bea
dC=-4dt\wedge d{\tilde A}_t=-d\left[\frac{1}{H_2}\right]\wedge dt\quad\Rightarrow\quad
C=-\frac{1}{H_2} dt.
\eea
Another example is \AdSS\ solution (\ref{GlobAdS}), which has
\bea
{\tilde F}=\frac{1}{4} d\cos\theta\wedge d{\tilde\phi}\quad\Rightarrow\quad 
C=-\cos\theta d{\tilde\phi}.
\eea
In this case, the coordinate $y$ corresponds to an $S^1$ Hopf fibration over $S^2$, and three coordinates $(\theta,{\tilde\phi},y)$ combine into $S^3$ in eleven dimensions \cite{Tseytl96,STWZ}.

\bigskip
\noindent
{\bf 3. D6--D2--D2--D2 intersection}

Finally, application of T dualities along $(Y_1,Y_2,X_3)$ to (\ref{F5flux}) leads to the fluxes
\bea
{\cal F}&=&2F\wedge\left[dX_{12}dY_{12}+dX_{13}dY_{13}+1+dX_{23}dY_{23}\right]\nn
&&-2{\tilde F}\wedge\left[-dX_3 dY_3+dY_{2}dX_{2}+dY_{123} dX_{123}+dY_{1}dX_{1}\right],
\eea
and electromagnetic duality gives the final answer:
\bea\label{LocalSolnIIA3}
ds^2&=&-h^{-2}(dt+V)^2+h^2 dx_adx_a+dz^{\dot a}d{\bar z}_{\dot a}\,,\\
F_2&=&4{F},\quad F_4=2i{\tilde F}\wedge dz^{\dot a}\wedge d{\bar z}_{\dot a}\,.
\nonumber
\eea
Interestingly, solution (\ref{LocalSolnIIA3}) can be obtained from (\ref{LocalSolnIIA2}) by swapping $F$ and 
${\tilde F}$, and since these fields appear on the same footing in (\ref{LocalSolnIIAfields}), our formalism does not distinguish between embeddings (\ref{LocalSolnIIA2}) and (\ref{LocalSolnIIA3}). The situation becomes rather different if one insists on using a real harmonic function $H$: as we already saw such restriction leads to a Poincare patch of the AdS space, making $F$ purely electric and ${\tilde F}$ purely magnetic. For such solutions 
(\ref{LocalSolnIIA2}) and (\ref{LocalSolnIIA3}) are interpreted as rather different brane configurations: 
(\ref{LocalSolnIIA2}) corresponds to (\ref{BraneScanIIA2}), while (\ref{LocalSolnIIA3}) is produced by
\bea\label{BraneScanIIA3}
\begin{array}{c|cccccccccc|}
&t&x_1&x_2&x_3&X_1&X_2&X_3&Y_1&Y_2&Y_3\\
\hline
D4_1&\bullet&&&&\bullet&\bullet&\sim&\bullet&\bullet&\sim\\
D4_2&\bullet&&&&\bullet&\sim&\bullet&\bullet&\sim&\bullet\\
D4_3&\bullet&&&&\sim&\bullet&\bullet&\sim&\bullet&\bullet\\
D0_4&\bullet&&&&\sim&\sim&\sim&\sim&\sim&\sim\\
\hline
\end{array}
\eea
These special cases were discussed in \cite{Tseytl96,KlebTseytl}. From our perspective, (\ref{LocalSolnIIA2}) and (\ref{LocalSolnIIA3}) should be viewed as the same embedding of two different {\it regular} bubbling solutions into type IIA supergravity.

\bigskip

To summarize, in this subsection we presented three alternative embeddings of the regular \AdSS\ solutions into type IIA supergravity and discussed lifts to eleven dimensions. The rest of this paper is focused on the type IIB solutions (\ref{LocalSoln})--(\ref{HarmMain}), but all results extend trivially to the embedding 
(\ref{LocalSolnIIA1}), (\ref{LocalSolnIIA2}), (\ref{LocalSolnIIA3}).

\section{Examples}
\renewcommand{\theequation}{4.\arabic{equation}}
\setcounter{equation}{0}
\label{SecExamples}

In this section we will consider several examples of regular geometries (\ref{LocalSoln})--(\ref{HarmMain}). To have complete solutions with all asymptotic regions, we will focus on planar curves, for which the analytic continuation is well understood. 

\subsection{\AdSS\ and its pp-wave limit}
\label{SecExAdS}

Embedding of \AdSS\  into the general solution (\ref{LocalSoln})--(\ref{HarmMain}) has been already discussed in section \ref{SecRegulAdS}, and here we will briefly mention some additional aspects of this embedding. Recall that the \AdSS\  geometry (\ref{GlobAdS}) can be expressed in the form (\ref{LocalSoln}) by defining new coordinates $(t,\phi)$ as
\bea
\phi={\tilde\phi}-{\tilde t},\quad t=L{\tilde t}
\eea
and rewriting (\ref{GlobAdS}) in terms of them:
\bea\label{AdSembPrm}
ds^2&=&-h^{-2}\left(dt-
\frac{L\sin^2\theta d\phi}{\rho^2+\cos^2\theta}\right)^2+h^2 ds_{base}^2+dz^ad{\bar z}_a\nonumber\\
F_5&=&d\rho\wedge \left[\frac{1}{4}
\left(dt-\frac{Ls^2_\theta d\phi}{\rho^2+\cos^2\theta}\right)+\frac{Lh^2}{4}\sin^2\theta d\phi\right]\wedge \mbox{Re}\Omega_3\\
&-&d\cos\theta\wedge \left[\frac{1}{4}\left(dt-\frac{L\sin^2\theta d\phi}{\rho^2+\cos^2\theta}\right)+
\frac{Lh^2}{4} (\rho^2+1)d\phi\right]\wedge \mbox{Im}\Omega_3
\nonumber
\eea
The flat metric on the base, $ds_{base}^2$ is given by (\ref{AdSbase}), and function $h$ is 
\bea
h=\frac{1}{\sqrt{\rho^2+c^2_\theta}}.
\eea
Extracting $A_t$ and ${\tilde A}_t$ from the time components of (\ref{AdSembPrm}),
\bea
A_t=-\frac{\rho}{4},\quad {\tilde A}_t=\frac{\cos\theta}{4},
\eea
we can verify the expressions for the magnetic components of $F_5$:
\bea
\star_3 dA_t=-\frac{L}{4}(\rho^2+1)\sin\theta d\theta\wedge d\phi,\quad
\star_3 d{\tilde A}_t=\frac{L}{4}\sin^2\theta d\rho\wedge d\phi
\eea
and for the harmonic function
\bea\label{AdSSharm}
H=H_1+iH_2=\frac{1}{4i({\tilde A}_t+iA_t)}=\frac{\rho-i\cos\theta}{\rho^2+\cos^2\theta}
\eea
Translation to cylindrical coordinates is given by (\ref{AdSRy}): 
\bea\label{AdSRyTwo}
r=L\sqrt{\rho^2+1}\sin\theta,\quad y=L\rho \cos\theta,\ \Rightarrow\
ds_{base}^2=dy^2+dr^2+r^2d\phi^2.
\eea
Notice that the standard coordinates $(\rho,\theta)$ of \AdSS\ can be viewed as oblate spheroidal  coordinates on the flat base (\ref{AdSbase}), and this seems to be a generic feature of all AdS spaces. As demonstrated in \cite{ChL}, in all known cases where AdS$\times$S space can be written as a fibration over a flat base, the standard parameterization of the global AdS is associated with the oblate spheroidal coordinates on the base. Moreover, supersymmetric geometries can have integrable geodesics if and only if the Hamilton--Jacobi equation separates in ellipsoidal coordinates \cite{ChL}, and the oblate spheroidal parameterization is a special case. 

As discussed in section \ref{SecRegulAdS}, parameterization (\ref{LocalSoln}) of \AdSS\ has a branch cut at $\rho=0$, which corresponds to a disk of radius $L$ in the $y=0$ plane. Expressions (\ref{H12AdS}) on the sheet $A$, which corresponds to $y>0$, satisfy the boundary conditions (\ref{BubblBC}),  and the analytic continuation (\ref{HonSheets}) gives the full \AdSS. Any 
two--dimensional surface surrounding the branch cut is non--contractible, and the flux through it is given by the first expression in (\ref{FluxAsheet}). The integrand,
\bea
&&\frac{1}{2} \mbox{Im}\left[V\wedge d(H^{-1})-i\star_3 dH
\right]_{\rho=0}=\frac{1}{2}\left[s_\theta d\theta\wedge \left(-\frac{Ls_\theta^2 d\phi}{c_\theta^2}\right)+\frac{L}{c_\theta^2}s_\theta d\theta\wedge d\phi
\right]\nn
&&\qquad=\frac{L}{2}s_\theta d\theta\wedge d\phi,
\eea
can be interpreted as an area form for the black droplet, and the integral (\ref{FluxAsheet}),
\bea
\oint_{\cal D} {\tilde F}=\frac{L}{2}\int_0^{\pi/2} s_\theta d\theta\int_0^{2\pi}d\phi=
\frac{L}{4}\int_0^{\pi} s_\theta d\theta\int_0^{2\pi}d\phi,
\eea
must be quantized in the units of $(2\pi^2 l_p^4)/l_T^3$, where $l_p$ is a ten--dimensional Planck length, and $l_T$ is a linear size of $T^6$. Since the \AdSS\ geometry (\ref{AdSembPrm}) does not have compact white droplets, it is impossible to form a topologically nontrivial manifold on the base that carries a nontrivial flux of $F$. 

\begin{figure}[t]
\centering
\includegraphics[width=0.4\textwidth]{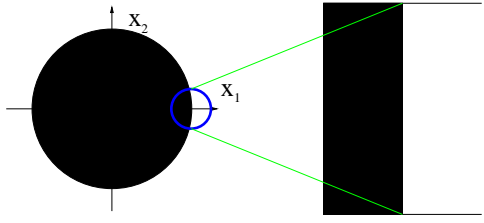} 
\caption{Relation between the droplet picture for \AdSS\ and the pp-wave: a graphical representation of the limit (\ref{ppScale}).}
\label{FigPPwave}
\end{figure}

The pp-wave limit of the geometry (\ref{AdSembPrm}) is obtained in a standard way \cite{ppWave} by zooming in on a vicinity of the singular curve. For the harmonic function (\ref{ComplAdS}) this implies a limit
\bea\label{ppScale}
L\rightarrow\infty \quad\mbox{with fixed}\quad {\tilde y}=yL,
\quad{\tilde x}=(r-L)L,\quad{\tilde\phi}=L^2\phi,\quad{\tilde t}=\frac{t}{L},
\eea
which leads to
\bea
H=\frac{L}{\sqrt{{\tilde x}-i{\tilde y}}},\quad 
{\tilde H}\equiv \frac{H}{L}=\frac{1}{\sqrt{{\tilde x}-i{\tilde y}}},
\eea
and the relevant coloring of the $y=0$ plane is depicted in figure \ref{FigPPwave}. The resulting geometry is
\bea
ds^2&=&-{\tilde h}^{-2}(d{\tilde t}+{\tilde V})^2+{\tilde h}^2 d{\tilde x}_ad{\tilde x}_a+dz^{\dot a}d{\bar z}_{\dot a}\nn
&=&2d{\tilde t}d{\tilde \phi}-[{\tilde\rho}^2+{\tilde\theta}^2]d{\tilde t}^2+d{\tilde\rho}^2+d{\tilde\theta}^2+dz^ad{\bar z}_a\\
F_5&=&\frac{1}{4}d{\tilde\rho}\wedge d{\tilde t}\wedge \mbox{Re}\Omega_3-
\frac{1}{4}d{\tilde\theta}\wedge d{\tilde t}\wedge \mbox{Im}\Omega_3
\nonumber
\eea
and as demonstrated in section \ref{SecRegul}, this is a generic behavior of the metric in a vicinity of the singular curve (see equation (\ref{SingMetrTmp1})). 

In the next subsection we will construct regular geometries corresponding to small perturbations of \AdSS, and light excitations of the pp-wave can be obtained from them by taking the limit (\ref{ppScale}). 

\subsection{Perturbative solution}

After discussing the \AdSS\ solution corresponding to the ground state of the system with a given amount of flux, we consider perturbations of this geometry. The light excitations are describes by the `gravitons', i.e., by combinations of the metric and fluxes, and coupled equations for such degrees of freedom have been extensively discussed in the literature for various AdS spaces and spheres \cite{gravSpectr,gravSpectrAdS2}. Although one can perform a similar analysis for \AdSS\ \cite{gravSpectrAdS2}, here we are interested in supersymmetric excitations, which are guaranteed to be covered by our ansatz (\ref{LocalSoln}), so the study of `gravitons' reduces to the analysis of small perturbations in the complex harmonic function $H$ parameterizing the bubbling solution (\ref{LocalSoln}). In this subsection we will expand $H$ around $H_0$ corresponding to \AdSS\ and construct the solutions describing small regular perturbations.

We will focus on the sector corresponding to planar curves, where `gravitons' correspond to small changes in the shape of the circles. Such ripples have been studied for the AdS spaces in higher dimensions, where geometries can be written explicitly in terms of functions parameterizing the curves \cite{Parad,LLM}. While in the present case it is difficult to solve the Laplace equation with arbitrary boundary conditions  (\ref{BubblBC}), small perturbations around \AdSS\ can be found explicitly. First we note that an arbitrary ripple on the circular shape with radius $L$ can be parameterized in polar coordinates as
\bea\label{RipplePolar}
x_1+ix_2=Le^{i\phi}+\sum a_m e^{im\phi+i\phi},
\eea
where the sum is assumed to be infinitesimal in comparison to the leading contribution. Every profile (\ref{RipplePolar}) generates a solution with harmonic function 
\bea
H=H_0+H',
\eea
where $H_0$ is given by (\ref{ComplAdS}), (\ref{AdSSharm}),
\bea\label{AdSSharm000}
H_0\equiv\frac{\rho-i\cos\theta}{\rho^2+\cos^2\theta}\,,
\eea
and every set of amplitudes $a_m$ in (\ref{RipplePolar}) translates into a particular mode expansion in $H'$:
\bea
H'=\sum h_m(\rho,\theta) e^{im\phi}. 
\eea
We will now determine the functional form of $h_m$ by solving the Laplace equation for $H'$ and imposing regularity conditions on the geometry (\ref{LocalSoln}). 

Writing  $H=H_0+H'$ and expanding the metric (\ref{LocalSoln}) to the first order in $H'$, we find  
\bea\label{PertMetr}
ds^2&=&\left[1+\frac{H'}{H_0}+\frac{\bar H'}{\bar H_0}\right]ds_0^2+
\frac{2}{h^{2}}\left[\frac{H'}{H_0}+\frac{\bar H'}{\bar H_0}\right](dt+V_0)^2-2h^{-2}dt V'
\eea
Here $ds_0^2$ is the metric (\ref{LocalSoln}) for  the \AdSS\ space, and $V'$ is the vector field corresponding to $H'$. To ensure regularity of (\ref{PertMetr}) in the vicinity of the singular curve, 
it is sufficient to require 
\bea\label{HprimeReg}
\frac{H'}{H_0}+\frac{\bar H'}{\bar H_0}\sim (\rho^2+\cos^2\theta)
\eea
for small $\rho$ and $\cos\theta$. This implies that $H'$ should vanish at least as $\rho$ or as $\cos\theta$. 

To construct the relevant solutions, we observe that the Laplace equation for function $H'$ on the flat base (\ref{AdSbase}) is equivalent to the wave equation on the \AdSS\ (\ref{AdSemb1}),
\bea\label{AdSSfive}
ds^2=-(\rho^2+\cos^2\theta)\left(dt-
\frac{L\sin^2\theta d\phi}{\rho^2+\cos^2\theta}\right)^2+L^2\left[\frac{d\rho^2}{\rho^2+1}+d\theta^2+
\frac{\sin^2\theta (\rho^2+a^2)d\phi^2}{\rho^2+\cos^2\theta}\right],
\eea
with an additional assumption of $t$--independence. Going to the standard coordinates of \AdSS\  by shifting and rescaling coordinates as
\bea
{\tilde\phi}=\phi+{\tilde t},\quad {\tilde t}=\frac{t}{L},
\eea
we conclude that $H'$ can depend only on three coordinates $(\rho,\theta,{\tilde\phi}-{\tilde t})$. The wave equation on the \AdSS
\bea\label{AdSmetrForPert}
ds^2=L^2\left[-(1+\rho^2)d{\tilde t}^2+\frac{d\rho^2}{\rho^2+1}\right]+
L^2\left[d\theta^2+\sin^2\theta d{\tilde\phi}^2\right]
\eea
separates between two subspaces, and, to ensure the $t$--independence of $H'$, we are looking for solutions which have the form
\bea\label{HprimePert}
H'=R(\rho)\Theta(\theta)e^{im{\tilde\phi}-im{\tilde t}}.
\eea
Function $R(\rho)$ must vanish at infinity to preserve the \AdSS\ asymptotics, and this implies that $R(0)\ne 0$. Then to ensure regularity at the singular curve (see (\ref{HprimeReg})), function $\Theta(\theta)$ must vanish at $\theta=\frac{\pi}{2}$. Since the wave equation separates between the AdS space and the sphere, the angular part of the function (\ref{HprimePert}) can be written as a superposition of spherical harmonics,
\bea\label{HprimePert2}
\Theta(\theta)e^{im{\tilde\phi}}=Y_{l,m}(\theta,{\tilde\phi}),
\eea
and the only harmonics that vanish at $\theta=\frac{\pi}{2}$ are 
\bea\label{HprimePert3}
Y_{|m|+1,m}(\theta,{\tilde\phi})\propto e^{im{\tilde\phi}}[\sin\theta]^{|m|}\cos\theta
\eea
Substituting (\ref{HprimePert2}) with $l=|m|+1$ into (\ref{HprimePert}) and writing the wave equation for $H'$ in the metric (\ref{AdSmetrForPert}), we arrive at an ordinary differential equation for $R(\rho)$,
\bea\label{PertRadEqn}
\frac{(l-1)^2R}{1+\rho^2}+\d_\rho[(1+\rho^2)\d_\rho R]-l(l+1) R=0,\quad l=|m|+1,
\eea
which can be solved in terms of the associated Legendre functions. In particular, the solution that vanishes at 
$\rho=\infty$ is
\bea\label{PertRl}
R_l(\rho)=(-1)^l\left[Q^{l-1}_l(i\rho)-\frac{\pi i}{2}P^{l-1}_l(i\rho)\right],
\eea
and it approaches zero as
\bea
R_l(\rho)\sim \frac{1}{\rho^{l+1}}.
\eea
As expected, all radial functions $R_l$ vanish faster than $H_0\sim \frac{1}{\rho}$. 
The first few cases of (\ref{PertRl}) are given by
\bea
R_1(\rho)&=&1+\rho\left[\arctan\rho-\frac{\pi}{2}\right],\nn
R_2(\rho)&=&\frac{2+3\rho^2+3\rho(1+\rho^2)[[\arctan\rho-\frac{\pi}{2}]}{\sqrt{\rho^2+1}},\\
R_3(\rho)&=&\frac{8+25\rho^2+15\rho^4+15\rho(1+\rho^2)^2[\arctan\rho-\frac{\pi}{2}]}{\rho^2+1}.
\nonumber
\eea

Rewriting the function (\ref{HprimePert}) in the original coordinates used in (\ref{AdSSfive}), we arrive at the final expression:
\bea\label{FinalHprime}
H_l^{\prime(\pm)}=(-1)^l\left[Q^{l-1}_l(i\rho)-\frac{\pi i}{2}P^{l-1}_l(i\rho)\right]\sin^{l-1}\theta\cos\theta e^{\pm i(l-1)\phi}
\eea
Harmonic function (\ref{FinalHprime}) gives rise to regular perturbations of the \AdSS\ geometry via (\ref{LocalSoln})--(\ref{HarmMain}), and it corresponds to exciting the $(l-1)$-st harmonic on a circle. Notice that (\ref{FinalHprime}) does not satisfy the boundary conditions (\ref{BubblBC}) since it corresponds to an infinitesimal perturbation, but a condensate of such modes would obey (\ref{BubblBC}). Some particular condensate deforms the circle into an ellipse, and an explicit solution for this case will be constructed in the next subsection.

\subsection{Elliptical droplet}

Although finding solutions with mixed boundary conditions (\ref{BubblBC}) is not easy, some examples can be constructed using separation of variables. It is well--known that Laplace equation in three dimensions separates only in ellipsoidal coordinates and in their degenerate cases \cite{MorseFesch}, and the standard coordinates of \AdSS\ defined by (\ref{AdSRyTwo}) correspond to such a degenerate case. We will now consider a more general situation involving generic ellipsoidal coordinates and use them to construct a harmonic function $H$ satisfying conditions  (\ref{BubblBC}) with an elliptical droplet.

We begin with recalling the ellipsoidal coordinates on ${\bf R}^3$ using the notation of \cite{LandLif}. Starting from the Cartesian coordinates 
$(x_1,x_2,y)$, one defines the ellipsoidal coordinates as solutions of a cubic equation for $u$:
\bea\label{DefElliptEqn}
\frac{(x_1)^2}{a^2+u}+\frac{(x_2)^2}{b^2+u}+\frac{y^2}{c^2+u}=1,
\eea
where $(a,b,c)$ are some positive constants. Without loss of generality, we assume that 
\bea
a\ge b\ge c\ge 0.\nonumber
\eea
Denoting the solutions of (\ref{DefElliptEqn}) by $(\xi,\eta,\zeta)$, one can find the explicit formulas for the Cartesian coordinates:
\bea\label{DefEllips}
&&x_1=\left[\frac{(\xi+a^2)(\eta+a^2)(\zeta+a^2)}{(b^2-a^2)(c^2-a^2)}\right]^{1/2},\quad
x_2=\left[\frac{(\xi+b^2)(\eta+b^2)(\zeta+b^2)}{(a^2-b^2)(c^2-b^2)}\right]^{1/2},\nonumber\\
&&y=\left[\frac{(\xi+c^2)(\eta+c^2)(\zeta+c^2)}{(c^2-a^2)(c^2-b^2)}\right]^{1/2},\quad
\xi\ge -c^2\ge \eta\ge -b^2\ge \zeta\ge -a^2
\eea
In the ellipsoidal coordinates the metric of the flat space becomes
\bea\label{EllpsMetr}
ds^2&=&(dx_1)^2+(dx_2)^2+dy^2\\
&=&(\xi-\eta)(\xi-\zeta)(\eta-\zeta)\left[\frac{d\xi^2}{4(\eta-\zeta)R(\xi)}-\frac{d\eta^2}{4(\xi-\zeta)R(\eta)}+\frac{d\zeta^2}{4(\xi-\eta)R(\zeta)}\right]\nn
R(t)&=&(t+a^2)(t+b^2)(t+c^2).\nonumber
\eea
In section \ref{SecExAdS} we used the oblate spheroidal coordinates, which are obtained by taking the limit $b\rightarrow a$ while keeping $\xi,\eta$ and
\bea
\frac{\zeta+a^2}{a^2-b^2}\equiv \cos^2\phi\nonumber
\eea
fixed. Then defining coordinates $(\rho,\theta)$ by 
\bea\label{AngCoordElpsd}
\xi=L^2\rho^2-c^2,\quad \eta=-c^2+(c^2-b^2)\cos^2\theta,\quad L^2\equiv a^2-c^2,
\eea
we recover the transformation (\ref{AdSRyTwo}):
\bea
x_1+ix_2=L\sqrt{\rho^2+1}\sin\theta e^{i\phi},\quad y=L\rho\cos\theta.
\eea
In particular, the disk $\rho=0$ corresponds to $\xi=-c^2$, and the rest of the $y=0$ plane corresponds 
to $\eta=-c^2$. This pattern persists for the elliptical droplet as well: as we will see, the interior of the ellipse corresponds to $\xi=-c^2$, its exterior corresponds to $\eta=-c^2$, and the ratio $a/b$ determines the eccentricity of the ellipse. 

Going back to the general ellipsoidal coordinates (\ref{DefEllips}) and setting $\xi=-c^2$ we find
\bea\label{ElpsIn}
x_1=\left[\frac{(\eta+a^2)(\zeta+a^2)}{(a^2-b^2)}\right]^{1/2},\quad
x_2=\left[\frac{(\eta+b^2)(\zeta+b^2)}{(b^2-a^2)}\right]^{1/2},\quad y=0
\eea
The range of $\zeta$ allows us to define a new angular coordinate $\phi$ by
\bea
\phi=\arccos\sqrt{\frac{\zeta+a^2}{a^2-b^2}},
\eea
then 
\bea
x_1=\sqrt{\eta+a^2}\cos\phi,\quad x_2=\sqrt{\eta+b^2}\sin\phi,
\eea
and for the allowed values of $\eta$ the coordinates in the plane satisfy an inequality
\bea\label{EllpsInt}
\frac{(x_1)^2}{a^2-c^2}+\frac{(x_2)^2}{b^2-c^2}\le 1. 
\eea
Thus (\ref{ElpsIn}) describes the interior of an ellipse. Similarly, the region $\eta=-c^2$ can be parameterized by $\xi$ and $\phi$ as
\bea
&&x_1=\sqrt{\xi+a^2}\cos\phi,\quad
x_2=\sqrt{\xi+b^2}\sin\phi,\quad
y=0,
\eea
and this describes the exterior of the same ellipse. 
 
To determine the harmonic functions $(H_1,H_2)$, we have to solve the Laplace equation on the flat base and impose the boundary conditions in the interior and exterior of the ellipse. The functions 
\bea\label{EllptHarm}
&&H_1=\frac{\sqrt{(\xi+a^2)(\xi+b^2)(\xi+c^2)}}{(\xi-\eta)(\xi-\zeta)},\quad
H_2=\frac{\sqrt{(\xi+a^2)(\xi+b^2)(\eta+c^2)(\zeta+c^2)}}{(\xi-\eta)(\xi-\zeta)\sqrt{a^2-c^2}}\nn
&&h^2=\frac{(\xi+a^2)(\xi+b^2)}{(\xi-\eta)^2(\xi-\zeta)^2(a^2-c^2)}[(a^2-c^2)(\xi-\eta)+(\zeta+a^2)(\eta+c^2)]
\eea
have the correct behavior in the $y=0$ plane, and a direct calculation shows that $H_1$ and $H_2$ are harmonic in the flat space with the metric (\ref{EllpsMetr}).
At large values of $\xi$, which correspond to infinity of ${\bf R}^3$, we also find the correct behavior:
\bea
&&H_1=\frac{1}{\sqrt{\xi}},\quad
H_2=\frac{\sqrt{(\eta+c^2)(\zeta+c^2)}}{\xi\sqrt{a^2-c^2}}=\frac{y\sqrt{b^2-c^2}}{\xi^{3/2}}\ll H_1
\eea
Recall that at large values of $\xi$, it is $\sqrt{\xi}$ that plays the role of the radial coordinate of 
${\bf R}^3$ (see (\ref{DefEllips})). An explicit expression for $V$ corresponding to the harmonic functions (\ref{EllptHarm}) can be found, but it is not very illuminating. 

We conclude this subsection by writing the approximate expressions for $H_1$ and $H_2$ in the vicinity of the singular curve. To do so, we introduce the counterparts of coordinates $(\rho,\theta,\phi)$ used for the circular droplet:
\bea
&&\xi=L^2\rho^2-c^2,\quad \eta=-c^2-L^2\cos^2\theta,\quad\zeta=-a^2+L^2q^2\cos^2\phi.
\eea
Here we defined two convenient constants:
\bea
L^2=b^2-c^2,\quad q^2=\frac{a^2-b^2}{b^2-c^2},
\eea
which control the size of the ellipse and its eccentricity $e$. Specifically, the relation (\ref{EllpsInt}) for the 
interior of the ellipse can be written as 
\bea 
(x_2)^2+\frac{(x_1)^2}{1+q^2}=L^2,
\eea
so the eccentricity of the ellipse is given by
\bea
e=\left[1+\frac{1}{q^2}\right]^{-1/2}.
\eea

In the vicinity of the singular curve we find the approximate expressions for the Cartesian coordinates,
\bea
x_1&\simeq&\sqrt{\frac{a^2-c^2}{b^2-c^2}}L\cos\phi,\quad x_2=L\sin\phi,
\nonumber\\
y&\simeq&\frac{L^2\rho\cos\theta\sqrt{a^2-c^2-L^2q^2\cos^2\phi}}{\sqrt{(a^2-c^2)(b^2-c^2)}},
\eea
and for the harmonic functions
\bea
H_1&\simeq&\frac{\sqrt{(a^2-c^2)(b^2-c^2)}L\rho}{L^2(\rho^2+\cos^2\theta)(a^2-c^2-L^2q^2\cos^2\phi)},\\
H_2&\simeq&\frac{\sqrt{(a^2-c^2)(b^2-c^2)}L\cos\theta}{L^2(\rho^2+\cos^2\theta)\sqrt{a^2-c^2}\sqrt{a^2-c^2-L^2q^2\cos^2\phi}}.\nonumber
\eea
As before, the singular curve is located at $\rho=0$, $\theta=\frac{\pi}{2}$. 

To summarize, in this subsection we have presented an interesting example of an explicit solution that goes beyond the circular droplet. The success in constructing this example is based on our ability to solve the boundary problem (\ref{BubblBC}) using separation of variables. Unfortunately, the boundary conditions (\ref{BubblBC}) for a generic droplet are not amenable to an analytical treatment, but for every shape the solution exists, and it is unique. 

\subsection{Asymptotically--flat solution}

So far we have been focusing on regular geometries which approach \AdSS\ at infinity, and it might be interesting to look for asymptotically flat solutions as well. One example of such solution is given by (\ref{D3stackGeom}), but this geometry has a singularity at $r=0$. This is an example of a general situation for AdS$_p$ with $p>3$: the regular solutions are described by bubbling geometries, which cannot be connected to flat space, while the asymptotically flat configurations of branes can only produce a singular Poincare patch of the AdS space. This dichotomy stems from different boundary conditions for the fermions on global AdS and on its Poincare patch, and only the latter can be glued to flat space. The situation is rather different in the AdS$_3$ case, where the global AdS can be connected to flat space via the spectral flow procedure developed in \cite{MMaoz}. Although the fermions on the global AdS and on the Poincare patch still have different boundary conditions (they correspond to the Neveu-Schwarz and to the Ramond sectors of the dual field theory), one can go from one description to another by performing a spectral flow on the boundary \cite{SpecFlow}, which corresponds to a diffeomorphism in the bulk\footnote{If a similar procedure existed in higher dimensions, it would have interpolated between the bubbling solutions of \cite{LLM} and the geometries corresponding to the Coulomb branch of the field theory constructed in \cite{Coulomb}.}. Specifically, starting from the NS vacuum described by the AdS$_3\times$S$^3$ geometry (\ref{AdS33}), one can go to one of the Ramond vacua by mixing the sphere and AdS coordinates as
\bea
{\tilde\phi}=\phi+t,\quad {\tilde\psi}=\psi+\chi
\eea
and using $(\theta,{\tilde\phi},{\tilde\psi})$ rather than $(\theta,{\phi},{\psi})$ to parameterize the sphere at infinity. As demonstrated in \cite{MMaoz}, coordinates $(\rho,\theta,{\tilde\phi},{\tilde\psi})$ can be extended to the asymptotically flat region, where they parameterize ${\bf R}^4$. We will now demonstrate that solutions (\ref{LocalSoln}) can accommodate a similar interpolation between a regular interior of the global AdS$_2$ and the flat space. 

To connect the global \AdSS\ (\ref{GlobAdS}) and the flat space, it is convenient to write the metric on the base in terms of the oblate spheroidal coordinates (\ref{AdSbase}):
\bea
ds_{base}^2&=&L^2\left[\frac{(\rho^2+\cos^2\theta)d\rho^2}{\rho^2+1}+(\rho^2+\cos^2\theta)d\theta^2+
{\sin^2\theta (\rho^2+1)d\phi^2}\right]\,.
\eea
The infinities of the two $(x_1,x_2,x_3)$ sheets correspond to $\rho=\pm\infty$, and the harmonic function (\ref{AdSSharm}) describing \AdSS\ approaches zero in both regions. For the flat space function $H$ should approach a constant, and since at large values of $\rho$ the real part of (\ref{AdSSharm}) dominates, it is natural to look for flat region where $H_1$ approaches a constant and adjust  $H_2$ accordingly\footnote{This may not be the only option, but here we are interested in constructing just one example.}. Writing the harmonic functions as
\bea
H_1=\frac{\rho}{\rho^2+\cos^2\theta}+h_1,\quad
H_2=\frac{\cos\theta}{\rho^2+\cos^2\theta}+h_2,
\eea
we conclude that function $h_1$ should vanish at $\rho=0$ to satisfy the boundary condition (\ref{BubblBC}) with a black circle in the $y=0$ plane, and it must approach a constant when $\rho$ goes to infinity. The easiest way to satisfy these requirements is to assume that $h_1$ depends only on one variable $\rho$, then the Laplace equation for $h_1$ has a unique solution with the desired properties:
\bea\label{FlatOne}
\d_\rho\left[(\rho^2+1)\d_\rho h_1\right]=0\quad\Rightarrow\quad h_1=\frac{2c_1}{\pi}\arctan(\rho)\,.
\eea
Here $c_1$ is the asymptotic value of $h_1$. To determine the function $h_2$, one can go to the Cartesian coordinate and ensure regularity by requiring that the complex harmonic function $H$ has the form (\ref{hhOne}) with function $f$ given by (\ref{hhTwo}). A simpler way of ensuring regularity is to notice that function
\bea\label{FlatTwo}
h_2=\frac{c_2\cos\theta}{1+\rho\arctan \rho}
\eea
is harmonic, it satisfies the correct boundary conditions in the $y=0$ plane ($\rho=0$ or $\theta=\frac{\pi}{2}$), and it vanishes at infinity. Enforcement of regularity on a circle, which amounts to imposing (\ref{hhTwo}), leads to a relation between $c_1$ and $c_2$. Specifically, near the singularity we find
\bea
H=H_1+H_2=\frac{1}{\rho-i\cos\theta}+\frac{2c_1}{\pi}\rho+ic_2\cos\theta+O(\rho^2+\cos^2\theta),
\eea
and this becomes a function of $\rho-i\cos\theta$ (an analog of $z+iy$ in (\ref{hhTwo})) if
\bea\label{C2AsC1}
c_2=-\frac{2c_1}{\pi}.
\eea
For this value the definition (\ref{HarmMain}) of the $V$ field gives
\bea\label{FlatVfield}
\frac{V_\phi}{Ls^2_\theta}=-\frac{1}{\rho^2+c^2_\theta}-\left\{
\frac{4c_1\rho[\rho+(1+\rho^2)\arctan\rho]}{\pi(\rho^2+c^2_\theta)}+
\frac{4c_1^2}{\pi^2}[(1+\rho^2)\arctan^2\rho-1]\right\}.
\eea
The leading term corresponds to the \AdSS\ space, and the expressions in the curly brackets remain finite in the vicinity of the 
circle $\rho^2+\cos^2\theta=0$, so the metric remains regular. To see this, it is sufficient to look at the $(t,\phi)$ sector:
\bea
ds_2^2&=&-\frac{1}{|H|^2}(dt+V_\phi d\phi)^2+|H|^2 L^2(1+\rho^2)\sin^2\theta d\phi^2\nonumber\\
&=&-\frac{1}{|H|^2}\left[dt+V_\phi d\phi-L|H|^2 \sqrt{1+\rho^2}\sin\theta d\phi\right]
\left[dt+V_\phi d\phi+L|H|^2 \sqrt{1+\rho^2}\sin\theta d\phi\right]\nonumber\\
&\simeq&2Ld\phi\left[dt+\frac{1}{2}d\phi+\frac{4c_1(c_1-\pi)}{\pi^2}d\phi\right]
\eea
If the relation (\ref{C2AsC1}) is not imposed, then $V_\phi$ has logarithmic singularities, e.g., $c_2=0$ gives
\bea
V_\phi&=&-\frac{L\sin^2\theta}{\rho^2+\cos^2\theta}\left[1+
\frac{4c_1}{\pi}\rho\arctan\rho\right]-\frac{2 Lc_1}{\pi}\ln\frac{1+\rho^2}{\rho^2+\cos^2\theta}
\eea
Such logarithms lead to geometries with shock waves \cite{AchSexl}. Similar singularities have been encountered in the AdS$_3\times$S$^3$ case \cite{LMpWave}, where it was shown that shock waves can be removed by perturbing the sources \cite{AdS3shock}. A similar resolution for solutions (\ref{FlatOne}), (\ref{FlatTwo}) violating (\ref{C2AsC1}) might also be possible, but we will not discuss this further.  

To summarize, we have constructed an example of an asymptotically flat regular geometry, and it is given by (\ref{LocalSoln}), (\ref{HarmMain}) with 
\bea\label{HarmFlatTrans}
H=H_1+iH_2=\frac{1}{\rho-i\cos\theta}+\frac{2c_1}{\pi}\arctan(\rho)-\frac{2i c_1}{\pi}\frac{\cos\theta}{1+\rho\arctan \rho}
\eea
and $V_\phi$ from (\ref{FlatVfield}). The resulting metric has two length scales: one determined the AdS radius, and the other one defines the scale of the transition between the AdS and flat regions. At sufficiently large values of $|\rho|$, the second term in (\ref{HarmFlatTrans}) dominates, and the geometry can be approximated by a flat metric. As expected, there are two such regions: they come from two copies of ${\bf R}^3$  in (\ref{LocalSoln}), and they correspond to positive and negative values of $\rho$. The transition to the near horizon regime happens when 
\bea\label{TransScale}
\frac{1}{\sqrt{\rho^2+\cos^2\theta}}\sim \frac{2c_1}{\pi}\arctan(\rho)
\eea
and the size of the AdS region, $\rho_{trans}$ depends on the value of $c_1$ through (\ref{TransScale}). On the other hand, the radius of the AdS space is $L$ (or one, if measured in $\rho$ coordinate), so a meaningful AdS region exists only if 
$\rho_{trans}\gg 1$, in other words, if parameter $c_1$ is small. 

In the AdS$_3$ case an extension of geometries from the AdS region to flat asymptotics was accomplished by adding one to the harmonic function \cite{Parad}, but now such procedure is more complicated even for the simplest state (\ref{HarmFlatTrans}).
It would be interesting to find the general algorithm for extending all solutions (\ref{LocalSoln}) from the near horizon geometry to an asymptotically flat space.

\section{Brane probes on bubbling geometries}
\renewcommand{\theequation}{5.\arabic{equation}}
\setcounter{equation}{0}
\label{SecBrane}

In section \ref{SecProbesAdS} we analyzed supersymmetric branes on \AdSS, and in section \ref{SecMain} we constructed geometries produced by such objects. Gravitational backreaction becomes important only when many branes are put on top of each other, and it might be interesting to study dynamics of one {\it additional} brane on the geometry produced by such stacks. This dynamics is governed by the DBI action for the probes placed on (\ref{LocalSoln})--(\ref{HarmMain}), and in this section we will analyze the behavior of such probes.

Supersymmetric D3 branes on (\ref{LocalSoln})--(\ref{HarmMain}) must wrap three directions on $T^6$, and it is convenient to introduce real coordinates $X_a$, $Y_a$ instead of complex $z_a$ used in  (\ref{LocalSoln}) by writing
\bea
z^a=X_a+iY_a
\eea
We begin with discussing a brane that wraps directions $(X_a,Y_a)$ is a specific way, and we will comment on the general situation in the end of this section. Let us assume that a brane appears as a point in the $(Y_2,Y_3)$ subspace and wraps $(X_2,X_3)$ as well as a line in the $(X_1,Y_1)$ plane. Then the following static gauge can be imposed
\bea\label{StaticGaugeGen}
t=\tau,\quad X_1=\xi_1\cos\beta,\quad Y_1=\xi_1\sin\beta,\quad X_2=\xi_2,\quad X_3=\xi_3,\quad Y_2=Y_3=0.
\eea
Assuming that $(x_1,x_2,x_3)$ are functions of $\tau$, we find the action for the D3 brane\footnote{The origin of the factor of four in the Chern--Simons term is explained in the footnote \ref{ftnOne} on page \pageref{ftnOne}.}:
\bea\label{BraneActFive}
S&=&-T\int d^4\xi\sqrt{-\mbox{det}\left[g_{mn}\frac{\d x^m}{\d\xi^a}\frac{\d x^n}{\d\xi^b}\right]}+
4T\int P[C_4]\\
&=&-T\int d^4\xi h^{-1}\sqrt{(1+V_i{\dot x}_i)^2-h^4{\dot x_i}{\dot x_i}}-
4T\int d^4\xi [c_\beta A_t+s_\beta {\tilde A}_t](1+V_i{\dot x}_i)
\nn
&&+4T\int d^4\xi \left(c_\beta [h^2\star_3 d{\tilde A}_t]_i+s_\beta [h^2\star_3 d{ A}_t]_i\right){\dot x}_i\nonumber
\eea
Equations of motion are solved by constant $(x_1,x_2,x_3)$, as long as the following constraints are satisfied:
\bea\label{BraneProj1}
\d_m\left[h^{-1}+4c_\beta A_t+4s_\beta {\tilde A}_t\right]=0.
\eea
Solution (\ref{LocalSoln}) has
\bea\label{AtProbe}
A_t=-\frac{1}{4h}s_\alpha,\quad {\tilde A}_t=-\frac{1}{4h}c_\alpha,
\eea 
so relation (\ref{BraneProj1}) can be rewritten as
\bea
\d_m\left[h^{-1}\{1-\sin(\alpha+\beta)\}\right]=0.
\eea
The easiest way to satisfy this constraint is to make $\beta$ a coordinate dependent quantity and to set
\bea\label{BetaAsAlpha}
\beta=\frac{\pi}{2}-\alpha.
\eea
Such configurations solve all equations of motion, moreover, the action (\ref{BraneActFive}) vanishes on the solutions, and this property often indicates an unbroken supersymmetry. 

To identify the supersymmetries preserved by the rotating branes, we recall the kappa--symmetry projection associated with a D3 brane \cite{KappaProj}:
\bea
\Gamma\eta=\eta,\quad \Gamma=i\sigma_2\otimes\left[{\cal L}^{-1}\left(\prod_{a=0}^3\frac{\d x^{m_a}}{\d\xi^a}\right)\gamma_{m_0\dots m_3}\right],\quad
{\cal L}=\sqrt{-\mbox{det}\left[g_{mn}\frac{\d x^m}{\d\xi^a}\frac{\d x^n}{\d\xi^b}\right]}
\eea
Rotating brane (\ref{StaticGaugeGen}) in the geometry (\ref{LocalSoln}) has\footnote{As in section \ref{SecProbesAdS}, $\Gamma_{\hat m}$ denote the gamma matrices with flat indices (see equation (\ref{FlatDirac})).}
\bea\label{GammaProjIntermG}
\Gamma&=&i\sigma_2\otimes\frac{1}{h^{-1}}\left[h^{-1}\Gamma_{\hat t}\right]
(c_\beta \Gamma_{\hat X_1}+s_\beta\Gamma_{\hat Y_1})\Gamma_{\hat X_2}\Gamma_{\hat X_3}\nn
&=&e^{-\h\beta\Gamma_{\hat X_1\hat Y_1}} 
\left[i\sigma_2\otimes  \Gamma_{\hat t} 
\Gamma_{\hat X_1}\Gamma_{\hat X_2}\Gamma_{\hat X_3}
\right]e^{\h\beta\Gamma_{\hat X_1\hat Y_1}}\nonumber
\eea
The last relation implies that the brane preserves supersymmetry satisfying a projection that depends on $\beta$:
\bea\label{GammaProjProbeG}
\Gamma\eta=\eta:\qquad 
\eta=e^{-\h\beta\Gamma_{\hat X_1\hat Y_1}}\eta_0,\quad
i\sigma_2\otimes  \Gamma_{\hat t} 
\Gamma_{\hat X_1}\Gamma_{\hat X_2}\Gamma_{\hat X_3}\eta_0=\eta_0.
\eea
The brane is supersymmetric if and only if the last relation is consistent with the projection imposed in (\ref{LocalSoln}),
\bea\label{4DprojMetr}
{\tilde\eta}&=&h^{-1/2}e^{i\alpha\Gamma_5/2}{\tilde\eps},\qquad \Gamma^{\bf t}\Gamma_5{\tilde\eps}={\tilde\eps},
\eea
in particular, the coordinate dependences of the projectors (\ref{GammaProjProbeG}) and (\ref{4DprojMetr}) must match.

Notice that relations (\ref{GammaProjProbeG}) are written for a spinor in ten dimensions, while (\ref{4DprojMetr}) are formulated in terms of a reduced four--dimensional object, and the relation between the two is described in Appendix \ref{AppReduction}:
\bea
\eta&=&{\tilde\eta}_{+++}\otimes\left(\begin{array}{c}1\\ 0\end{array}\right)
\otimes\left(\begin{array}{c}1\\ 0\end{array}\right)
\otimes\left(\begin{array}{c}1\\ 0\end{array}\right)+
{\tilde\eta}_{---}\otimes\left(\begin{array}{c}0\\ 1\end{array}\right)
\otimes\left(\begin{array}{c}0\\ 1\end{array}\right)\otimes\left(\begin{array}{c}0\\ 1\end{array}\right),\nn
{\tilde\eta}&=&{\tilde\eta}_{+++}+{\tilde\eta}_{---},\quad 
\gamma_5 {\tilde\eta}_{+++}=-{\tilde\eta}_{+++},\quad
\gamma_5 {\tilde\eta}_{---}={\tilde\eta}_{---}
\eea 
Writing similar relations for ${\eps}$ and ${\tilde\eps}$, we find
\bea
\eps&\equiv& {\tilde\eps}_{+++}\otimes\left(\begin{array}{c}1\\ 0\end{array}\right)
\otimes\left(\begin{array}{c}1\\ 0\end{array}\right)
\otimes\left(\begin{array}{c}1\\ 0\end{array}\right)+
{\tilde\eps}_{---}\otimes\left(\begin{array}{c}0\\ 1\end{array}\right)
\otimes\left(\begin{array}{c}0\\ 1\end{array}\right)\otimes\left(\begin{array}{c}0\\ 1\end{array}\right),\nn
&=&h^{1/2}\left[
e^{\frac{i}{2}\alpha}{\tilde\eta}_{+++}\otimes\left(\begin{array}{c}1\\ 0\end{array}\right)
\otimes\left(\begin{array}{c}1\\ 0\end{array}\right)
\otimes\left(\begin{array}{c}1\\ 0\end{array}\right)\right.\nn
&&\qquad\left.+
e^{-\frac{i}{2}\alpha}{\tilde\eta}_{---}\otimes\left(\begin{array}{c}0\\ 1\end{array}\right)
\otimes\left(\begin{array}{c}0\\ 1\end{array}\right)\otimes\left(\begin{array}{c}0\\ 1\end{array}\right)
\right]\nn
&=&h^{1/2}\left[I_4\otimes e^{i\alpha\sigma_3/2}\otimes 1\otimes 1\right]\eta=
h^{1/2}e^{\frac{i}{2}\Gamma_{\hat X_1\hat Y_1}\alpha}\eta.
\eea
Here we used the explicit expressions (\ref{GammaXYZ}) for the gamma matrices $\Gamma_{\hat X_1}$ and 
$\Gamma_{\hat X_2}$ and for their product:
\bea
\Gamma_{\hat X_1\hat Y_1}=I_4\otimes (i\sigma_3)\otimes 1\otimes 1,
\eea
To summarize, we found that a ten--dimensional spinor $\eta$ for the solution  (\ref{LocalSoln}) can be expressed in terms of a constant spinor $\eps$ as
\bea\label{KillSpinMetr}
\eta=h^{-1/2}e^{\frac{i}{2}\alpha\Gamma_{\hat X_1\hat Y_1}}\eps=
h^{-1/2}e^{-\frac{i}{2}\beta\Gamma_{\hat X_1\hat Y_1}}\left[e^{-\frac{i\pi}{4}\Gamma_{\hat X_1\hat Y_1}}\eps\right]
\eea
Here we used the relation (\ref{BetaAsAlpha}) to express $\alpha$ in terms of $\beta$. Comparing the projection (\ref{GammaProjProbeG}) coming from the brane and projection (\ref{KillSpinMetr}) coming from the geometry, we find a perfect match in the functional dependence of the two spinors\footnote{An extra normalization factor $h^{-1/2}$ in (\ref{KillSpinMetr}) is irrelevant since projection is a linear relation.}, and the two coordinate--independent restrictions on $\eps$ and $\eta_0$ are also consistent. We conclude that the brane (\ref{StaticGaugeGen}) does not break any supersymmetry of the background, as long as it is placed at an appropriate point, i.e., as long as relation (\ref{BetaAsAlpha}) is satisfied. In other words, orientation of the D branes on the torus (angle $\beta$) must be adjusted to match the {\it known} function of coordinates $\alpha$, which comes from (\ref{LocalSoln}). 

\bigskip

Although our argument were made for the brane (\ref{StaticGaugeGen}) that does not stretch in $(Y_2,Y_3)$ directions, it can be easily generalized to branes with generic orientation on the torus. We conclude this section by presenting such generalization for the action (\ref{BraneActFive}), and extension of the supersymmetry analysis is straightforward, although the notation becomes cumbersome. 

Any supersymmetric D3 brane wrapping three directions of $T^6$ can be described in a static gauge that generalizes (\ref{StaticGaugeGen}):
\bea\label{StaticGaugeGenZ}
t=\tau,\quad 
\left(\begin{array}{c}
z_1\\z_2\\z_3
\end{array}\right)=M
\left(\begin{array}{c}
\xi_1\\ \xi_2\\ \xi_3
\end{array}\right),\quad x_m=x_m(\tau),
\eea
where $M$ is a $3\times 3$ complex matrix with a non--zero determinant. The induced metric on the brane is
\bea
ds_{ind}^2=g_{\mu\nu}{\dot x}^\mu{\dot x}^\nu d\tau^2+d\xi^T M^\dagger M d\xi,
\eea
where index $\mu$ goes over four non--compact directions, including time. The determinant of this metric is
\bea\label{IndMetr}
\mbox{det}\, g_{ind}=g_{\mu\nu}{\dot x}^\mu{\dot x}^\nu\left[\mbox{det}(M^\dagger M)\right].
\eea
We can always normalize coordinate $\xi^m$ to ensure that $\mbox{det}(M^\dagger M)=1$, then the DBI action coming from (\ref{IndMetr}) is identical to the first term in (\ref{BraneActFive}). Next we look at the pullback of the gauge potential that appears in the Chern--Simons term:
\bea\label{ChernPull}
P[C_4]&=&\frac{1}{2}P[F+i{\tilde F}]\wedge P[\Omega_3]+
\frac{1}{2}P[F-i{\tilde F}]\wedge P[{\bar\Omega}_3]\nonumber\\
&=&\frac{1}{2}P\left[F(\mbox{det}M+\mbox{det}M^\dagger)+
i{\tilde F}(\mbox{det}M-\mbox{det}M^\dagger)\right]\wedge d^3\xi\\
&=&P\left[\cos\beta F+\sin\beta{\tilde F}\right]\wedge d^3\xi\nonumber
\eea
Here we defined angle $\beta$ by
\bea\label{MixedBeta}
\cos\beta\equiv \frac{1}{2}(\mbox{det}M+\mbox{det}M^\dagger),\qquad 
\sin\beta\equiv \frac{i}{2}(\mbox{det}M-\mbox{det}M^\dagger).
\eea
The consistency condition,
\bea
\cos^2\beta+\sin^2\beta=\mbox{det}M\ \mbox{det}M^\dagger=\mbox{det}(M^\dagger M)=1,
\eea
is satisfied due to normalization of $M$. It is clear that the pullback (\ref{ChernPull}) gives rise to a Chern--Simons term, which is identical to the one used in (\ref{BraneActFive}), so the entire action (\ref{BraneActFive}) is recovered for an arbitrary complex matrix $M$ in (\ref{StaticGaugeGen}). The angle $\beta$, which translates into the location of a brane in the non--compact direction and into the kappa projection via (\ref{AtProbe}) and (\ref{BetaAsAlpha}) is determined for every normalized matrix $M$ by (\ref{MixedBeta}).

\section{Discussion}

In this paper we have constructed regular BPS geometries with \AdSS$\times$T$^6$ asymptotics and demonstrated that such solutions of supergravity are parameterized by one complex harmonic function on ${\bf R}^3$ with sources distributed along arbitrary curves. To construct a geodesically complete space, one has to glue several copies of ${\bf R}^3$ through a series of branch cuts, and we have presented the explicit procedure for the analytic continuation in the case when all curves belong to one plane. Although the geometric data paramaterizing the new solutions is analogous to its conterparts for the bubbling geometries in ten dimensions (where one specifies the white and black regions in a plane) and for the six--dimensional 1/2--BPS fuzzballs (where one specifies contours in a 4-dimensional base), the mechanism of resolving the singularity in the \AdSS\ case is very peculiar, and it is based on existence of several copies of the base and on analytic continuation. 

Another peculiar feature of the new solutions is the lack of a clear connection between the gravity picture and a theory on the boundary, which was present in the six-- and ten--dimensional cases. For example, the 1/2--BPS D1--D5 geometries of \cite{Parad} corresponded to chiral primaries in the dual field theory, and this connection could be visualized via an effective multiwound string \cite{DasMath}. The ten--dimensional bubbling solutions of \cite{LLM} were mapped to a quantum mechanics of a matrix model on the boundary \cite{Beren} via a very explicit correspondence. Unfortunately, the field theory dual to \AdSS$\times$T$^6$  is not well-understood, and this impedes the construction of an explicit map between the boundary and the bulk, but perhaps one can use the gravity side to get some insights into the dynamics of fields theory using the methods developed in \cite{Rych}. This may also allow one to count the bubbling states in supergravity and extend the fuzzball proposal \cite{Fuzz} to the four--dimensional black holes constructed from intersecting D3 branes. 

\section*{Acknowledgements}
It is a pleasure to thank Yuri Chervonyi, Emil Martinec, Radu Roiban, Ashoke Sen, and Arkady Tseytlin for useful discussions. This work
is supported in part by NSF grant PHY-1316184.

\appendix

\section{Giant gravitons on AdS$_3\times$S$^3$ and on fuzzballs}
\renewcommand{\theequation}{A.\arabic{equation}}
\setcounter{equation}{0}\label{AppAdS3}

In this article we study supersymmetric branes on \AdSS\ and their gravitational backreaction, and we find that such branes are rather different from their counterparts on AdS$_5\times$S$^5$. It turns out that branes on AdS$_3\times$S$^3$ share some of these peculiar properties, but to see this one has to go beyond the standard giant gravitons discussed in \cite{SusskGiant,MyersGiant}. In AdS$_5\times$S$^5$ giant gravitons exhaust all 1/2--BPS configurations, and their counterparts with lower supersymmetry have also been classified in \cite{Mikhailov}. In this section we will analyze the probe branes on AdS$_3\times$S$^3$ and on its supersymmetric excitations. 

We begin with discussing  branes on AdS$_3\times$S$^3$:
\bea\label{AdS33}
ds^2&=&L^2\left[-(\rho^2+1)d{t}^2+\frac{d\rho^2}{\rho^2+1}+\rho^2 d\chi^2+d\theta^2+
s^2_\theta d{\phi}^2+c^2_\theta d{\psi}^2\right]+dz^ad{\bar z}_a\nn
C_2&=&L\rho^2  dt\wedge d\chi-Lc^2_\theta \wedge d\phi\wedge d\psi
\eea
The standard giant graviton \cite{SusskGiant} and the dual giant \cite{MyersGiant} are obtained by imposing the following  ansatz for the worldvolume of the D1 brane:
\bea\label{AdS3giantStand}
\mbox{giant}:&&\quad t=\tau,\quad {\dot\phi}=\mbox{const}, \quad \psi=\sigma,\quad \rho=0;\nn
\mbox{dual giant}:&&\quad t=\tau,\quad {\dot\phi}=\mbox{const}, \quad \chi=\sigma,\quad 
\theta=\frac{\pi}{2}.
\eea
However, in the AdS$_3\times$S$^3$ case, one can introduce a more general ansatz,
\bea
t=\tau,\quad {\dot\phi}=\mbox{const}, \quad \psi=a \sigma,\quad \chi=b \sigma,
\eea
which leads to the following combination of the DBI action and the Chern--Simons term:
\bea
S=-TL\int d\sigma d\tau\sqrt{(b^2 \rho^2+ a^2 c_\theta^2)(\rho^2+1-s_\theta^2{\dot\phi}^2)}
+TL\int d\tau d\sigma\left[\rho^2b-c_\theta^2 a{\dot\phi}\right]
\eea
Equations of motion for cyclic variables $(\phi,\chi,\psi)$ are satisfied automatically, while equations for $\rho$ and $\theta$ give 
\bea
\frac{\rho[a^2 c_\theta^2+2b^2\rho^2+b^2-b^2s_\theta^2{\dot\phi}^2]}{\sqrt{(b^2 \rho^2+ a^2 c_\theta^2)(\rho^2+1-s_\theta^2{\dot\phi}^2)}}=2\rho b,\
\frac{s_{2\theta}[a^2(1+\rho^2)+b^2\rho^2{\dot\phi}^2+a^2 {\dot\phi}^2 c_{2\theta}]}{
2\sqrt{(b^2 \rho^2+ a^2 c_\theta^2)(\rho^2+1-s_\theta^2{\dot\phi}^2)}}
=-s_{2\theta} b\dot\phi\,.\nonumber
\eea
These equations are solved by 
\bea\label{AdS3giant}
{\dot\phi}=-1,\quad a=b,\quad \psi=\sigma,\quad \chi=\sigma
\eea
and arbitrary $(\rho,\theta)$. Moreover, configurations (\ref{AdS3giant})  have vanishing Lagrangian density so their energy and angular momentum are equal up to a sign,
\bea
J=\frac{\d{\cal L}}{\d\dot\phi},\qquad {\cal E}=J{\dot\phi}-{\cal L}=-J,
\eea
and they also preserve supersymmetries, as we will see below. 

Giant gravitons (\ref{AdS3giantStand}) wrapping the sphere or the AdS space have counterparts in higher dimensions \cite{SusskGiant,MyersGiant}, but their ``mixed'' generalization (\ref{AdS3giantStand}) exists only 
in AdS$_3\times$S$^3$ and \AdSS. This is related to another peculiar property of giant gravitons observed in \cite{MyersGiant}: in sharp contrast to higher dimensional cases, where the size of the giant graviton is fixed by its angular momentum, the branes  (\ref{AdS3giantStand}) on AdS$_3\times$S$^3$ can wrap arbitrary cycles on AdS or on a sphere since the potential for their size is flat. Now we see that not only the size of a cycle is arbitrary, but a mixture between AdS and sphere is also allowed. 

We will now demonstrate that configurations (\ref{AdS3giant}) are supersymmetric. Rather than proving this only for giant gravitons on AdS$_3\times$S$^3$, we will show that counterparts of (\ref{AdS3giant}) on any 1/2--BPS geometry preserve SUSY. First we recall that all 1/2--BPS geometries with  AdS$_3\times$S$^3$ asymptotics are known explicitly \cite{Parad}\footnote{For simplicity we focus only on profiles in the non--compact space. Extension to the torus modes constructed in \cite{lmm,SkenTayMom} is straightforward.}, and they are given by
\bea\label{FuzzBall}
ds^2&=&\frac{1}{\sqrt{H_1H_5}}\left[-(d{\tilde t}-A)^2+(dy+B)^2\right]+\sqrt{H_1H_5}dx_idx_i+\sqrt{\frac{H_1}{H_5}}dz_adz_a\nn
C^{(2)}&=&\frac{1}{H_1}[d{\tilde t}-A]\wedge [dy+B]+{\cal C},\quad e^{2\Phi}=\frac{H_1}{H_5}\\
&&d{\cal C}=-\star_4 dH_5,\qquad dB=-\star_4 dA\nonumber
\eea
Here $dx_idx_i$ denotes the metric on a flat four--dimensional base, and $dz_adz_a$ denotes a metric on $T^4$. All functions can depend on $(x_1,x_2,x_3,x_4)$.
Harmonic functions $(H_1,H_5)$ and the gauge field $A$ are determined from microscopic analysis, and expressions 
\bea
H_5=\frac{Q_5}{L}\int_0^L\frac{dv}{|{\bf x}-{\bf F}|^2},\quad 
H_1=\frac{Q_5}{L}\int_0^L\frac{|{\bf \dot F}|^2dv}{|{\bf x}-{\bf F}|^2},\quad 
A_i=-\frac{Q_5}{L}\int_0^L\frac{{ \dot F}_idv}{|{\bf x}-{\bf F}|^2}
\eea
give rise to regular solutions \cite{Parad,lmm}. To recover the AdS$_3\times$S$^3$ (\ref{AdS33}) from this construction, one has to choose a circular profile in the four dimensional space $(x_1,x_2,x_3,x_4)$ and to perform a spectral flow \cite{MMaoz,LMwound,Parad}. Specifically, the relevant harmonic functions are given by \cite{LMwound}
\bea
H_1=H_5=\frac{Q}{r^2+a^2c^2_\theta},\quad A=-\frac{aQs^2_\theta}{r^2+a^2c^2_\theta} d{\tilde\phi},\quad 
B=\frac{aQc^2_\theta}{r^2+a^2c^2_\theta} d{\tilde\psi},\nonumber
\eea
and combining this with flat metric on the base, 
\bea
ds_4^2=(r^2+a^2c_\theta^2)\left[\frac{dr^2}{r^2+a^2}+d\theta^2\right]+
(r^2+a^2)s_\theta^2 d{\tilde\phi}^2+r^2c_\theta^2 d{\tilde\psi}^2,\nonumber
\eea
we find
\bea
ds^2=Q\left[\frac{dr^2}{r^2+a^2}+d\theta^2+s_\theta^2 (d{\tilde\phi}-\frac{a}{Q}dt)^2+
c_\theta^2 (d{\tilde\psi}+\frac{a}{Q}dy)^2-(r^2+a^2)dt^2+r^2 dy^2\right]+dz^2\nonumber
\eea
This relation reduces to (\ref{AdS33}) after identification $Q=L^2$ and a change of coordinates 
\bea
\phi={\tilde\phi}-\frac{a}{Q}{\tilde t},\quad \psi={\tilde\psi}+\frac{a}{Q}y,\quad \rho=\frac{r}{a},\quad 
t=\frac{\tilde t}{a},\quad \chi=\frac{y}{a}
\eea
In particular, the shift of the angular coordinates corresponds to a spectral flow from the Ramond to the NS sector in the dual CFT \cite{MMaoz}. Rewriting this spectral flow as
\bea
{\tilde\phi}=\phi+t,\quad {\tilde\psi}=\psi-\chi,
\eea
we conclude that configurations (\ref{AdS3giant}) correspond to constant values of 
$(r,\theta,{\tilde\phi},{\tilde\psi})$, i.e., to one point on the base. This observation suggests a simple ansatz generalizing (\ref{AdS3giant}) to branes on an arbitrary supersymmetric geometry (\ref{FuzzBall}): one should put a D1--brane at one point on the base, while stretching it along ${\tilde t}$ and $y$:
\bea\label{StraightGG}
{\tilde t}=\tau,\quad y=\sigma.
\eea
Assuming that coordinates $x_i$ on the base depend only on time, we find the action governing the dynamics of D1 branes on (\ref{FuzzBall}):
\bea
S=-T\int d\tau d\sigma e^{-\Phi}\frac{1}{\sqrt{H_1H_5}}\sqrt{(1-A_i{\dot x}_i)^2-H_1H_5({\dot x_i}^2)}+T\int d\tau d\sigma\frac{1}{H_1}[1-A_i{\dot x}_i]\nonumber
\eea
It is clear that all equations of motion are solved by constant $x_i$, moreover, the action vanishes on such solutions.

To verify that configuration (\ref{StraightGG}) preserve supersymmetry, we recall the expression for the kappa--symmetry projection associated with a D1 brane \cite{KappaProj}:
\bea
\Gamma\eps=\eps,\quad \Gamma=i\sigma_3\sigma_2\otimes\left[{\cal L}^{-1}\left(\prod_{a=0}^1\frac{\d x^{m_a}}{\d\xi^a}\right)\gamma_{m_0m_1}\right],\quad
{\cal L}=\sqrt{-\mbox{det}\left[g_{mn}\frac{\d x^m}{\d\xi^a}\frac{\d x^n}{\d\xi^b}\right]}
\eea
For configurations (\ref{StraightGG}) this expression  reduces to a very simple  projection
\bea
\Gamma\eps=\eps,\quad \Gamma=i\sigma_3\sigma_2\otimes \Gamma_{\hat t}\Gamma_{\hat y},
\eea
which is consistent with supersymmetries preserved by the background geometry (\ref{FuzzBall}). An analogous projection for the D5 branes wrapping $(t,y,z_a)$ is
\bea
\Gamma\eps=\eps,\quad \Gamma=i\sigma_3\sigma_2\otimes 
\Gamma_{\hat t}\Gamma_{\hat y}\Gamma_{z_1}\Gamma_{z_2}\Gamma_{z_3}\Gamma_{z_4}.
\eea
Backreaction of such D1 and D5 branes modifies the geometry, while leaving it in the general class (\ref{FuzzBall}).

To summarize, in this appendix we reviewed some properties of supersymmetric branes on AdS$_3\times$S$^3$  and on fuzzball geometries constructed in \cite{Parad,lmm}. We demonstrated that such branes are much more general than the giant gravitons in higher dimensions \cite{SusskGiant,MyersGiant}, but they are very similar to the branes on \AdSS\ discussed in sections \ref{SecProbesAdS}.

\section{Derivation of the solution}
\renewcommand{\theequation}{B.\arabic{equation}}
\setcounter{equation}{0}
\label{AppMain}

In this appendix we derive the geometry (\ref{LocalSoln}) by solving equations for the Killing spinors and self--duality conditions for $F_5$ after imposing the ansatz (\ref{Ansatz}):
\bea\label{AnsatzApp}
ds^2&=&g_{mn}dx^mdx^n+dz^ad{\bar z}_a\nonumber\\
F_5&=&
\frac{1}{2}(F+i{\tilde F})\wedge \Omega_3+\frac{1}{2}(F-i{\tilde F})\wedge {\bar\Omega}_3\\
\Omega_3&=&dz_{123},\qquad \star_6\Omega_3=i\Omega_3,\quad \star_4 F={\tilde F}.\nonumber
\eea
Here index $m$ refers to four non--compact directions, and index $a$ runs from one to three.

\subsection{Reduction to four dimensions}
\label{AppReduction}

Supersymmetry of the geometry (\ref{AnsatzApp}) implies an existence of a Killing spinor $\eta$. Since only the five--form is excited, the variations of dilatino under supersymmetry transformations vanish trivially, 
and we only have to solve the gravitino equations \cite{SchwWest}
\bea
\nabla_M\eta+\frac{i}{480}{\not F}_5\Gamma_M\eta=0
\eea
It is convenient to separate the gravitino equation into its torus components,
\bea\label{SuTorus}
(\not F+i{\tilde{\not F}})\Gamma^{123}\Gamma_{a}\eta=0,\qquad (\not F-i{\tilde{\not F}})\Gamma^{\overline{123}}\Gamma_{\bar a}\eta=0.
\eea
and the remaining projections,
\bea\label{SuSpace}
\nabla_m\eta+\frac{i}{16}\left[(\not F+i{\tilde{\not F}})\Gamma^{123}+(\not F-i{\tilde{\not F}})\Gamma^{\overline{123}}\right]
\Gamma_m\eta=0
\eea
To proceed we choose a convenient basis of gamma matrices\footnote{To avoid unnecessary clutter, in this appendix we don't use hats to label the frame indices of the Dirac matrices. Throughout this paper we use capital gamma to denote Dirac matrices with flat indices (see equation (\ref{FlatDirac})), but in this appendix we also write $\Gamma_x$ instead of the proper $\Gamma_{\hat x}$.}:
\bea
&&\Gamma_m=\gamma_m\otimes 1\otimes 1\otimes 1,\quad
\Gamma_{z_1}=\gamma_5\otimes \sigma_-\otimes 1\otimes 1,\quad
\gamma_m=e_m^A{\tilde\Gamma}_A,\quad g_{mn}=e_m^Ae_n^B \eta_{AB},
\nonumber\\
&&\Gamma_{z_2}=\gamma_5\otimes \sigma_3\otimes \sigma_-\otimes 1,\quad
\Gamma_{z_3}=\gamma_5\otimes \sigma_3\otimes \sigma_3\otimes \sigma_-,\quad
\gamma_5=i{\tilde\Gamma}_0{\tilde\Gamma}_{123}.
\eea
Noticing that in this basis
\bea\label{GammaXYZ}
\Gamma^{x_1}=\gamma_5\otimes \sigma_1,\quad 
\Gamma^{y_1}=\gamma_5\otimes \sigma_2,\quad 
\Gamma_{z_1}=\gamma_5\otimes \sigma_-,\quad
\Gamma^{{\bar z}_1}=2
\gamma_5\otimes \sigma_-\,,
\eea
we can compute several useful products:
\bea
&&\Gamma^{\overline{123}}=8\Gamma_{123}=
8\gamma_5\otimes\sigma_-\otimes\sigma_-\otimes\sigma_-,\quad
\Gamma^{{123}}=
-8\gamma_5\otimes\sigma_+\otimes\sigma_+\otimes\sigma_+
\nonumber\\
&&\Gamma^{torus}=
-i\,1\otimes \sigma_3\otimes \sigma_3\otimes\sigma_3,\quad \Gamma_{11}=-\gamma_5\otimes \sigma_3\otimes \sigma_3\otimes\sigma_3
\eea
Recall that the Killing spinor must satisfy the chiral projection of type IIB SUGRA:
\bea\label{ChiralProj}
\Gamma_{11}\eta=\eta
\eea
Duality between $F$ and ${\tilde F}$ implies a relation\footnote{To verify the signs, one can look at a particular component in frames, where ${\hat \eps}^{0123}=1$, e.g.,
$$
F_{01}{\tilde\Gamma}^{01}+i{\tilde F}_{23}{\tilde\Gamma}^{23}=
F_{01}{\tilde\Gamma}^{01}(1-i{\tilde\Gamma}_{01}{\tilde\Gamma}^{23})=
F_{01}{\tilde\Gamma}^{01}(1-\gamma_5).
$$}
\bea\label{DualRelGam5}
(\not F+i{\tilde{\not F}})=\not F(1-\gamma_5),\quad
(\not F-i{\tilde{\not F}})=\not F(1+\gamma_5)
\eea
and equations (\ref{SuTorus}), (\ref{SuSpace}) become
\bea\label{IntOne}
&&\not F(1-\gamma_5)\Gamma^{123}\Gamma_{a}\eta=0,\qquad 
\not F(1+\gamma_5)\Gamma^{\overline{123}}\Gamma_{\bar a}\eta=0,\\
\label{IntTwo}
&&\nabla_m\eta+\frac{i}{16}\left[\not F(1-\gamma_5)\Gamma^{123}+\not F(1+\gamma_5)\Gamma^{\overline{123}}\right]
\Gamma_m\eta=0
\eea
Next we decompose the spinor $\eta$ into eight components $\eta_{\pm\pm\pm}$:
\bea
&&[1\otimes \sigma_3\otimes 1\otimes 1]\eta_{\pm\bullet\bullet}=\pm\eta_{\pm\bullet\bullet},\quad 
[1\otimes 1\otimes \sigma_3\otimes 1]\eta_{\bullet\pm\bullet}=\pm\eta_{\bullet\pm\bullet},\nonumber\\
&&[1\otimes 1\otimes  1\otimes \sigma_3]\eta_{\bullet\bullet\pm}=\pm\eta_{\bullet\bullet\pm}\nonumber
\eea
The first equation in (\ref{IntOne}) with $a=1$ and $(+--)$ projection of (\ref{IntTwo}) give
\bea
0&=&\not F(1-\gamma_5)\Gamma^{123}\Gamma_{1}\eta
=8\not F(1-\gamma_5)\eta_{+--}=
16\not F\eta_{+--},\\
0&=&\nabla_m\eta_{+--}\nonumber
\eea
These two equations appear to be inconsistent for nontrivial $F$ and $\eta_{+--}$, and since we are looking for solution with flux, we will set $\eta_{+--}=0$. Other mixed components of $\eta$ vanish for the same reason, and we end up with two non--vanishing projections, $\eta_{+++}$ and $\eta_{---}$, which trivially satisfy  (\ref{IntOne}) and mix in equation (\ref{IntTwo}). 

To factorize the torus, we define two four--component objects ${\tilde\eta}$,
\bea
\eta={\tilde\eta}_{+++}\otimes\left(\begin{array}{c}1\\ 0\end{array}\right)
\otimes\left(\begin{array}{c}1\\ 0\end{array}\right)
\otimes\left(\begin{array}{c}1\\ 0\end{array}\right)+
{\tilde\eta}_{---}\otimes\left(\begin{array}{c}0\\ 1\end{array}\right)
\otimes\left(\begin{array}{c}0\\ 1\end{array}\right)\otimes\left(\begin{array}{c}0\\ 1\end{array}\right),
\eea 
which are subject to constraints 
\bea
\gamma_5 {\tilde\eta}_{+++}=-{\tilde\eta}_{+++},\qquad
\gamma_5 {\tilde\eta}_{---}={\tilde\eta}_{---}
\eea
due to the chirality condition  (\ref{ChiralProj}). Equation (\ref{IntTwo}) reduces to a system
\bea
&&\nabla_m{\tilde\eta}_{+++}+\frac{i}{2}\left[\not F(1-\gamma_5)(-\gamma_5)\right]
{\gamma}_m{\tilde\eta}_{---}=0,\nonumber\\
&&\nabla_m{\tilde\eta}_{---}+\frac{i}{2}\left[\not F(1+\gamma_5)(\gamma_5)\right]
{\gamma}_m{\tilde\eta}_{+++}=0,\nonumber
\eea
which can be written in compact form using (non--chiral) spinor  
\bea
{\tilde\eta}={\tilde\eta}_{+++}+{\tilde\eta}_{---}.
\eea
The final equation for the four dimensional spinor ${\tilde\eta}$ is
\bea\label{4DspinorTil}
\nabla_m{\tilde\eta}+{i}\not F
{\gamma}_m{\tilde\eta}=0,
\eea
and it will be analyzed in the remaining part of this appendix. 

\subsection{Spinor bilinears}

Existence of Killing spinors severely constrains the geometry, and a powerful technique for extracting the constraints is based on analyzing spinor bilinears \cite{bilinears}. We will now use this technique to explore the consequences of equation (\ref{4DspinorTil}). To simplify the notation, we will drop tildes in (\ref{4DspinorTil}), then equations for the Killing spinor and its conjugate become
\bea\label{4Dspinor}
&&\nabla_m\eta+i\not F
\gamma_m\eta=0\\
&&\nabla_m{\bar\eta}-i{\bar\eta}
\gamma_m\not F=0\nonumber
\eea
From now on $\Gamma^A=e^A_m\gamma^m$ and $\eta$ denote four dimensional gamma matrices 
and Killing spinor\footnote{These quantities were accompanied by tildes in section \ref{SecLocal}.}. 
Combining the duality relations (\ref{DualRelGam5}),
\bea
i{\tilde{\not F}}=-\not F\gamma_5,\nonumber
\eea
and expressing ${F}$ in terms of ${\tilde F}$, we obtain an alternative form of (\ref{4Dspinor}) :
\bea\label{4DspinorTilda}
&&\nabla_m\eta+\not \tilde F\gamma_5
\gamma_m\eta=0\\
&&\nabla_m{\bar\eta}-{\bar\eta}
\Gamma_m\gamma_5\not \tilde F=0\nonumber
\eea
Using an identity
\bea
F^{ab}(\gamma_m\gamma_{ab}-\gamma_{ab}\gamma_m)=
4F_m{}^a\gamma_a,
\eea
we find equations for the spinor bilinears:
\bea
&&\nabla_m[{\bar\eta}\eta]+4iF_{am}{\bar\eta}\gamma^a\eta=0\\
&&\nabla_m[{\bar\eta}\gamma_5\eta]+4{\tilde F}_{am}{\bar\eta}\gamma^a\eta=0\\
&&\nabla_m[{\bar\eta}\gamma_n\eta]+i{\bar\eta}\left[\gamma_n\not F\gamma_m
-\gamma_m\not F\gamma_n\right]\eta=0\\
&&\nabla_m[{\bar\eta}\gamma_n\gamma_5\eta]-
i{\bar\eta}\left[
\gamma_m\not F\gamma_n+
\gamma_n\not F\gamma_m\right]\gamma_5\eta=0
\eea
The last two equations imply an existence of a Killing vector $K$ and an exact form $L$:
\bea\label{KillVect}
K_m=-{\bar\eta}\gamma_m\eta,\quad
L_mdx^m={\bar\eta}\gamma_m\gamma_5\eta dx^m\equiv dy.
\eea
Using the Fierz identities,
\bea\label{Fierz}
-K_mK^m=L_mL^m=({\bar\eta}\gamma_5\eta)^2-({\bar\eta}\eta)^2,\qquad
K_mL^m=0
\eea
we conclude that $K^m$ cannot be space--like\footnote{Recall that $({\bar\eta}\eta)=\eta^\dagger\Gamma^{\bf t}\eta$ is imaginary.}. 
In this paper we will focus on time--like $K^m$ and choose coordinate $t$ along this vector. The exact form $L_m dx^m$ selects a second coordinate $y$, and we can choose the remaining two coordinates $(x^1,x^2)$ to be orthogonal to $y$. Notice that due to the second Fierz identity (\ref{Fierz}), $t$ is also orthogonal to $y$, so we arrive at the most general metric consistent with (\ref{KillVect}) and (\ref{Fierz}):
\bea\label{BasePreMetr}
ds_4^2=-h^{-2}(dt+V)^2+h^2[dy^2+{\hat q}_{\alpha\beta}dx^\alpha dx^\beta], \quad
\alpha,\beta=1,2.
\eea
Furthermore, equations for the bilinears ensure that we can choose a gauge where
\bea\label{Bilone}
{\bar\eta}\eta=-4iA_t,\qquad
{\bar\eta}\gamma_5\eta=-4{\tilde A}_t,
\eea
then (\ref{Fierz}) implies a relation
\bea\label{HSquare}
h^{-2}=16[(A_t)^2+({\tilde A}_t)^2].
\eea

We will now determine the functional form of $\eta$ by
combining the last two equations with relations following from definitions of $K$ and $L$:
\bea\label{Biltwo}
L_y=h{\eta}^\dagger\Gamma^{\bf ty}\gamma_5\eta=1,\quad
K^t=h{\eta}^\dagger\eta=1,
\eea
 After solving equation (\ref{HSquare}) by introducing a real angle $\alpha$,
\bea
A_t=-\frac{1}{4}h^{-1}\sin\alpha,\qquad {\tilde A}_t=-\frac{1}{4}h^{-1}\cos\alpha,
\eea
we define a new spinor $\eps$ by
\bea
\eta=h^{-1/2}e^{i\alpha\Gamma_5/2}\eps.
\eea
Substitution of this expression into (\ref{Bilone}), (\ref{Biltwo}) gives relations for the bilinears involving $\eps$:
\bea\label{PreProjSpin}
&&\eps^\dagger\eps=h\eta^\dagger\eta=1,\nonumber\\
&&\eps^\dagger\Gamma^{\bf t}\Gamma_5\eps=
h\eta^\dagger\Gamma^{\bf t}\Gamma_5
[\cos\alpha-i\sin{\alpha}\Gamma_5]\eta=\cos^2\alpha+\sin^2\alpha=1,\\
&&\eps^\dagger\Gamma^{\bf t}\eps=
h\eta^\dagger\Gamma^{\bf t}
[\cos\alpha-i\sin{\alpha}\Gamma_5]\eta=i\cos\alpha\sin\alpha-i\sin\alpha\cos\alpha=0,
\nonumber\\
&&{\eps}^\dagger\Gamma^{\bf ty}\Gamma_5\eps=h{\eta}^\dagger\Gamma^{\bf ty}\Gamma_5\eta=1
\nonumber
\eea
Taking a difference of the first two relations,
\bea
\eps^\dagger[1-\Gamma^{\bf t}\Gamma_5]\eps=0,
\eea
and observing that $\frac{1}{2}[1-\Gamma^{\bf t}\Gamma_5]$ is a projector, we conclude that 
\bea
[1-\Gamma^{\bf t}\Gamma_5]\eps=0.
\eea
Then the first and the last relations in (\ref{PreProjSpin}) imply that 
\bea
\eps^\dagger[1+\Gamma^{\bf y}]\eps=0,\quad\Rightarrow\quad [1+\Gamma^{\bf y}]\eps=0.
\eea
We conclude that a four--component spinor $\eps$ satisfies two independent projections:
\bea\label{Projectors}
\Gamma^{\bf t}\Gamma_5\eps=\eps,\qquad \Gamma^{\bf y}\eps=-\eps,
\eea
so it effectively reduces to a one--component complex object. Moreover, the normalization condition
\bea
\eps^\dagger\eps=1
\eea
determines $\eps$ up to a pure phase. 

To restrict the form of the two--dimensional metric ${\hat q}_{\alpha\beta}$ in (\ref{BasePreMetr}), we consider equation for a new bilinear that does not involve complex conjugation of spinors\footnote{We work in the representation where $\Gamma_{\bf t}$, $\Gamma_{\bf y}$ and $\Gamma_1$ are symmetric, while $\Gamma_2$ is antisymmetric. Then $\Gamma^2\gamma_m^T\Gamma_2=-\gamma_m$ and $\Gamma_5^T=-\Gamma_5$.}:
\bea
\nabla_m[\eta^T\Gamma^2\gamma_n\eta]+i\eta^T\Gamma^2
\left[\gamma_n{\not F}\gamma_m+\gamma_m{\not F}\gamma_n\right]\eta=0.
\eea
This equation implies a relation, which can be written either in terms of $\eta$ or in terms of $\eps$: 
\bea\label{EpsEpsTdif}
d\left[\eta^T\Gamma^2\gamma_n\eta dx^n\right]=0\quad\Rightarrow\quad
d\left[h^{-1}\eps^T\Gamma^2\gamma_n\eps\, dx^n\right]=0.
\eea
Choosing the frames for the metric (\ref{BasePreMetr}):
\bea
\gamma_n dx^n=\Gamma_{\bf t}h^{-1}(dt+V)+\Gamma_{\bf y}hdy+
\Gamma_1 h {\hat e}^1_\alpha dx^\alpha+\Gamma_2 h {\hat e}^2_\alpha dx^\alpha
\eea
and using projectors (\ref{Projectors}), the one--form entering (\ref{EpsEpsTdif}) can be simplified:
\bea
h^{-1}\eps^T\Gamma^2\gamma_n\eps\, dx^n=\eps^T\eps\left[{\hat e}^2_\alpha-
i{\hat e}^1_\alpha \right]dx^\alpha,
\eea
then the resulting equation reads
\bea
d\left[\eps^T\eps\left[{\hat e}^2_\alpha-
i{\hat e}^1_\alpha \right]dx^\alpha\right]=0.
\eea
Using this relation to define a complex coordinate $w$:
\bea
dw\equiv \eps^T\eps\left[{\hat e}^2_\alpha-
i{\hat e}^1_\alpha \right]dx^\alpha,
\eea
we find
\bea
dwd{\bar w}=| \eps^T\eps|^2{\hat q}_{\alpha\beta}dx^\alpha dx^\beta=
(\eps^\dagger\eps)^2{\hat q}_{\alpha\beta}dx^\alpha dx^\beta=
{\hat q}_{\alpha\beta}dx^\alpha dx^\beta.
\eea
In other words, the two dimensional metric ${\hat q}_{\alpha\beta}$ must be flat. 

To summarize, we have used some spinor bilinears to determine the bosonic fields:
\bea\label{AnswApp}
ds^2&=&-h^{-2}(dt+V)^2+h^2[dy^2+dwd{\bar w}]+dz^ad{\bar z}_a\nonumber\\
F_5&=&
\frac{1}{2}(F+i{\tilde F})\wedge \Omega_3+\frac{1}{2}(F-i{\tilde F})\wedge {\bar\Omega}_3
\nonumber\\
\Omega_3&=&dz_{123},\qquad \star_6\Omega_3=i\Omega_3,\quad \star_4 F={\tilde F}.\\
A_t&=&-\frac{1}{4}h^{-1}\sin\alpha,\qquad {\tilde A}_t=-\frac{1}{4}h^{-1}\cos\alpha,
\nonumber
\eea
and the Killing spinor:
\bea\label{AnsAppSpin}
\eta=h^{-1/2}e^{i\alpha\Gamma_5/2}\eps,\quad \Gamma^{\bf t}\Gamma_5\eps=\eps,\quad \Gamma^{\bf y}\eps=-\eps,\quad \eps^\dagger\eps=1.
\eea
In the next two subsections we will analyze the remaining equations for the Killing spinor and Bianchy identities for $F_5$ to find the relations between $h$, $\alpha$, and vector $V$.

\subsection{Remaining equations for the Killing spinor}

Although introduction of $y$ coordinate was helpful in deriving the metric (\ref{AnswApp}), from now on it is convenient to treat all three coordinate on a flat base of (\ref{AnswApp}) uniformly\footnote{To simplify the notation, in this subsection we reserve indices $a,b,\dots$ for coordinates on the base and use dotted indices for the holomorphic coordinates on ${\bf C}^3$.}:
\bea\label{A3metric}
ds^2&=&-h^{-2}(dt+V)^2+h^2 dx_adx_a+dz^{\dot a}d{\bar z}_{\dot a}\nonumber\\
F_5&=&F\wedge\mbox{Re}\,\Omega_3-
{\tilde F}\wedge\mbox{Im}\,\Omega_3\\
F&=&dA_t+\frac{1}{2}F_{ab}dx^adx^b,\quad 
{\tilde F}=d{\tilde A}_t+\frac{1}{2}{\tilde F}_{ab}dx^adx^b,\quad
{\tilde A}_t+iA_t=-\frac{1}{4h}e^{i\alpha}
\nonumber
\eea    
and impose only one of the projections from  (\ref{AnsAppSpin})
\bea\label{Gam5Proj}
\eta=h^{-1/2}e^{i\alpha\Gamma_5/2}\eps,\qquad \Gamma^{\bf t}\Gamma_5\eps=\eps.
\eea
In this subsection we will verify the equation (\ref{4Dspinor}) for the Killing spinor. To do so, we introduce the frames
\bea
e^{\bf n}_{\ m}=\left(\begin{array}{cc}
h^{-1}&h^{-1}V_a\\
0&h
\end{array}\right),\qquad
e^{m}_{\ \ \bf n}=\left(\begin{array}{cc}
h&-h^{-1}V_a\\
0&h^{-1}
\end{array}\right).
\eea
and simplify the expression for ${\not F}$:
\bea\label{SlashF}
{\not F}&=&
2F_{\bf ta}\Gamma^{\bf ta}+F_{\bf ab}\Gamma^{\bf ab}=
2F_{\bf ta}\Gamma^{\bf ta}-
\eps_{\bf abc}{\tilde F}_{\bf tc}\Gamma^{\bf ab}=
2(F_{ta}-i{\tilde F}_{ta}\gamma_5)\Gamma^{\bf ta}
\nonumber\\
&=&
-2h(\Gamma^{\bf t}{\not\d}A_t-i\Gamma_5\Gamma^{\bf t}{\not\d}{\tilde A}_t)
\eea
We used a relation
\bea
\eps_{\bf abc}\Gamma^{\bf ab}=2\Gamma^{\bf c}\Gamma_{\bf 123}=
2\Gamma^{\bf ct}\Gamma_{\bf t123}=2i\Gamma^{\bf tc}\gamma_5\,.\nonumber
\eea
Substitution of (\ref{SlashF}) into (\ref{4Dspinor}) gives
\bea\label{4DspinorA}
&&\nabla_m\eta-2{i}h(\Gamma^{\bf t}{\not\d}A_t-i\Gamma_5\Gamma^{\bf t}{\not\d}{\tilde A}_t)
\gamma_m\eta=0
\eea
Using expressions (\ref{AnswApp}) for $A_t$ and ${\tilde A}_t$, the last equation can be rewritten as 
\bea
\nabla_m\eta+\frac{i}{2}e^{i\alpha\Gamma_5}(
-ih\Gamma_5\Gamma^{\bf t}{\not\d}h^{-1}+\Gamma^{\bf t}{\not\d}\alpha)
\gamma_m\eta=0\nonumber
\eea
Further rewriting of $\eta$ in terms of $\eps$ (see (\ref{AnsAppSpin})), we arrive at the equation for the $\eps$, which is equivalent to (\ref{4Dspinor}):
\bea\label{KSeqnEps}
\nabla_m\eps+
\frac{1}{2}\d_m\left[i\alpha\Gamma_5-\ln h\right]\eps-
\frac{1}{2}{\not\d}(\ln h+i
\Gamma_5\alpha)\Gamma_5\Gamma^{\bf t}
\gamma_m\eps=0
\eea
The rest of this subsection will be devoted to proving that this equation is satisfied, as long as the vector 
field $V$ obeys a duality relation (\ref{Vdual}).

To verify equation (\ref{KSeqnEps}), we need to compute the spin connection $\omega_{\bf p,mn}$:
\bea
de^{\bf p}&=&\frac{1}{2}\Gamma^{\bf p}_{\ ,{\bf mn}}e^{\bf m}\wedge e^{\bf n},\quad
\omega_{\bf p,mn}=-
\frac{1}{2}\left[\Gamma_{\bf p,mn}+\Gamma_{\bf n,mp}+\Gamma_{\bf m,pn}\right]
\eea
Taking derivatives,
\bea
de^{\bf t}&=&\d_a h^{-1}e^{\bf a}\wedge e^{\bf t}+\frac{1}{2h^3}W_{ab}e^{\bf a}\wedge e^{\bf b},
\qquad W\equiv dV,\\
de^{\bf b}&=&-\d_a h^{-1}e^{\bf a}\wedge e^{\bf b},\nonumber
\eea
we find the nontrivial components of $\omega$
\bea
\omega_{\bf t,ab}&=&-
\frac{1}{2}
\left[\Gamma_{\bf t,ab}+\Gamma_{\bf b,tc}-\Gamma_{\bf a,bt}\right]
=\frac{1}{2h^3}W_{ab}\nonumber\\
\omega_{\bf c,at}&=&
-\frac{1}{2}
\left[\Gamma_{\bf c,at}+\Gamma_{\bf t,ac}-\Gamma_{\bf a,tc}\right]
=\frac{1}{2h^3}W_{ac}\nonumber\\
\omega_{\bf t,at}&=&
-\frac{1}{2}
\left[\Gamma_{\bf t,at}+\Gamma_{\bf t,at}-\Gamma_{\bf a,tt}\right]
=\d_a h^{-1}\nonumber\\
\omega_{\bf c,ab}&=&
\delta_{bc}\d_a h^{-1}-\delta_{ac}\d_b h^{-1}\nonumber
\eea
This leads to the following expressions for the derivatives in the orthonormal frame:
\bea
\nabla_{\bf t}\eps&=&
\frac{1}{4}\omega_{\bf t,mn}\Gamma^{\bf mn}\eps=
\frac{1}{8h^{3}}W_{ab}\Gamma^{\bf ab}\eps+
\frac{1}{2}\d_{a}h^{-1}\Gamma^{\bf at}\eps\nonumber\\
\nabla_{\bf a}\eps&=&\frac{1}{h}\d_a \eps+\frac{1}{4}\omega_{\bf a,mn}\Gamma^{\bf mn}\eps
=
\frac{1}{h}\d_a \eps
-\frac{1}{4h^{3}}W_{ab}\Gamma^{\bf bt}\eps-
\frac{1}{2}\d_{b}h^{-1}\Gamma^{\bf ab}\eps\nonumber
\eea
Using projection (\ref{Gam5Proj}) and relations 
\bea
{\not\d}h=h^{-1}\Gamma^{\bf a}\d_a h,\quad
W_{ab}\Gamma^{\bf ab}\eps=-i\varepsilon_{abc}W_{ab}\Gamma^{\bf c}\eps=
-2i(\star_3 W)_c\Gamma^{\bf c}\eps
\eea
the ${\bf t}$ component of (\ref{KSeqnEps}) can be simplified:
\bea
&&\frac{1}{8h^{3}}W_{ab}\Gamma^{\bf ab}\eps+
\frac{1}{2}\d_{a}h^{-1}\Gamma^{\bf at}\eps
-\frac{1}{2}{\not\d}(\ln h+i
\Gamma_5\alpha)\Gamma_5\eps=0\nonumber\\
&&
\left[\frac{1}{8h^{3}}W_{ab}\Gamma^{\bf ab}
-\frac{i}{2}{\not\d}\alpha\right]\eps
+
\frac{1}{2}\d_{a}h^{-1}\Gamma^{\bf a}(\Gamma^{\bf t}+\Gamma_5)\eps=0\nonumber\\
&&\left[h^{-2}(\star_3 W)_c
+2{\d}_c\alpha\right]\Gamma^{\bf c}\eps=0
\eea
Since three spinors $\Gamma^{\bf c}\eps$ are independent, the last equation is equivalent to the relation
\bea\label{Vdual}
dV=-2h^2 \star_3 d\alpha.
\eea
Spacial components of (\ref{KSeqnEps}) give
\bea
&&\nabla_{\bf a}\eps+
\frac{1}{2}h^{-1}\d_a\left[i\alpha\Gamma_5-\ln h\right]\eps-
\frac{1}{2}{\not\d}(\ln h+i
\Gamma_5\alpha)\Gamma_5\Gamma^{\bf t}
\Gamma_{\bf a}\eps=0\nonumber\\
&&\nabla_{\bf a}\eps+\frac{1}{2h^2}\d_b h\Gamma^{\bf ba}\eps+
\frac{i}{2h}\d_b\alpha\Gamma_5\left[\delta_a^b-\Gamma^{\bf b}\Gamma_{\bf a}\right]\eps
=0\nonumber\\
&&
\frac{1}{h}\d_a \eps
-\frac{1}{4h^{3}}W_{ac}\Gamma^{\bf ct}\eps-\frac{i}{2h}\d_b\alpha\Gamma_5\Gamma^{\bf ba}\eps=0
\nonumber\\
&&
\label{temp61}
\d_a \eps
-\frac{1}{4h^{2}}\left[W+2h^2\star_3 d\alpha\right]_{ac}\Gamma^{\bf ct}\eps=0
\eea
We used the relation
\bea
\Gamma_5\Gamma^{\bf ba}\d_b\alpha=i\vep_{abc}\Gamma_{\bf t}\Gamma_{\bf c}\d_b\alpha=
-i(\star_3 d\alpha)_{ac}\Gamma^{\bf ct}\nonumber
\eea
Substitution of the duality relation (\ref{Vdual}) into (\ref{temp61}) leads to the conclusion that $\eps$ is a constant spinor. 

To summarize, we have demonstrated that the geometry (\ref{A3metric}) admits a Killing spinor $\eta$ 
given by (\ref{AnsAppSpin}) with constant $\eps$, as long as the duality relation (\ref{Vdual}) is satisfied. In the next subsection we will show that the self--duality condition for $F_5$ leads to additional relations between $h$ and $\alpha$.

\subsection{Equations for the field strength}

To find supersymmetric solutions of type IIB supergravity it is sufficient to solve the equations for the Killing spinors and the Bianchy identities for various fluxes \cite{bilinears}. Equation (\ref{Vdual}) gives the condition for existence of a Killing spinor in the geometry  (\ref{A3metric}), and now we will analyze the Bianchi identity for $F_5$. 

To determine the $F_{ab}$ components in (\ref{A3metric}), it is convenient to go back to (\ref{SlashF}),
\bea\label{temp62}
{\not F}=
-2(\Gamma^{\bf ta}{\d}_aA_t-i\Gamma_5\Gamma^{\bf ta}{\d}_a{\tilde A}_t),
\eea
and use a ``duality relation" for gamma matrices:
\bea
\Gamma_5\Gamma^{\bf ta}=-i\Gamma_{\bf 123}\Gamma^{\bf a}=-\frac{i}{2}
\varepsilon_{\bf abc}\Gamma^{\bf bc}.
\eea
Removing $\Gamma_5$ from (\ref{temp62}),
\bea
{\not F}=
-2\Gamma^{\bf ta}{\d}_aA_t+\varepsilon_{\bf abc}\Gamma^{\bf bc}{\d}_a{\tilde A}_t\,,
\eea
and reading off various components of $F$,
\bea
F_{\bf ta}=-\d_a A_t,\quad 
F_{\bf bc}=\varepsilon_{\bf abc}\d_a {\tilde A}_t,
\eea
we arrive at the final expression for the field strength $F$:
\bea
F=-\d_a A_t (dt+V)\wedge dx^a+h^2\star_3 d{\tilde A}_t\,.
\eea
The Bianchi identity implies a relation
\bea
d[dA_t\wedge V]+d[h^2\star_3 d{\tilde A}_t]=0,
\eea
and vector field $V$ can be eliminated from this equation using the duality condition (\ref{Vdual}):
\bea
&&dA_t\wedge 2h^2 \star_3 d\alpha+d[h^2\star_3 d{\tilde A}_t]=0,\nonumber\\
&&d\left[h^2\star_3(2A_t d\alpha+ d{\tilde A}_t)\right]=0.\nonumber
\eea
Substitution of the expressions (\ref{AnswApp}) for $A_t$ and ${\tilde A}_t$ leads to the final equation:
\bea
d\star_3d[h\cos\alpha]=0,
\eea
and the Bianchi identity for ${\tilde F}$ leads to a similar relation:
\bea
d\star_3d[h\sin\alpha]=0.
\eea
The last two equations imply that functions 
\bea
H_1=h\sin\alpha,\quad H_2=h\cos\alpha
\eea
are harmonic:
\bea
d\star_3d H_a=0
\eea
and it is convenient to parameterize geometry in terms of them. Once the harmonic functions are found, the vector field $V$ can be determined using the duality condition  (\ref{Vdual}):
\bea
dV=-2h^2\star_3d\alpha=-2\star_3[H_2 dH_1-H_1 dH_2]
\eea
Integrability condition for this equation is trivially satisfied:
\bea
d[H_2\star_3dH_1-H_1 \star_3dH_2]=dH_2\wedge \star_3dH_1-dH_1\wedge \star_3dH_2=0
\eea

\bigskip

\bigskip

To summarize, we have demonstrated that geometry (\ref{A3metric}) admits a Killing spinor (\ref{Gam5Proj}) and solves the Bianchi identities for $F$ and ${\tilde F}$ as long as
\bea
h=\sqrt{H_1^2+H_2^2},\quad \tan\alpha=\frac{H_1}{H_2},\quad dV=-2\star_3[H_2 dH_1-H_1 dH_2],
\eea
where $H_1$ and $H_2$ are harmonic functions on the three--dimensional base. 
The full expressions for the fields strength are given by
\bea\label{Strenght}
F&=&-\d_a A_t (dt+V)\wedge dx^a+h^2\star_3 d{\tilde A}_t\,,\nonumber\\
{\tilde F}&=&-\d_a {\tilde A}_t (dt+V)\wedge dx^a-h^2\star_3 d{A}_t\,.
\eea
The results of this appendix are summarized in equations (\ref{LocalSoln})--(\ref{HarmMain}). 

\section{Regularity analysis: lessons from \AdSS}
\renewcommand{\theequation}{C.\arabic{equation}}
\setcounter{equation}{0}
\label{AppRegul}

As we saw in section \ref{SecRegulAdS}, \AdSS\ geometry remains regular in spite of singularities in the harmonic functions. Of course, 
generic sources in $H$ would lead to singular geometries, and in this appendix we will use the insights from the \AdSS\ example to identify the allowed sources. The result of this heuristic analysis is summarized in equation (\ref{HarmIntAp}) and in  section \ref{SecRegul} we demonstrate that this setup indeed leads to regular solutions.

We begin this appendix with recalling the harmonic function (\ref{ComplAdS}) for \AdSS, rewriting it in Cartesian coordinates:
\bea\label{ComplAdSApp}
H=\frac{L}{\sqrt{x_1^2+x_2^2+(y-iL)^2}},
\eea
and writing this function as an integral over the sources of $H$. It is clear that such sources are located on the circle of radius $L$ in the $y=0$ plane, and we introduce a parameter $v$ along this circle. Then the singular curve can be written as
\bea
x_1\equiv F_1(v)=L\cos \frac{v}{L},\quad  x_2\equiv F_2(v)=L\sin \frac{v}{L},\quad  y\equiv F_3(v)=0,\quad  
0\le v<2\pi L.
\eea 
To simplify further analysis, we fixed the freedom in selecting the parameter $v$ by choosing the natural parameterization of the curve, where
\bea\label{Fdot2}
{\dot{\bf F}}^2=1.
\eea
Direct calculation demonstrates that function (\ref{ComplAdSApp}) can be written as
\bea\label{IntAdS}
H=\frac{1}{2\pi}\int 
\frac{\sqrt{({\bf r}-{\bf F})\cdot ({\bf r}-{\bf F}+{\bf A})}}{
({\bf r}-{\bf F})^2}dv\,,
\eea
where
\bea\label{AvecAdS}
{\bf A}=2L(\cos\frac{v}{L},\sin\frac{v}{L},i).
\eea
We will now generalize (\ref{IntAdS}) and (\ref{AvecAdS}) to an arbitrary profile ${\bf F}(v)$ and find the conditions on 
${\bf A}$ that make the solution (\ref{LocalSoln})--(\ref{HarmMain}) regular.  

Using the intuition from the AdS$_3\times$S$^3$ case \cite{Parad,lmm}, where harmonic functions for all 1/2--BPS states were given in terms of the same integrals involving string profiles, we will look for a similar construction here. 
Any one--dimensional curve in three dimensions $(x_1,x_2,x_3)$ can be parameterized by a profile ${\bf F}(v)$ that satisfies the normalization condition (\ref{Fdot2})\footnote{We assume that the curve is smooth, it has no cups and self--intersections.}. Motivated by the \AdSS\ example, we associate such a curve with a (complex) harmonic function
\bea\label{IntGen}
H=\frac{1}{2\pi}\int 
\frac{\sqrt{({\bf r}-{\bf F})\cdot ({\bf r}-{\bf F}+{\bf A})}}{
({\bf r}-{\bf F})^2}\sigma(v)dv\,,
\eea
where the complex vector ${\bf A}(v)$ is yet to be determined. At large values of $r$ the harmonic function approaches
\bea
H_\infty=\frac{1}{2\pi r}\int \sigma(v)dv
\eea 
so it is natural to interpret an arbitrary function $\sigma(v)$ as a dimensionless charge density. In the \AdSS\ example this function was equal to one. 

The Laplace equation for function the $H$ given by (\ref{IntGen}) implies a relation
\bea
\nabla^2\left[\frac{\sqrt{({\bf r}-{\bf F})\cdot ({\bf r}-{\bf F}+{\bf A})}}{
({\bf r}-{\bf F})^2}\right]=0,\nonumber
\eea
which is satisfied if and only if 
\bea\label{ConstrProf1}
({\bf A}\cdot {\bf A})=0,
\eea
so vector ${\bf A}$ must be complex. Regularity of the geometry (\ref{LocalSoln}) imposes additional constraints on this vector.

Let us consider a small vicinity of a point ${\bf F}(v_0)$ on the profile. To simplify notation, we shift 
parameter $v$ to set $v_0=0$ and choose a new coordinate system with the origin at ${\bf F}(0)$ and  $x_3$ axis pointing along ${\dot{\bf F}}(0)$. Further, we introduce polar coordinates $(R,\zeta)$ in the $(x_1,x_2)$ plane:
\bea\label{localCoord}
{\bf F}(0)=0,\quad {\dot F}_1(0)={\dot F}_2(0)=0,\quad x_1+ix_2=Re^{i\zeta}.
\eea 
The four--dimensional part of the metric (\ref{LocalSoln}) becomes
\bea
ds^2=-h^{-2}(dt+V)^2+h^2 \left[dR^2+R^2 d\zeta^2+dx_3^2\right].\quad
h^2=H{\bar H}
\eea 
To reproduce the regularization mechanism encountered for \AdSS, we require the leading contributions to $V$ and $h$ to have the form:
\bea\label{temp116}
V\simeq V_3 dx^3, \quad h^2\simeq \frac{Q^2}{R},\quad h^{-2}V_3\simeq {\tilde P},
\quad 
h^{-2}(V_3)^2-h^2\simeq P,
\eea
where $(Q,P,{\tilde P})$ approach constants as $R\rightarrow 0$. Clearly this requires 
${\tilde P}=-1$. Expression for $h^2$ determines $H$ up to a phase,
\bea
H\simeq \frac{Q}{\sqrt{R}}e^{i\Phi(R,\zeta)},\nonumber
\eea
and since $\zeta$--dependence of $\Phi$ must be linear, the Laplace equation completely fixes this phase:
\bea\label{leadH}
H\simeq \frac{Q}{\sqrt{R}}e^{i(\zeta-\zeta_0)/2}
\eea
We will now relate the complex parameter $Qe^{-i\zeta_0/2}$ (which generically depends on a point on the singular curve) with $\sigma$ and ${\bf A}$ which enter (\ref{IntGen}). 

To analyze the behavior of function (\ref{IntGen}) near the singular curve, we first focus on the $x_3=0$ plane, where
\bea\label{Hint}
{\tilde H}\equiv H|_{x_3=0}\simeq\frac{1}{2\pi}\int 
\frac{\sigma\sqrt{R^2-{R}(2{\bf F}-{\bf A}){\bf n}+{\bf F}({\bf F}-{\bf A})}}{
R^2-2{R}({\bf F}{\bf n})+{\bf F}^2}dv,\qquad {\bf n}\equiv \frac{\bf R}{R}.
\eea
To extract the leading contribution to the last expression, one is tempted to replace the profile by a straight line and integrate over $v$ from minus infinity to infinity. Unfortunately such replacement leads to unphysical divergences at large values of $|v|$. To cure this problem, we take a derivative of (\ref{Hint}) with respect to $R$,
\bea\label{eqn119}
\d_R{\tilde H} &\simeq&\frac{1}{2\pi}\int 
\frac{\sigma[2R-(2{\bf F}-{\bf A}){\bf n}]}{2\sqrt{R^2-{R}(2{\bf F}-{\bf A}){\bf n}+{\bf F}({\bf F}-{\bf A})}
[R^2-2{R}({\bf F}{\bf n})+{\bf F}^2]}dv\nonumber\\
&&-\frac{1}{2\pi}\int 
\frac{\sigma\sqrt{R^2-{R}(2{\bf F}-{\bf A}){\bf n}+{\bf F}({\bf F}-{\bf A})}}{
[R^2-2{R}({\bf F}{\bf n})+{\bf F}^2]^2}[2R-2({\bf F}{\bf n})]dv
\eea
 before replacing the profile by a straight line. Introducing expansions
\bea\label{temp120}
{\bf F}(v)={\dot{\bf F}}_0 v+\frac{v^2}{2}{\ddot{\bf F}}_0 v+\dots,\quad 
{\bf A}={\bf A}_0+{\dot{\bf A}}_0v+\dots
\eea
and recalling that ${\dot{\bf F}}^2=1$, we find the leading contribution to (\ref{eqn119}) at small values of $R$ by extending the integral over $v$ from minus infinity to infinity:
\bea\label{LeadDrH}
\d_R {\tilde H}&\simeq&\frac{1}{2\pi}\int_{-\infty}^\infty
\frac{\sigma({\bf A}_0{\bf n})}{2\sqrt{{R}({\bf A}_0{\bf n})-({\dot{\bf F}}_0{\bf A}_0)v}\,
[R^2+ v^2]}dv
\nonumber\\
&&-\frac{1}{2\pi}\int_{-\infty}^\infty
\frac{\sigma\sqrt{{R}({\bf A}_0{\bf n})-({\dot{\bf F}}_0{\bf A}_0)v}}{
[R^2+v^2]^2}2Rdv
\eea
The fact that all additional terms in the expansion (\ref{temp120}) lead to subleading contributions becomes especially obvious if one introduces a new integration variable $u=v/R$. 
Equation (\ref{LeadDrH}) has to match the $R$--derivative of (\ref{leadH}):
\bea\label{leadHdR}
\d_R H\simeq -\frac{Q}{2R^{3/2}}e^{i(\zeta-\zeta_0)/2}\,,
\eea
and this matching leads to a constraint on ${\bf A}_0$. Indeed, when ${\bf n}$ changes sign, $\zeta$ shifts by $\pi$, and expression (\ref{leadHdR}) is multiplied by $i$. Equation (\ref{LeadDrH}) reproduces this transformation law only if 
\bea\label{ConstrProf}
({\dot{\bf F}}_0{\bf A}_0)=0,
\eea 
i.e., if vector ${\bf A}$ belongs to a plane transverse to the profile. In this case the integral (\ref{LeadDrH}) simplifies,
\bea
\d_R {\tilde H}&\simeq&\frac{1}{2\pi}\int_{-\infty}^\infty
\frac{\sigma\sqrt{({\bf A}_0{\bf n})}[R^2+v^2-4R^2]}{2\sqrt{{R}}\,
[R^2+v^2]^2}dv=
\frac{1}{4\pi}\frac{\sigma\sqrt{({\bf A}_0{\bf n})}}{R^{3/2}}\int_{-\infty}^\infty \frac{u^2-3}{(u^2+1)^2}du
\nonumber\\
&=&\frac{1}{4\pi}\frac{\sigma\sqrt{({\bf A}_0{\bf n})}}{R^{3/2}}\left[-\frac{2u}{1+u^2}-\arctan u\right]_{-\infty}^\infty=-\frac{1}{4}\frac{\sigma\sqrt{({\bf A}_0{\bf n})}}{R^{3/2}},
\eea
and we recover (\ref{leadHdR}) with a particular value of $Q$. Substitution into (\ref{leadH}) gives the leading contribution to $H$ in terms of ${\bf A}$:
\bea\label{H124}
H\simeq\frac{\sigma\sqrt{({\bf A}_0{\bf n})}}{2\sqrt{R}}
\eea
We conclude that complex vector ${\bf A}$ is subject to two conditions (\ref{ConstrProf1}) and (\ref{ConstrProf}):
\bea\label{ConstrProfFin}
({\bf A}{\dot {\bf F}})=0,\qquad ({\bf A}{\bf A})=0.
\eea
Notice that these are the necessary conditions for regularity which come only from the analysis of the leading divergence, so we have proved that function $H$ can give rise to a regular geometry 
(\ref{LocalSoln})--(\ref{HarmMain}) only if it has the form
\bea\label{HarmIntAp}
H&=&H_1+iH_2=\frac{1}{2\pi}\int 
\frac{\sigma\sqrt{({\bf r}-{\bf F})\cdot ({\bf r}-{\bf F}+{\bf A})}}{
({\bf r}-{\bf F})^2}dv+H_{reg},\\
&&({\bf A}{\dot {\bf F}})=0,\qquad ({\bf A}{\bf A})=0.\nonumber
\eea
Here $H_{reg}$ is a regular part of the harmonic function, which is not fixed by our analysis of the divergent terms in a vicinity of the singular curve. 
In section \ref{SecRegul} we will start with imposing (\ref{HarmIntAp}), derive some additional constraints on $H$, and prove that the resulting harmonic functions always produce regular geometries via (\ref{LocalSoln})--(\ref{HarmMain}). 

We conclude this appendix by rewriting the constraints (\ref{ConstrProfFin}) on a complex vector field 
${\bf A}$ in terms of a real vector. Breaking ${\bf A}$ into a real and imaginary part as
\bea
{\bf A}={\bf S}+i{{\bf T}},
\eea
we conclude that ${\bf S}$ and ${{\bf T}}$ must be orthogonal to ${\dot{\bf F}}$ and 
\bea
({\bf S}{{\bf T}})=0.
\eea
Then, without loss of generality, we can set\footnote{There is only one alternative: 
${{\bf T}}={\dot{\bf F}}\times {\bf S}$, and it can be eliminated by reverting orientation of the curve. Notice that the normalization condition (\ref{Fdot2}) plays an important role in simplifying (\ref{temp128}).}
\bea\label{temp128}
{{\bf T}}={\dot{\bf F}}\times {\bf S}:\qquad {\bf A}={\bf S}+i{\dot{\bf F}}\times {\bf S}
\eea
Introducing $\zeta$ as an angle between ${\bf S}$ and ${\bf n}$, we can rewrite (\ref{H124}) in a form that matches (\ref{leadH}):
\bea\label{Hzeta}
H\simeq\frac{\sigma\sqrt{({\bf A}{\bf n})}}{2\sqrt{R}}=\frac{\sigma\sqrt{S}}{2\sqrt{R}}e^{i\zeta/2}\,.
\eea

\end{document}